\documentclass[aps,prb,twocolumn]{revtex4-2}

\usepackage{graphicx}
\usepackage{amsmath}
 \usepackage{amssymb}
\usepackage{amsfonts}
\usepackage{comment}
\usepackage{color}
\usepackage{hyperref}
\usepackage[makeroom]{cancel}
\usepackage[normalem]{ulem}
\usepackage{mathtools}
\usepackage{tikz}
\usepackage{bbold}

\usetikzlibrary{decorations.pathreplacing,decorations.markings, decorations.pathmorphing,arrows.meta}
\tikzset{snake it/.style={decorate, decoration=snake}}

\newcommand{\fsl}[1]{\ensuremath{\mathrlap{\!\not{\phantom{#1}}}#1}}% \fsl{<symbol>}

\renewcommand{\Re}{\operatorname{Re}}
\renewcommand{\Im}{\operatorname{Im}}

\tikzset{
  on each segment/.style={
    decorate,
    decoration={
      show path construction,
      moveto code={},
      lineto code={
        \path [#1]
        (\tikzinputsegmentfirst) -- (\tikzinputsegmentlast);
      },
      curveto code={
        \path [#1] (\tikzinputsegmentfirst)
        .. controls
        (\tikzinputsegmentsupporta) and (\tikzinputsegmentsupportb)
        ..
        (\tikzinputsegmentlast);
      },
      closepath code={
        \path [#1]
        (\tikzinputsegmentfirst) -- (\tikzinputsegmentlast);
      },
    },
  },
  mid arrow/.style={postaction={decorate,decoration={
        markings,
        mark=at position .57 with {\arrow[scale=1.8]{stealth}}
      }}},
}

\begin{document}

% \title{Non-equilibrium systems with long-ranged interactions}
% \title{Non-equilibrium diagrammatic approach to strongly interacting polaritons}
\title{Non-equilibrium diagrammatic approach to strongly interacting photons}
 
\author{Johannes Lang$^{1,2}$}
\author{Darrick E. Chang$^{3,4}$}
 \author{Francesco Piazza$^{2}$}

\affiliation{$^{1}$\small Physik Department, Technische Universit\"at M\"unchen, 85747 Garching, Germany}
\affiliation{$^{2}$\small Max-Planck-Institut f\"ur Physik komplexer Systeme, 01187 Dresden, Germany}
\affiliation{$^{3}$\small ICFO-Institut de Ciencies Fotoniques, The Barcelona Institute of Science and Technology, 08860 Castelldefels (Barcelona), Spain}
\affiliation{$^{4}$\small ICREA-Instituci\'o Catalana de Recerca i Estudis Avan\c{c}ats, 08015 Barcelona, Spain}

\begin{abstract}
 We develop a non-equilibrium field-theoretical approach, based on a systematic diagrammatic expansion, for strongly interacting photons in optically dense atomic media.
We consider the case where the characteristic photon-propagation range $L_P$ is much larger than the interatomic spacing $a$ and where the density of atomic excitations is low enough to neglect saturation effects. In the highly polarizable medium the photons experience nonlinearities through the interactions they inherit from the atoms. If the atom-atom interaction range $L_E$ is also large compared to $a$, we show that scattering processes with momentum transfer between photons are suppressed by a factor $a/L_E$. We are then able to perform a self-consistent resummation of a specific (Hartree-like) diagram subclass and obtain 
quantitative results in the highly non-perturbative regime of large single-atom cooperativity. Here we find important, conceptually new collective phenomena emerging due to the dissipative nature of the interactions, which even give rise to novel phase transitions. The robustness of these is investigated by inclusion of the leading corrections in $a/L_E$.
We consider specific applications to photons propagating under EIT conditions along waveguides near atomic arrays as well as within Rydberg ensembles.
\end{abstract}

\maketitle

\section{Introduction}
The possibility to implement interactions between photons in the quantum regime is recently attracting a lot of interest \cite{chang_vuletic_lukin_2014}. One reason is technological, as photon-photon interactions are essential for quantum information processing and would allow to build quantum networks exploiting the ability of photons to efficiently carry information over long distances \cite{quantum_internet}. Interacting photons are also promising for the creation of synthetic quantum matter, like superfluids \cite{carusotto_rev_2013}, or gapped \cite{hartmann_2008,hafezi_chempot_2015,lebreuilly_2017,simon_mott_2018} and even topological \cite{topol_photonics_rev_2018} phases. 

From a more fundamental, many-body perspective, an ensemble of strongly interacting photons shows crucial differences from any condensed-matter counterpart and is therefore likely to show novel collective phenomena which have no analog in conventional materials. The first such difference is that the photon number is never conserved so that repumping is needed to compensate losses and reach a driven-dissipative steady state, the latter thus generically being far away from thermal equilibrium. Moreover, photons do not interact in vacuum and need a material to mediate their mutual interactions. 
The electromagnetic (EM) modes hybridize with the material giving rise to polaritonic excitations.
Here we concentrate on materials made of uncharged but polarizable atoms, where the polaritons (and therefore the photons) inherit their interactions from the latter. This implies a second important feature, namely that the interaction between two photons is a higher order process, requiring the intermediate excitation of the atomic dipoles. Interactions between polaritons in such systems are also  naturally long-ranged (as the relevant electromagnetic modes typically extend over many atoms) and retarded (as the characteristic time scales of photons and atoms can be respectively tuned to be comparable). 
Finally, interactions inherited from atomic dipoles can be strongly dissipative due to the spontaneous decay of excited atomic levels. This feature in particular has been shown to be capable of introducing novel many-body phenomena, whereby correlations can be induced by dissipation \cite{carusotto_2009,hartmann_2011,zeuthen_EFT,firstenberg_review_2016,Peyronel2012}.

The implementation of strong interactions between photons in the quantum regime typically requires significant single-photon nonlinearities induced by a large interaction cross-section between a single photon and a single atom \cite{chang_vuletic_lukin_2014}, which poses an experimental challenge. It can be overcome by light-confinement via evanescent waves or optical resonators, and/or by providing the atoms with strong, long-ranged interactions preventing multiple atoms to be excited within a large radius, as done by using Rydberg levels\cite{zeuthen_EFT,firstenberg_review_2016,Peyronel2012}.

The theoretical description of such a strongly-interacting, driven-dissipative system of photons in the many-body regime constitutes a challenging task as well. In particular, the large interaction cross sections prevent a perturbative treatment, the driven-dissipative nature does not allow to exploit fluctuation-dissipation relations and prevents for instance the application of Monte Carlo methods, while the long-range interactions additionally hinder an efficient employment of tensor network methods, even in one spatial dimension.
A few theoretical approaches have been developed for the few-body regime \cite{pohl_ryd_old,bienias2014scattering,caneva2015quantum,shi2015multiphoton,moos_2015}, while effective field theories have been applied in the many-body regime \cite{gullans_EFT,zeuthen_EFT,photon_prop_ryd_authors_2018, Nikoghosyan2010,Mahmoodian2018}.

Here, we introduce a systematic, diagrammatic approach for the computation of non-equilibrium correlators for a many-body system of strongly interacting photons in an optically dense medium. If the characteristic photon propagation range $L_P$ in the medium is much larger than the spacing $a$ between the atoms, we show that a controlled diagrammatic expansion in powers of $a/L_P$ can be performed, even if the collective light-matter coupling $g_P$ within the mode volume of the photon is large. 
This perturbative expansion in $a/L_P$ is always valid when the single-atom cooperativity $C_P^{\rm sa}=(g_P^2/\gamma\kappa)(a/L_P) $ is much smaller than unity, where $\gamma,\kappa$ are the characteristic dissipation rates of excited atomic levels and photons, respectively. 
The quantitative validity of our approach can however even be extended to a regime of large single atom cooperativities $C_P^{\rm sa}\gtrapprox 1$, provided that the density of atomic excitations is low enough to neglect saturation effects. In such a situation, photons would not experience any nonlinearity or interactions, unless the atoms experience additional, mutual interactions which the photons can inherit. If inter-atomic interactions are present and if their range $L_E/a\gg 1$ is large, we show that the subclass of diagrams describing scattering processes with momentum transfer between photons is suppressed by a factor $\sim a/L_E$ with respect to the remaining Hartree-like diagrams. In this case we are able to perform a self-consistent resummation of the Hartree-like diagram subclass and obtain quantitative results in a strongly non-perturbative regime, which indeed shows important collective behavior and even phase transitions (see also \cite{lang_EIT_short} for a discussion focusing on a specific example).

From a quantum-field-theory perspective, this work constitutes a first attempt to develop a non-relativistic version of Quantum Electrodynamics (QED) where the matter degrees of freedom are dipoles instead of charged electrons, with two further important differences: i) the photons are driven and (partially) confined in space, and ii) the light-matter coupling is far away from the perturbative regime. We therefore believe that our work establishes the critical framework that will enable the application of diagrammatic techniques to a wide variety of problems of interest within many-body quantum optics.

In the following, we illustrate specific applications to experiments involving interactions mediated through waveguide photons, for example in photonic-crystal-waveguides \cite{kimble_2014_crystal,Douglas2015a}, as well as Rydberg interactions \cite{grangier_2012,Dudin887,weide_2013,firstenberg_2013, weath_rybist_exp_2013,firstenberg_review_2016}. For concreteness, we consider atomic level structures allowing the photons to propagate under electromagnetically-induced-transparency conditions \cite{Fleischhauer2002,Fleischhauer2005}.

The paper is structured as follows: In section \ref{sec:intro} we introduce the $a/L$-expansion in general terms, which is then formalized in the language of non-equilibrium field theory and exemplified using a minimal model of two-level atoms in section \ref{sec:formalism}. After briefly revisiting the phenomenon of electromagnetically induced transparency using our diagrammatic approach (section \ref{sec:EIT}) and the general structure of interactions between polaritons (section \ref{sec:1oL_all}), the implications of strong interactions are discussed, first on the Hartree level (sections \ref{sec:Gerry} and \ref{sec:LE_inf}) and finally including all scattering effects to order $a/L_E$ in section \ref{sec:LE_finite}. In section \ref{sec:Rydberg}, we conclude with a short comparison between the case of waveguide-mediated interactions and the case of Rydberg inter-atomic interactions, demonstrating the wide applicability of the presented approach.

\section{A controlled expansion for strong light-matter interactions}\label{sec:intro}
\begin{figure}[htp]
\begin{center}
\includegraphics[width=\columnwidth]{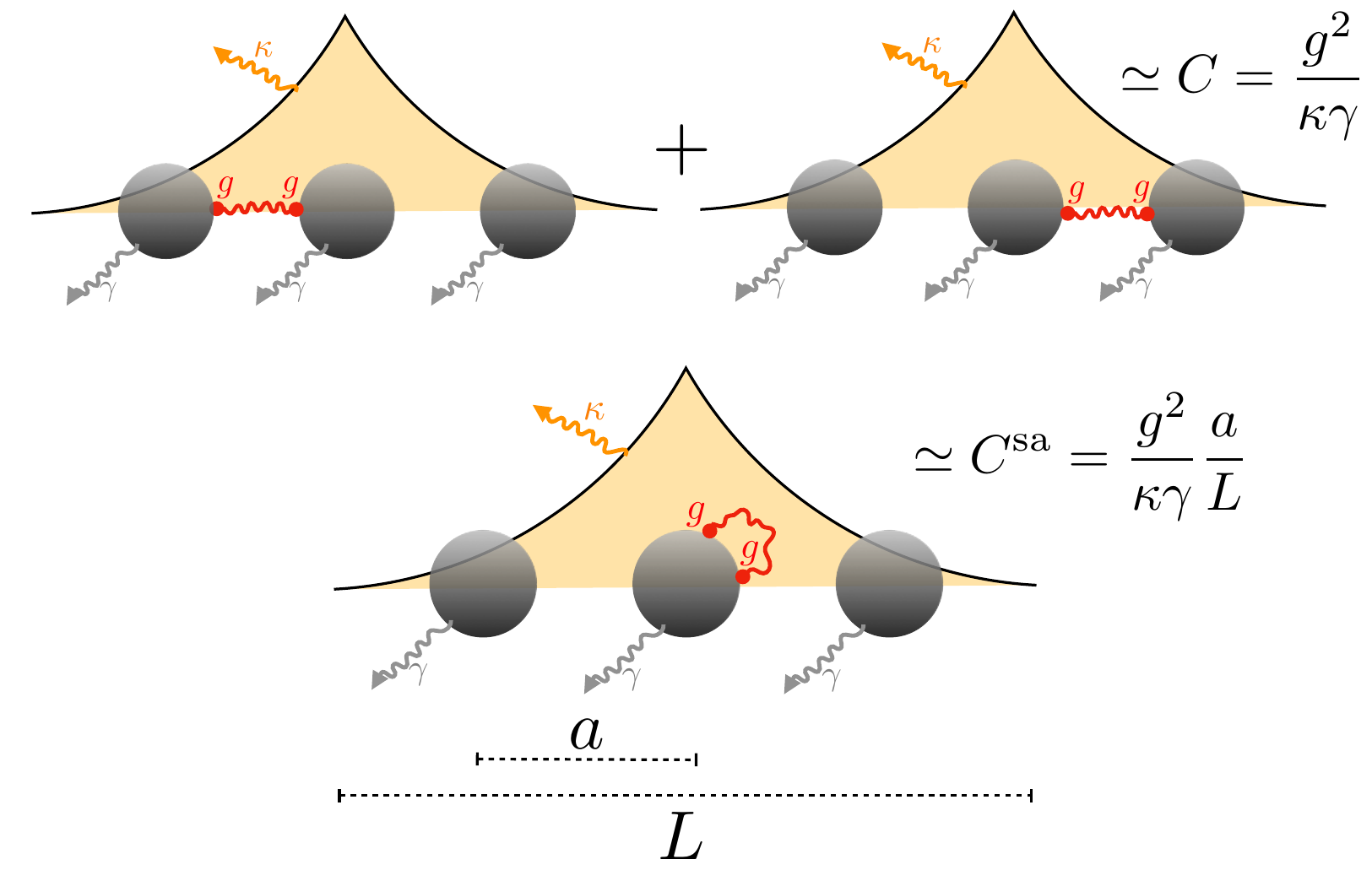}
\caption{Basic idea underlying our diagrammatic expansion. Exchanges of excitations between atoms are preferred over self-interactions of individual atoms. Red arrows indicate the propagation of the photonic excitation emitted by the central atom into the exponentially localized wave function illustrated in yellow. Decay channels are shown as orange and gray arrows.}
\label{fig:basic_idea}
\end{center}
\end{figure}

The basic idea underlying our diagrammatic approach can be understood in quite general terms.
Let us consider a system of two completely different types of particles, which we will for later convenience call photons and atoms. For now we will keep these particles as generic as possible and only fix their mass: Photons are very light (or even massless) and therefore propagate very fast and over long distances, whereas atoms are considered as comparatively heavy, localized and thus slowly moving. Furthermore, neither atoms nor photons shall interact among themselves, that is, atoms can only interact via the exchange of photons and photons only via the non-linear susceptibility of the atomic medium, which results in a Yukawa-type coupling. The stark contrast between the two free theories of atoms and photons allows for a controlled expansion, even in the case of strong collective light-matter interactions. This is due to the large effective mode volume of the photon, suppressing the coupling rate between photons and individual atoms, thereby providing a useful expansion parameter.\\
To make this argument more concrete, let us for simplicity consider the specific case of a quasi-1D continuum of photon field modes (as defined by a physical waveguide, or a focused beam). These modes have a group velocity $c$ and couple at a rate $g$ to the collection of all atoms within an effective mode volume $L$. We furthermore assume that the atoms are confined to fixed positions in a one-dimensional chain with characteristic spacing $a$. Moreover, photons are lost out of the one-dimensional medium at a rate $\kappa$, and an excited atom can independently decay into channels other than the 1D continuum of interest at a rate $\gamma$. In this case the photon effective mode volume is given by the group velocity $c$ times the characteristic time $1/\kappa$ spent by the photon in the medium i.e. $L=c/\kappa$. Note that from the point of view of the atoms $L$  corresponds to the effective interaction range. A pictorial representation of this simple model is given in Fig.~\ref{fig:basic_idea}.
In the first process shown in the figure, a photon is exchanged between an arbitrary pair of atoms. The importance of this process as a modification to the non-interacting dynamics can be estimated from collective cooperativity $C=g^2/(\gamma\kappa)$, which compares the rate of the coherent photon exchange with the competing single-particle dissipative processes. If this dimensionless quantity becomes of order unity, the naive expansion in powers of $g$ breaks down. In the second process the photon is emitted and absorbed by the same atom. Compared to the first process, this one is thus less probable for any finite $L$. Correspondingly, the figure of merit for this type of self-interaction is the single atom cooperativity $C^\text{sa}=g^2 a/(c \gamma)$, which describes the branching ratio of single-atom emission into the waveguide versus free space as determined by Fermi's golden rule. Compared to $C$ in $C^\text{sa}=g^2a/(\gamma\kappa L)$ the collective coupling rate $g$ has been replaced by the single atom coupling $g/\sqrt{L/a}$. We thus see that even at strong collective coupling, when an expansion in the coupling constant $g$ is not applicable, an expansion in the inverse interaction range $a/L$ (which at order $a/L$ is equivalent to an expansion in the single-atom cooperativity) can still be possible.

Clearly this very basic argument can be extended to include all types of processes and interactions, provided the atoms are well localized and slow compared to their exchange particles. In every order of the expansion in the coupling rate, those processes involving the maximal number of atoms will be most important. In the following, we will elevate this argument to a formal level, making it amenable to the use of Feynman diagrams. We will then see, that it is actually possible to switch from an expansion in $g$ to a self-consistent description in powers of $a/L$. In other terms, strongly coupled theories are accessible to a controlled field-theoretic treatment, given that the interactions are sufficiently long-ranged.\\
% \fc{I would strongly suggest to remove the paragraph below. It is more confusing than not and kind of immediately kills the positive climax we built so far!}

Before we press on, however, a word of caution is in order, as there are several scenarios where the simple arguments presented here either break down or need to be refined. The most important of these cases are closed systems. % , where in low dimensions Luttinger liquids or Peierls instabilities are ubiquitous, independent of the range of interactions.}\jlc{If this sentence is removed entirely, the next sentence makes no sense.}
The reason being that without the presence of dissipation no meaningful equivalent of the mode volume $L$ can be defined. In fact, the ostensible equivalent of a mean free path is fallacious, since at its end the photon isn't lost, but merely scattered and therefore still available for further interactions. Additionally, fine tuning at critical points in open systems can also give rise to vanishing losses, which can result in effectively increased cooperativites. Finally, special care has to be taken when treating systems where conservation laws of the non-interacting system are broken by interactions. Since the non-interacting degrees of freedom have no loss rates, the expansion has to be extended to include at least the lowest order at which those arise.

\section{Diagrammatic approach to non equilibrium Green's functions}\label{sec:formalism}
Building on the newly gained understanding that a physical system is suitable for a $1/L$ expansion as long as it exclusively couples degrees of freedom that are well localized in position space to others that are tightly confined in the conjugate momentum space, we will now be more concrete and apply this approach to photons in optical waveguides coupled to an array of two-level atoms. This will allow us to give a pedagogical introduction to the concepts and techniques required to treat more complex systems. 
Furthermore, the configuration considered here can be directly extended to scenarios where large cooperativities are experimentally accessible. Such setups include for instance atoms trapped within the evanescent wave of photonic crystal waveguides (PCWs) \cite{kimble_2014_crystal,Douglas2015a} or tapered-nanofiber waveguides (TNWs) \cite{rauschenb_2010,appel_2016,laurat_2016,rauschenbeutel_2018}. A quantitative description of these setups will be provided in Sec.~\ref{sec:four_level}.
The concepts introduced in this section are however far more general and can be applied in similar ways to any system of interacting polaritons. We will demonstrate this on the example of a gas of Rydberg atoms in Sec.~\ref{sec:Rydberg}.\\

\subsection{Minimal model: A chain of two-level atoms}
\label{sec:appl_twolevel}
\begin{figure}[htp]
\begin{center}
\includegraphics[width=\columnwidth]{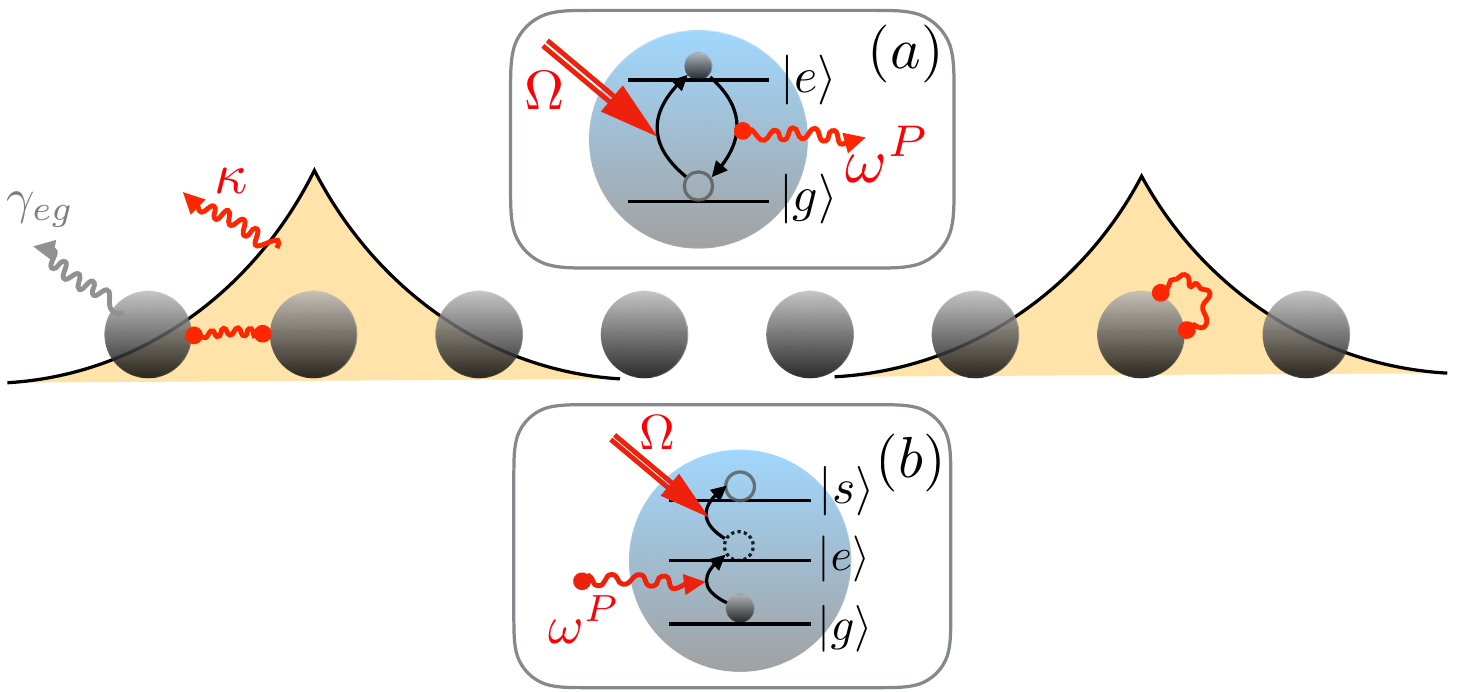}
\caption{Schematic illustration of the system of atoms coupled to the field of a waveguide, which will be used to demonstrate the formalism and the expansion in the inverse interaction range. Probe-photons with characteristic propagation range indicated by the yellow-shaded area interact with atoms and are dissipated at a rate $\kappa$. Insets show the atomic level structure. (a) Two-level atoms considered in section \ref{sec:formalism}: the $g-e$ transition is driven by an external laser $\Omega$ and also couples to the probe photons $\omega^P$. The excited state $|e\rangle\rangle$ decays spontaneously by emitting a photon outside the waveguide with a rate $\gamma_{eg}$ and is also subject to dephasing with a rate $\gamma_{e_{\rm deph}}$ (not shown). (b) Three-level atoms implementing the EIT considered in section \ref{sec:EIT}: the external laser $\Omega$ couples $|e\rangle$ to a third metastable state $|s\rangle$.}
\label{fig:setup}
\end{center}
\end{figure}

We consider a system of atoms fixed in a periodic one-dimensional arrangement and coupled to the propagating photon mode of a waveguide with dispersion $\omega^P_k$ as shown in Fig.~\ref{fig:setup}. The ground state of each atom will be denoted by $|g\rangle$ and the excited, unstable state with energy $\omega_e$ by $|e\rangle$. To compensate the inevitable emission of photons (parametrized by the decay rate $\kappa$), the atomic transition is driven by a laser with energy $\omega_L$ and Rabi amplitude $\Omega$. Since we also allow for dephasing and decay of the excited atomic state, the full dynamics of the system is described by
\begin{align}
\frac{d\rho}{dt}=-\frac{i}{\hbar}[\hat{H},\rho]+\mathcal{L}_{\gamma_{eg}}\rho+\mathcal{L}_{\gamma_\text{deph}}+\mathcal{L}_{\kappa}\rho
\end{align}
with Hamiltonian
\begin{align}\label{eq:2Level_Hamiltonian}
\begin{split}
\hat{H}=&\hbar \left[\sum_z \left\{\omega_e \sigma_{ee}(z)+\left(\Omega e^ {-i\omega_L t}\sigma_{eg}+h.c.\right)\right.\right.\\ &\left.\left.+\int_{-\pi}^\pi \frac{dk}{2\pi} \left(\omega^P_k\hat{b}_P(k)^\dagger \hat{b}_P(k)\right.\right.\right.\\ &\left.\left.\left.\qquad+g\left(\hat{b}_P(k) e^{i k z} u_k(z) \sigma_{eg} + h.c.\right)\right)\right\}\right]
\end{split}
\end{align}
and Lindblad operators
\begin{subequations}
\begin{align}\label{eq:quad_losses}
\mathcal{L}_{\gamma_{eg}}\rho&=-\hbar\sum_z \frac{\gamma_{eg}}{2}\left(\left\{\sigma_{ee},\rho\right\}-2\sigma_{ge}\rho\sigma_{eg}\right)\\
\mathcal{L}_{\gamma_\text{deph}}\rho&=-\hbar\sum_z \frac{\gamma_{e_\text{deph}}}{2}\left(\left\{\sigma_{ee},\rho\right\}-2\sigma_{ee}\rho\sigma_{ee}\right)\\
\mathcal{L}_{\kappa}\rho&=-\hbar\!\int_k\!\frac{\kappa}{2}\left(\!\left\{\hat{b}_P(k)^\dagger \hat{b}_P(k),\rho\right\}\!-2\hat{b}_P(k)\rho\hat{b}_P(k)^\dagger\!\right)\!,
\end{align}
\end{subequations}
which account only for independent emission from each atom, neglecting collective effects \cite{asenjo_radiance_2017}.
Here $u_k(z)$ represents the periodic part of the photonic Bloch function with quasi-momentum $k$. We make use of the standard convention for the thermodynamic limit in a crystal with lattice constant $a=1$, namely $\sum_z e^{ikz}=2\pi\delta(k)$ and introduce the notation $\int_k = L \int \frac{dk}{2\pi}$. Note, that the rotating wave approximation employed in the derivation of the Hamiltonian \eqref{eq:2Level_Hamiltonian} is highly justified throughout this article. The extremely small shift of the atomic resonance frequency resulting from counter-rotating terms $\sim 1\,\text{MHz}$ has been determined by realistic numerical calculations \cite{chang_njp_photcrys_2013} and can be absorbed into the transition frequency $\omega_e\sim 400\,\text{THz}$. Furthermore all inverse lifetimes resulting from photon emission and interactions alike will remain small compared to the transition frequency.\\
%The (complex) energy shifts that nearby states $|s\rangle$ can induce in one another via the coupling to the non-propagating mode $\omega_k^E$ with adjustable range, shifts the EIT window for the propagating mode depending on the position of all other polaritonic excitations. This unconventional type of interaction will be treated as self energy insets to next to leading order in the photonic mode volumes, modifying the RPA diagrams that describe photon-atom-mixing.

Our diagrammatic approach will be formulated within a non-equilibrium functional-integral formalism. However, since
for each atom the Hilbert space is finite, more precisely the occupation of both states sums up to one, the representation of atomic operators in a form that is convenient for the functional integral formulation has to be given some thought. Here we will restrict ourselves to the limit of a small density of excited atoms, where the saturation effects associated with the finiteness of the local Hilbert space of the medium can be neglected. As a result, the Schwinger-boson representation without explicit restriction of the boson number of each atomic transition will suffice. Moreover, a distinction between decay and dephasing will no longer be necessary, the leading effect of both processes being the linewidth $\gamma_e=\gamma_{e_\text{deph}}+\gamma_{eg}$ acquired by state $|e\rangle$ \cite{Fleischhauer2005}. In particular, we will use the approximate expression
\begin{align}
\sigma_{\mu\nu}=\hat{a}^\dagger_\mu\hat{a}_\nu\,,
\end{align}
where $\{\mu,\nu\}\in\{g,e\}$ and $\hat{a}$ and $\hat{a}^\dagger$ are bosonic annihilation and creation operators respectively.
Clearly this approximation allows for an unrestricted occupation of any state of any atom -- a shortcoming which will later be mitigated by the application of non-linear Feynman rules.
Since treating spins within a functional integral formulation is considerably more complicated than bosons \cite{DallaTorre2016,Shchadilova2018}, this transformation is crucial for the tractability of the calculations that lie ahead. Within this linear regime the Hamiltonian part of the system is given by
\begin{align}\label{eq:H}
\begin{split}
\hat{H}=\hbar &\left[\sum_z \left\{\omega_e \hat{a}^\dagger_e(z)\hat{a}_e(z)\right.\right.\\&\left.\left.+\int_{-\pi}^\pi \frac{dk}{2\pi} \left(\omega^P_k\hat{b}_P(k)^\dagger \hat{b}_P(k)\right.\right.\right.\\&\left.\left.\left.+g\left(\hat{b}_P(k) e^{i k z} u_k(z) \hat{a}^\dagger_e(z) \hat{a}_g(z) + h.c.\right)\right)\right\}\right],
\end{split}
\end{align}
while the atomic losses are treated in a simplified manner that reproduces the correct linewidth
\begin{align}\label{eq:atomloss}
\mathcal{L}_{\gamma_e}\rho&=-\hbar\sum_z \frac{\gamma_e}{2}\left(\left\{\hat{a}_e^\dagger(z)\hat{a}_e(z),\rho\right\}-2\hat{a}_e(z)\rho\hat{a}_e^\dagger(z)\right).
\end{align}
% \jlc{We could add a discussion of losses here or in the appendix, however with bosons instead of fermions this actually doesn't work particularly well and a proper derivation has to be replaced by some rather crude limits.}
The fact that the linearized description of decay of excited atoms violates atom number conservation is an unphysical feature of this approximation. Since a more rigorous modeling of spontaneous decay, e.g. via the Lindblad operator $\hat{a}_e\hat{a}^\dagger_g$, is diagrammatically equivalent to a two-body interaction, which significantly complicates a systematic treatment, we compensate these spurious atom losses by fixing the density of atoms in the ground state. As we will see later, as long as saturation effects are negligible, this description of the incoherent dynamics of the atoms in combination with a specific selection rule for the Feynman diagrams becomes exact (see Sec.~\ref{sec:NLFR}).

\subsection{Keldysh formulation}\label{sec:Keldysh}
In order to recast our non-equilibrium problem into a functional-integral form, we choose the real-time Keldysh contour (c.f.~\cite{kamenev_book} and specifically for driven-dissipative systems~\cite{sieberer_Keldysh_review}). This contour is directly obtained by writing the expectation value of an operator $\mathcal{\hat{O}}$ at a time $t$ by time-evolving the system from the distant past:
\begin{align}\label{eq:expval}
\langle \mathcal{\hat{O}} \rangle(t)=\frac{\text{Tr}\left(\hat{\mathcal{U}}_{-\infty,t}\hat{\mathcal{O}}\hat{\mathcal{U}}_{t,-\infty}\hat{\rho}(-\infty)\right)}{\text{Tr}\left(\hat{\rho}(-\infty)\right)}\,.
\end{align}
Here $\text{Tr}(\cdot)$ is the trace, $\hat{\mathcal{U}}_{t,t'}$ is the time evolution operator from time $t'$ to $t$ and $\hat{\rho}(-\infty)$ is the density matrix of the system in the distant past.\\
Our goal is to compute the single-particle Green's functions or propagators.
Due to our system being driven-dissipative, we cannot assume thermal equilibrium i.e. detailed balance, such that there are in principle two independent propagators, the retarded
\begin{align}
i\left[G_a^R\right]_{ij}(x,x')=\theta(t-t')\left\langle\left[\hat{a}_i(x),\hat{a}_j^\dagger(x')\right]\right\rangle\,,
\end{align}
and the Keldysh Green's function
\begin{align}
i\left[G_a^K\right]_{ij}(x,x')=\left\langle\left\{\hat{a}_i(x),\hat{a}_j^\dagger(x')\right\}\right\rangle\,,
\end{align}
with $i,j\in\{g,e\}$ labeling the atomic states and
$x=(t,z)$ being the space-time coordinate. The same construction and in fact all the following steps apply also to the definition of the photonic equivalents $G_\text{ph}^R(x,x')$ and $G_\text{ph}^K(x,x')$. We therefore focus on the atomic sector and treat the time-evolution of these expectation values by means of the coherent-state functional integral. In doing so one inserts resolutions of unity in terms of coherent states spaced in infinitesimal timesteps along the time-evolution \cite{Altland_book}. Evaluation of the resulting matrix elements then replaces the operators $\hat{a}_j(x)$ and $\hat{a}_j(x)^\dagger$ by the field $a_j(x)$ and its complex conjugate $\bar{a}_j(x)$. However, according to \eqref{eq:expval}, one has to evolve the system both forward and backward in time, which requires us to split each field into a part on the forward branch (denoted with a superscript $+$) and one on the backward branch (labeled by a $-$), whereby the Green's functions are now given by
\begin{subequations}
\begin{align}
i\left[G_a^R\right]_{ij}(x,x')=&\theta(t-t')\nonumber\\&\times\left(\langle a_i^-(x)\bar{a}_j^+(x')\rangle-\langle a_i^+(x)\bar{a}_j^-(x')\rangle\right)\\
i\left[G_a^K\right]_{ij}(x,x')=&\langle a_i^-(x)\bar{a}_j^+(x')\rangle+\langle a_i^+(x)\bar{a}_j^-(x')\rangle\,.
\end{align}
\end{subequations}
Once one performs the so called Keldysh rotation to \emph{quantum} and \emph{classical} fields
\begin{subequations}
\begin{align}
a^\text{q}_j(x)&=\frac{1}{\sqrt{2}}\left(a_j^+(x)-a_j^-(x)\right)\\
a^\text{cl}_j(x)&=\frac{1}{\sqrt{2}}\left(a_j^+(x)+a_j^-(x)\right)\,,
\end{align}
\end{subequations}
of which the former have identically vanishing correlations: $\langle a^\text{q}_i(x)\bar{a}^\text{q}_j(x')\rangle\equiv 0$, the retarded and Keldysh Green's functions take the much simpler forms
\begin{subequations}
\begin{align}
i G_{ij}^R(x,x')\equiv i\left[G_a^R\right]_{ij}(x,x')&=\langle a^\text{cl}_i(x)\bar{a}^\text{q}_j(x')\rangle\\
i G_{ij}^K(x,x')\equiv i\left[G_a^K\right]_{ij}(x,x')&=\langle a^\text{cl}_i(x)\bar{a}^\text{cl}_j(x') \rangle\,.
\end{align}
\end{subequations}
Since additionally the advanced Green's function $G^A(x,x')=\langle a^\text{q}_i(x)\bar{a}^\text{cl}_j(x')\rangle$ satisfies $G^A_{ij}(x,x')=\left[G^R_{ij}\right]^*(x',x)$, where $(\cdot)^*$ denotes the complex conjugation, no further independent propagators exist.
For the non-interacting atoms coupled to the coherent laser fields, the inverse retarded Green's function reads
\begin{align}
\begin{split}
&\left[\tilde{G}_{a,0}^R\right]^{-1}(\omega,\omega')=\\&\begin{pmatrix}
\left(\omega-\omega_e+i\frac{\gamma_e}{2}\right)\delta(\omega-\omega')&&-\Omega\delta(\omega-\omega'+\omega_L)\\
-\Omega\delta(\omega-\omega'-\omega_L)&&(\omega+i\frac{\epsilon}{2})\delta(\omega-\omega')\\
\end{pmatrix}\,,
\end{split}
\end{align}
where we used the basis
\begin{align}
\mathbf{a}^\text{(q,cl)}(\omega,z)=\begin{pmatrix}
a_e(\omega,z)\\
a_g(\omega,z)\\
\end{pmatrix}^\text{(q,cl)}\,.
\end{align}
Note, that we use a lower index $0$ to indicate bare Green's functions, i.e. those without self-energy corrections induced by interactions (see below). As it turns out, the explicit time-dependence of $\hat{H}$ caused by the external laser field breaks time translation invariance in the retarded Green's function. One can, however, overcome this obstacle by transforming into a rotating frame, where the state $|e\rangle$ rotates at the frequency of the laser $\omega_L$. 
Within this frame the atomic Green's function is once again time translation invariant, that is $G_{a,0}^{-1}(\omega,\omega')=G_{a,0}^{-1}(\omega)\delta(\omega-\omega')$ with
\begin{align}
  \label{eq:bare_2level_atom_GF}
\left[G_{a,0}^R\right]^{-1}(\omega)=\begin{pmatrix}
\omega-\Delta_e+i\frac{\gamma_e}{2}&&-\Omega\\
-\Omega&&\omega+i\frac{\epsilon}{2}\\
\end{pmatrix}\,,
\end{align}
and the fields shifted accordingly in frequency:
\begin{align}
\mathbf{a}^\text{(q,cl)}(\omega,z)=\begin{pmatrix}
a_e(\omega+\omega_L,z)\\
a_g(\omega,z)\\
\end{pmatrix}^\text{(q,cl)}\,.
\end{align}
Here the detuning $\Delta_e=\omega_e-\omega_L$ between the laser frequency and atomic transition has been introduced. In order to avoid confusion, throughout the remainder of this manuscript we will exclusively work in the rotating frame.
The corresponding Keldysh component of the inverse Green's function within the same frame of reference is then given by
\begin{align}
D_{a,0}^K(\omega)=\begin{pmatrix}
i\gamma_e&&0\\
0&&(3-2n_V)i\epsilon\\
\end{pmatrix}\,.
\end{align}
It should be pointed out, that the factor $3-2n_V$ in the ground-state sector accounts for the occupation of this mode with a homogeneous number density of lattice defects or vacancies $n_V \in [0,1]$. Thus for $n_V = 0$ the ground-state is homogeneously occupied with one atom per site, as can be seen from $n_g=-1/2+i\int \frac{d\omega}{4\pi}G_{gg,0}^K=1-n_V=1$.

Applying the same rotation to the inverse photon Green's function the simple expressions 
\begin{subequations}
\begin{align}
\left[G_{\text{ph},0}^R\right]^{-1}(\omega,k)&=
\omega-\Delta(k)+i\frac{\kappa}{2}\\
D_{\text{ph},0}^K(\omega,k)&=i\kappa\,
\end{align}
\end{subequations}
with the detuning $\Delta(k)=\omega(k)-\omega_L$ are obtained.\\
% Independent of all light sources and loss channels
Making use of the above notation, the non-interacting part of the action $S=S_0+S_\text{int}$ can be fully expressed in terms of the bare atomic (subscript $a$) and photonic (subscript $\text{ph}$) Green's functions as
\begin{subequations}
\begin{align}
S_0=&\hbar\!\int\!\frac{d\omega}{2\pi}\left(\!\sum_{z}\mathbf{a}^*\!(\omega,z)\mathcal{G}_{a,0}^{-1}(\omega)\mathbf{a}(\omega,z)\right.\\&\qquad\quad\left.+\!\!\int\!\frac{dk}{2\pi}\mathbf{b}_P^*\!(\omega,k)\mathcal{G}_{\text{ph},0}^{-1}(\omega,k)\mathbf{b}_P(\omega,k)\!\!\right).
\end{align}
\end{subequations}
Here $\mathbf{a}=\left\{\mathbf{a}^\text{cl},\mathbf{a}^\text{q}\right\}$ and $\mathbf{b}_P=\left\{b_P^\text{cl},b_P^\text{q}\right\}$ are the vectors of classical and quantum fields with the corresponding inverse Keldysh matrix Green's functions given by
\begin{align}
\mathcal{G}_{\mu,0}^{-1}=
\begin{pmatrix}
0&&\left[G^A_{\mu,0}\right]^{-1}\\
\left[G^R_{\mu,0}\right]^{-1}&&D^K_{\mu,0}\\
\end{pmatrix}\,
\end{align}
with $\mu\in\{a,\text{ph}\}$.\\
Finally, the interaction part of the action reads
\begin{widetext}
\begin{align}\label{eq:Sint_two_level}
\begin{split}
S_\text{int}=&\int\frac{d\omega}{2\pi}\int\frac{dk}{2\pi}\sum_z\\ & \left(\frac{1}{\sqrt{2}}g e^{ikz} u_k(z)\left[b_P^\text{q}(k) \left(\bar{a}_e^\text{q}(z) a_g^\text{q}(z)+\bar{a}_e^\text{cl}(z)a_g^\text{cl}(z)\right)+b_P^\text{cl}(k)\left(\bar{a}_e^\text{cl}(z)a_g^\text{q}(z)+\bar{a}_e^\text{q}(z) a_g^\text{cl}(z)\right)\right]+c.c.\right).
\end{split}
\end{align}
\end{widetext}
As the atoms are fixed at positions commensurate with the Bloch wave, we can use the periodicity of the dimensionless Bloch function $u_k(z)$ to replace it by $u_k(0)$. In general, careful engineering of the waveguides allows some control over the momentum dependence of $u_k(0)$ \cite{chang_njp_photcrys_2013,kimble_2014_crystal}. Here, we choose the simplest approximation of a constant, which we then absorb into the coupling via the replacement $g|u_k(0)|\to g$.\\
As in equilibrium theory, one can apply Wick's theorem to find the dressed Green's functions
\begin{subequations}
\begin{align}\label{eq:Gfull}
\mathcal{G}_a^{\alpha \beta}(x,x')&=\langle \mathbf{a}^\alpha(x) *\bar{\mathbf{a}}^\beta(x')\rangle_S\\
\mathcal{G}_\text{ph}^{\alpha \beta}(x,x')&=\langle b_P^\alpha(x)\bar{b}_P^\beta(x')\rangle_S
\end{align}
\end{subequations}
with $*$ the outer product and $\alpha,\beta=\text{cl},\text{q}$.
Here, as opposed to the bare propagators, the expectation value is taken with respect to the full action $S$. Expanding the exponent $e^{i S_\text{int}}$ under the functional integral, one obtains the infinite Dyson series
\begin{align}
\begin{split}
\mathcal{G}=\mathcal{G}_{\mu,0}&+\mathcal{G}_{\mu,0}\circ\Sigma_\mu\circ\mathcal{G}_{\mu,0}\\&+\mathcal{G}_{\mu,0}\circ\Sigma_\mu\circ\mathcal{G}_{\mu,0}\circ\Sigma_\mu\circ\mathcal{G}_{\mu,0}+\dots\,,
\end{split}
\end{align}
where $\circ$ denotes the convolution in space and time with a simultaneous matrix product in the Keldysh index $\left\{\text{cl},\text{q}\right\}$ as well as the field components $g,e$ of the atomic propagator ($\mu=a$). Summation of this geometric series for the retarded Green's function gives the same result as in equilibrium theory
\begin{align}
G^R_\mu=\left(G^R_{\mu,0}-\Sigma_\mu^R\right)^{-1}\,.
\end{align}
For the Keldysh component on the other hand one finds
\begin{align}
G^K_\mu=G^R_\mu \circ \left(\Sigma^K_\mu - D^K_{\mu,0}\right)\circ G^A_\mu\,,
\end{align}
which is conveniently parametrized in terms of the hermitian distribution function $F_\mu$, defined via
\begin{align}\label{eq:F}
G^K_\mu=G^R_\mu\circ F_\mu - F_\mu \circ G^A_\mu\,.
\end{align}
As the self-energies $\Sigma^{R,K}_\mu$ in general depend on Keldysh and retarded components, the two Dyson equations are coupled and have to be solved simultaneously. Similar to equilibrium theory, self-energies are generically a sum of convolutions of a number of Green's functions. % \fm{The number of convolutions and their order depend on the order at which we truncate the expansion in $S_{\rm int}$. Such a truncation (or an alternative approximation) is obviously necessary for the solution of the Dyson equation.} \jlc{The Dyson equation is independent of the approximation scheme used. Only the self-energy is affected.}
However, due to the many terms in the interaction part of the action it is easy to over- or undercount certain combinations. In this respect, Feynman diagrams and the corresponding Feynman rules turn out to be very helpful. These we will summarize in the next section.

\subsection{Representation via Feynman diagrams}\label{sec:Feynman}
\begin{figure}[htp]
\begin{center}
\includegraphics[width=\columnwidth]{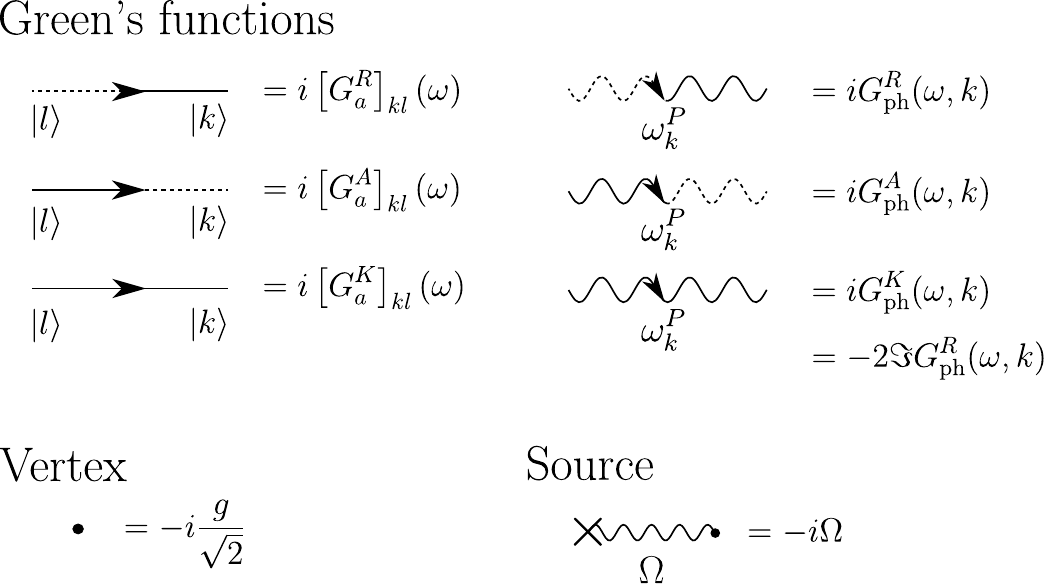}
\caption{Translation table for Feynman diagram. In order to avoid cluttering, the Keldysh structure will never be shown explicitly and contractions over the corresponding indices at each vertex are instead implied according to Eq.~\eqref{eq:Sint_two_level}.}
\label{fig:Feynman}
\end{center}
\end{figure}

% \jlc{In my opinion, an actual introduction into Feynman diagrams in this section would be too basic. It is something you can look up in any textbook and has no place in a research paper.}\\
Feynman diagrams mimic the propagation of excitations in an intuitive way: Lines connecting two space-time points $x$ and $x'$ correspond to Green's functions, with an arrow pointing in the direction of propagation. To distinguish between mobile and immobile particles, we draw atoms with straight and photons with wavy lines (see Fig.~\ref{fig:Feynman}). As opposed to equilibrium theory each propagator has an additional causality index $(R,A,K)$ arising from the Keldysh structure. In Feynman diagrams it is customary to account for this by drawing quantum fields as dashed lines (i.e. a retarded propagator starts as a dashed line that turns into a full line, while the opposite is the case for an advanced Green's function). An interaction vertex is drawn as a dot and connects one photon propagator with an incoming and outgoing atom propagator. Due to causality, coherent interactions require that the number of quantum fields joined at each vertex must be odd (see also Eq.~\eqref{eq:Sint_two_level}).\\
Apart from this additional structure the derivation of Feynman rules proceeds completely analogously to equilibrium theory \cite{kamenev_book}. We therefore only state the resulting Feynman rules. Self-energies at order $g^n$ are obtained according to the following recipe:
\begin{itemize}
\item[1.]Using straight lines for atoms and wavy lines for photons, draw all topologically distinct, fully connected diagrams with $n$ vertices and the same external legs as the bare inverse propagator which is going to be corrected by the self-energy we want to compute.
\item[2.]Allowing for each propagator to take any causality index $(R,A,K)$, keep only those diagrams where each vertex connects to an odd number of dashed lines and where either an incoming photon excites a ground state or an atom decaying from the excited state emits a photon. 
\item[3.]Following the translation table in Fig.~\ref{fig:Feynman} associate each line with a factor $iG_\mu^{R/A/K}(\omega,k)$ and each vertex with $-ig/\sqrt{2}$.
\item[4.]Conserve energy at each vertex by equating the sums of incoming and outgoing frequencies.
\item[5.]Integrate over all internal momenta and frequencies with $\int\frac{dk}{2\pi}\int
\frac{d\omega}{2\pi}$.
\item[6.]Multiply each diagram with $i$.
\end{itemize}
Note that, since we explicitly distinguish between the different atomic states all symmetry factors are equal to one. Also, due to causality, the integral in the second to last step will always evaluate to zero for all diagrams that involve loops with counter-propagating retarded or advanced Green's functions as well as those with a retarded Green's function co-propagating with an advanced Green's function.\\
Despite this simplification the Keldysh structure nevertheless gives rise to a large number of topologically equivalent diagrams that differ only in the causality indices. However, as we explain in Appendix~\ref{app:KK}, many of these can be expressed through one another by the use of Kramers-Kronig relations. Given the large amount of cancellations, we will in the following suppress the Keldysh structure in all drawings of Feynman diagrams, implicitly assuming the sum over all allowed causality indices.

As is apparent from Fig.~\ref{fig:Feynman}, we have two kinds of processes coupling different atoms: interactions with dynamical photons and coupling to the external field described as a source term. These lead in principle to two separate expansions in the corresponding coupling constants $g$ and $\Omega$. However, since the laser $\Omega$ itself is not treated as a quantum field, it acts as a quadratic term in the action. Consequently, the infinite series of diagrams at all orders in $\Omega$ is readily accounted for in the matrix Green's function Eq.~\eqref{eq:bare_2level_atom_GF}. However, a clearer physical picture often emerges if the lowest order is drawn explicitly. Each interaction with the Rabi laser is associated with a factor $-i\Omega$ and transfers between the two eigenstates of the isolated atom. In the rotating frame this process conserves the energy of the atom.

Instead of explicitly writing all diagrams in terms of bare propagators it is convenient to introduce the concept of bold lines, which indicate Green's functions with self-energy corrections that have to be specified in a separate equation.

\subsection{Expansion in the inverse propagation range}
\begin{figure}[htp]
\begin{center}
\includegraphics[width=\columnwidth]{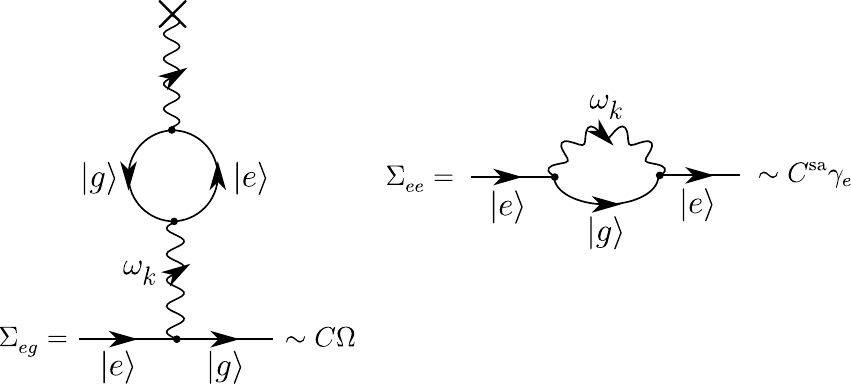}
\caption{Lowest order self-energy corrections to the atomic propagator in a simultaneous expansion in $\Omega$ and $g$. Both terms are of order $g^2$, however the Hartree diagram in $\Sigma_{eg}$ couples to the polarization of all atoms within the volume $L$ around the atom under consideration. The Fock diagram in $\Sigma_{ee}$ on the other hand describes a Lamb shift requiring the photon to be emitted and absorbed by the same atom.}
\label{fig:perturbative}
\end{center}
\end{figure}

% \jlc{With the current implementation of the losses the atom density vanishes once all orders of $\Omega$ are included. If instead we were to drive the photons incoherently Hartree diagrams are topologically forbidden. I don't see any way around this that does not require an additional atomic level.}
The formalism is now ready to quantify the statements made in the introduction. To do so, we will consider a simultaneous expansion in $\Omega$ and $g$ and treat the two lowest order self-energy corrections to the retarded atomic propagator, which are shown in Fig.~\ref{fig:perturbative}. There, as will be the case throughout the remainder of this manuscript, the Keldysh structure will not be made explicit. Note, that using bare Green's functions one has $\Sigma_{gg}^R(\omega)=0$, which can be traced back to a more general symmetry of vacuum Green's functions (see App.~\ref{app:KK}). Both diagrams in Fig.~\ref{fig:perturbative} are of second order in the coupling constant, however the Hartree diagram $\Sigma_{eg}^R$  in Fig.~\ref{fig:perturbative}a) involves two  atoms instead of just one, as is the case for the Fock diagram $\Sigma_{ee}^R$ in Fig.~\ref{fig:perturbative}b). Based on the arguments in the introduction we expect to find $\Sigma_{eg}^R\sim C$ and $\Sigma_{gg}^R\sim C^\text{sa}$. Using bare propagators the Feynman diagrams can be directly evaluated and for the Hartree self-energy one finds
\begin{align}
\begin{split}
\Sigma_{eg}^R(\omega)=&i\frac{g^2\Omega}{2}G_\text{ph}^A(\omega=0,k=0)\\&\!\times\!\int\!\frac{d\omega}{2\pi}\!\left[G_{gg}^K(\omega)G_{ee}^A(\omega)+G_{gg}^R(\omega)G_{ee}^K(\omega)\right]\\=&\frac{\Omega g^2(1-n_V)}{(\Delta_e-i\gamma_e/2)(\Delta(0)-i\kappa/2)}\,.
\end{split}
\end{align}
This becomes significant, when the intensity of the field re-scattered by other atoms $\sim\Sigma_{eg}^R$ becomes comparable to the external drive given by the bare inverse propagator $[G_{eg}^R]^{-1}=-\Omega$. This is the case if the collective coupling strength $g$ satisfies $|g^2(1-n_V)/(\Delta_e+i\gamma_e/2)(\Delta(0)+i\kappa/2))|\gtrsim 1$. In the case of small detunings $\Delta_e\lesssim \gamma_e$, $\Delta(0)\lesssim \kappa$ and without defects ($n_V=0$) this indeed simplifies to $C\gtrsim 1$. For the Fock diagram we have to fix the photon dispersion and choose $\Delta(k)=\Delta_0-J \cos(k)$, as it allows to consider the two relevant cases of ballistic photons obtained for $|\Delta_0\pm J|\gg \kappa$ and diffusive behavior in case of $|\Delta_0\pm J|\lesssim \kappa$. Employing the Feynman rules introduced in the last section, one finds
\begin{align}
\begin{split}
\Sigma_{ee}^R(\omega)\!=&i\frac{g^2}{2}\int\frac{d\omega'}{2\pi}\int\frac{dk}{2\pi}\\&\!\!\left(G^R_{gg}(\omega-\omega')G_\text{ph}^K(\omega',k)\!+\!G_{ee}^K(\omega-\omega')G_\text{ph}^A(\omega',k)\right)\\=&\frac{g^2(2-n_V)}{\sqrt{\omega-\Delta_0-J+i\kappa/2}\sqrt{\omega-\Delta_0+J+i\kappa/2}}\,,
\end{split}
\end{align}
which for a large bandwidth $J$ becomes
\begin{align}
\Sigma_{ee}^R(\omega)\approx\begin{cases} -i\frac{g^2(2-n_V)}{\sqrt{J^2-\Delta_0^2}} & \text{ballistic photons}\\
(1-i) \frac{g^2(2-n_V)}{\sqrt{2J\kappa}} & \text{diffusive photons.}\\
\end{cases}
\end{align}
These have to be compared with the bare loss rate $\gamma_e$.
Since in the first case one can identify $\sqrt{J^2-\Delta_0^2}$ with the group velocity of the photons on resonance, one has $L=2\sqrt{J^2-\Delta_0^2}/\kappa$. In the second case the photon localizes on a length scale $L=\sqrt{J/\kappa}$. Hence, in both cases one recovers the initial claim that the self-interaction becomes relevant only if $C^\text{sa}=g^2/(\kappa \gamma_e L)\gtrsim 1$.\\
While the toy model considered here suffices to explain the basic formalism and illustrate the expansion in the inverse propagation range, it is also plagued by large photon losses caused by excited state emission. In a two-level system these can only be avoided via a large detuning, which then severely limits the maximal attainable interactions. In the following we will therefore increase the complexity of the internal level structure of each atom, while maintaining the same basic setup.

\section{Application: Electromagnetically induced transparency}\label{sec:EIT}
While near-resonant interactions between photons and two-level atoms result in strong dissipation, the interaction can be made largely coherent by introducing an additional metastable atomic state $|s\rangle$, which is coupled to $|e\rangle$ by an auxiliary laser field with frequency $\omega_L^{(1)}$ and Rabi amplitude $\Omega$ (see Fig.~\ref{fig:setup}b).
%Strong interactions between light and matter require small loss rates. The latter can be reduced in two ways: firstly, to avoid stimulated emission from the propagating photon mode, the laser mode coupled to the $g-e$ transition has to be removed. Secondly, a simple, yet very effective method to reduce photon losses caused by spontaneous emission from the excited state $|e\rangle$ is realized by adding a further, metastable atomic state $|s\rangle$, which can be reached from $|e\rangle$ by stimulated emission of a photon with energy $\omega_L^{(1)}$ into a laser mode driven at Rabi amplitude $\Omega$. 
Three level systems of this type have been investigated extensively \cite{Fleischhauer2005, Vitanov_Review}. Since photons can be converted into atomic excitations, the EM modes of the waveguide hybridize with the two atomic transitions and give rise to three polariton branches. If the condition $\omega_k^P=\omega_s+\omega_L^{(1)}$ is satisfied one of these linear combinations of photons and atoms contains no contribution from the excited state, instead forming a lossless ''dark-state polariton'' involving the waveguide photon, and atomic states $|g\rangle$ and $|s\rangle$. Consequently, the electromagnetic field on the $e-s$ transition has rendered the probe photons robust against decay and dephasing of state $|e\rangle$. The mechanism for lossless propagation has therefore been named electromagnetically induced transparency (EIT). At its heart lies a destructive quantum interference between the direct excitation pathway from $|g\rangle$ to $|e\rangle$ and an indirect process via $|s\rangle$ \cite{Harris1990}.\\
In the following, we will demonstrate how the expansion in $1/L$ can be used to reproduce the hallmark results of EIT. In doing so we can check the validity of our approximations and lay the foundation for the subsequent discussion on interacting polaritons.\\
We note that the formation of the EIT state may take a long time \cite{Jyotsna1995, Choi2014} or require careful engineering of the laser drive \cite{Vitanov_Review}. The dynamics of the lossy transient state poses interesting questions \cite{Finkelstein2019} even more so in the presence of long-range interactions where they remain accessible to the formalism presented here. For now, however, we will focus on the steady state and defer all discussions on the dynamics to future research.\\
Following the changes with respect to the system described in Sec.~\ref{sec:appl_twolevel}, the Hamiltonian is now given by
\begin{widetext}
\begin{align}
\begin{split}
\hat{H}_\text{EIT}=&\hbar \left[\sum_z \left\{\omega_e \hat{a}^\dagger_e(z)\hat{a}_e(z) + \omega_s \hat{a}^\dagger_s(z) \hat{a}_s(z) + \left(\Omega e^{-i \omega_L^{(1)} t}\hat{a}^\dagger_e(z) \hat{a}_s(z) + h.c.\right)\right.\right.\\&\left.\left.+\frac{1}{L}\int_k \left(\omega_k^P\hat{b}_P^\dagger(k) \hat{b}_P(k)+g_P \left(e^{i k z} u_k^P(z)\hat{b}_P(k)\hat{a}^\dagger_e(z) \hat{a}_g(z) + h.c.\right)\right)\right\}\right],
\end{split}
\end{align}
\end{widetext}
while the decay rates remain the same. Note that, for later convenience we have renamed $g$ to $g_P$ and $u_k(z)$ to $u_k^P(z)$. In the absence of a laser coupling to the ground state, the system will instead be excited by an incoherent and homogeneous pumping of the propagating modes with a transverse light source. Without affecting the EIT physics, one could simply describe this light source by a Markovian bath:
\begin{align}
\begin{split}
\mathcal{L}_{\kappa_s}\rho=-\hbar\!\int_k\!\frac{\kappa_s}{2}&\left(\left\{\hat{b}_P(k)\hat{b}_P^\dagger(k)+h.c.,\rho\right\}\right.\\&\left.-2\hat{b}_P^\dagger(k)\rho\hat{b}_P(k)\!-\!2\hat{b}_P(k)\rho\hat{b}_P^\dagger(k)\right).
\end{split}
\end{align}
The only disadvantage of this description is a large population of non-interacting photons propagating through the system at frequencies far detuned from any atomic resonances. In fact, a transversal light source will not couple to all modes equally well, but due to frequency dependencies of the mode matching, will predominantly couple to a certain frequency interval. We will model this with a frequency dependent rate $\kappa_s(\omega)=\kappa_s/((\omega-\omega_0)^2+\kappa_0^2)$ centerd near the EIT condition ($\omega_0\approx\omega_s+\omega_L^{(1)}$). To satisfy the Markov approximation $\kappa_0$ will be chosen large compared to the the relevant frequency scales of EIT polaritons. For a derivation of the specific form of $\kappa_s(\omega)$ see App.~\ref{app:Bath}.\\
Despite the modifications relative to the model discussed in Sec.~\ref{sec:appl_twolevel}, the interaction part of the action $S_\text{int}$ is unchanged and also the general form of the quadratic part of the action $S_0$ remains the same with the new fields and Green's functions in the rotating frame given in App.~\ref{app:matrix_GF}. We will merely simplify notation from here on by using the shorter $G_{k}^{R/A/K}(\omega)\equiv G_{kk}^{R/A/K}(\omega)$ for diagonal entries of the atomic Green's function.

\subsection{Nonlinear Feynman rules}\label{sec:NLFR}

Before we continue with the specific application, we notice that, when expanding the Keldysh action order by order in the coupling rate $g_P$, one applies bosonic Feynman rules to 
atoms, which should instead have a restricted Hilbert space with $\sum_{j={g,e,s,d}}\langle\hat{a}_j^\dagger\hat{a}_j\rangle=1$. This implements the physical constraint that each atom occupies either only one level or in general a properly restricted superposition. One therefore has to be careful not to overcount diagrams by simultaneously placing an atom in the same state twice (which would be allowed for bosons). This means that, at every point in time and in every diagram, two counter propagating atomic lines belonging to the same atom have to be found in distinct levels, or must otherwise be identified with one another, i.e. their lines in the Feynman diagram have to be contracted.\\
In general it is very hard to fully enforce these conditions, as one would need to implement increasingly complicated restrictions in real-time on each and every perturbation to the bare scalar Green's functions. Doing so for all diagrams would eventually restore the exact, finite Fock space of the atoms. Here, we instead limit ourselves to impose restrictions allowing to
exactly compute the fully dressed, single-probe-photon propagator -- i.e. all modifications in the regime of linear optics.\\
\begin{figure}[htp]
\begin{center}
\includegraphics[width=\columnwidth]{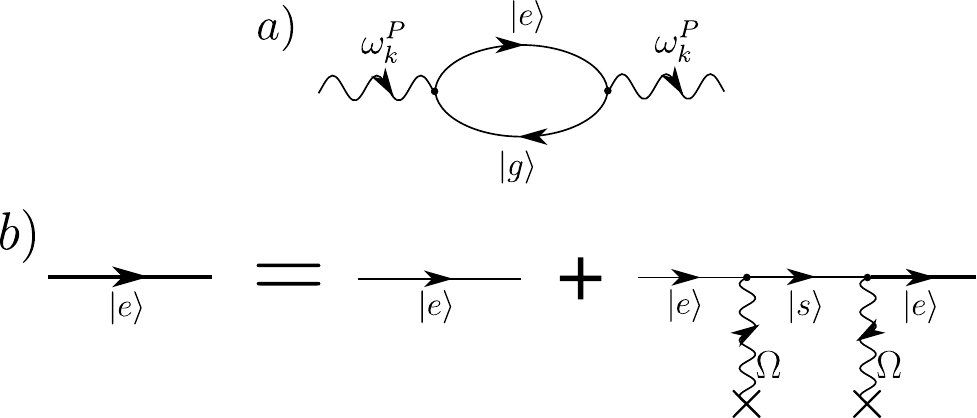}
\caption{The polarization bubble in a) gives rise to polaritons by hybridizing photons with atomic excitations. b) shows the Dyson equation for atomic propagator $G_e$, appearing in the polarization bubble. The interference between the direct excitation of an atom to state $|e\rangle$ and the indirect path via $|s\rangle$ gives rise to EIT.}
\label{fig:polbubble}
\end{center}
\end{figure}
As we will see, the insertion of self-energies in the form of polarization bubbles -- which are diagrams of the type shown in Fig.~\ref{fig:polbubble}a) -- into the bare probe photon Green's function will hybridize this propagating photon mode with stationary atoms, forming polaritons in the process. Without saturation effects these polaritons will not interact among each other. When eventually introducing polariton-polariton interactions in Sec.~\ref{sec:four_level}, it will be of paramount importance to expand around the correct limit of non-interacting polaritons, which will only by ensured by the implementation of the above restrictions imposed by non-linear Feynman rules.\\ 
In the non-interacting regime, where the polariton self-energy is given by a polarization bubble with the external laser fields mixing states $|e\rangle$ and $|s\rangle$ and the probe photons mixing $|g\rangle$ and $|e\rangle$, it suffices to demand that any two counter-propagating Green's functions of the same atom have to involve disjoint sets of states. All diagrams where this is not the case are simply set to zero.\\
We now show that these simplified non-linear selection rules correctly capture the retarded polariton Green's function. The latter reads
\begin{align}\label{eq:GP_LO}
G_P^R(\omega,k)&=\left(G_{P_0}^R(\omega,k)^{-1}-\Sigma_P^R(\omega)\right)^{-1}\,,
\end{align}
with the self-energy given by
\begin{align}
\begin{split}
\Sigma_P^R(\omega)=\frac{i g_P^2}{2}\int \frac{d\omega'}{2\pi}&\left(G_e^K(\omega+\omega')G_{g}^A(\omega')\right.\\ &\quad\left.+G_e^R(\omega+\omega')G_{g}^K(\omega')\right)\,.
\end{split}
\end{align}
We now make use of the Kramers-Kronig relations (see \eqref{eq:KKSigmaR} in the appendix) and realize that only diagrams with either $F_g(\omega)\neq 1$ or $F_e(\omega)\neq1$ are finite, and thus
\begin{align}
\begin{split}
\Sigma^R_P(\omega)=\frac{i g_P^2}{2}\int_{-\infty}^\infty\frac{d\omega'}{2\pi}&\left(\delta G_g^K(\omega')G_e^R(\omega+\omega')\right.\\&\quad\left.+G_g^A(\omega')\delta G_e^K(\omega+\omega')\right)\;,
\end{split}
\end{align}
where $\delta G^K(\omega)=G^K(\omega)-2i\Im{G^R(\omega)}$ is related to the spectral number density by $n(\omega)=i \delta G^K/2$. However, as the atomic medium without probe photons is entirely in the ground state and no atoms are being created, the only way to get $\delta G_e^K(\omega)\neq 0$ is by coupling to $\delta G_g^K(\omega)$. On the other hand, corrections to the bare ground-state propagator all inevitably have to involve the excited state $|e\rangle$.\\
To compare the effect of the exact and simplified non-linear Feynman rules, consider the perturbative insertion of corrections into the bare retarded Green's functions:
\begin{align}\label{eq:Dyson_time}
\begin{split}
G^R(t,t')=&G^R_0(t,t')\\&+\int d t_1\int d t_2 G^R_0(t,t_1)\Sigma^R(t_1,t_2)G^R_0(t_2,t')\\&+\dots\,,
\end{split}
\end{align}
which due to causality are non-zero only if $t>t_1>t_2>t'$. Consequently, none of the Green's functions and self-energies under the integral need to be evaluated simultaneously and no cancellations due to the non-linear Feynman rules are required. Similarly, the Keldysh component of the interacting Green's function is given by
\begin{align}
\begin{split}
\delta G^K(t,t')=\int d t_1\int d t_2 &G^R(t,t_1)(\delta\Sigma^K(t_1,t_2)\\&-\delta D_0^K(t_1,t_2))G^A(t_2,t'),
\end{split}
\end{align}
where $\delta D_0^K=D_0^K-2i\Im{\left(\left[G_0^R\right]^{-1}\right)}$ and $\delta\Sigma^K(t_1,t_2)=\Sigma^K(t_1,t_2)-2i\Im{\Sigma^R(t_1,t_2)}$ have been introduced. Due to the retarded and advanced Green's functions one has $t>t_1$ and $t'>t_2$. Clearly those insertions with $t_1<t'$ have to be discarded, as then, between these times, the retarded and advanced Green's function of the same state counter-propagate. With this restriction in place $\delta\Sigma^K(t_1,t_2)$ has to be evaluated at $t'$, which is necessarily simultaneous with the retarded Green's function of the other state in the polarization bubble, and the diagram again has to be removed. In the end, as only the ground-state satisfies $\delta D_0^K\neq0$, we are left with the simple result
\begin{align}\label{eq:Sigma_LO}
\Sigma^R_P(\omega)=\frac{i g_P^2}{2}\int_{-\infty}^\infty\frac{d\omega'}{2\pi}\delta G_{g,0}^K(\omega')G_e^R(\omega+\omega')\,,
\end{align}
where in $\Sigma_e^R$ no dependence on $G_g$ is allowed.
For the Keldysh component of the polariton self-energy one has, due to the Kramers-Kronig relations \eqref{eq:KKSigmaK},
\begin{align}\label{eq:Sigma_LO_K}
\delta\Sigma^K_P(\omega)\!=\!\frac{i g_P^2}{2}\!\!\int_{-\infty}^\infty\!\!\frac{d\omega'}{2\pi}\!\left(2G_g^{K_0}+\delta G_{g,0}^K(\omega')\!\right)\!\delta G_e^K(\omega+\omega')\,.
\end{align}
Following a similar argument as above, one can show that this contribution vanishes once either the full or the simplified non-linear selection rule is applied. As these arguments can be continued order by order in the coupling constants, we find that for non-interacting polaritons both selection rules coincide. For an alternative proof that, in the limit of low polariton densities, EIT is exactly recovered by the simplified non-linear Feynman rules see Appendix \ref{app:Feynman}.\\

In summary, the nonlinear Feynman rules outlined here partially compensate the unphysical tendency of the bosonized atomic excitations to bunch together with the photons. As long as the number density of excited atoms is small compared to that of the ground-state atoms, saturation effects of the atomic medium can be neglected and no further selection rules have to be implemented.
While this restriction to the selection of diagrams might seem complicated to enforce consistently, we will see that it actually simplifies the Feynman diagrams. To avoid confusion, we will label all atomic states and explicitly show all couplings to external sources in every graphical representation of a Dyson equation.

\subsection{Self-consistency and conserving approximations}
\label{sec:SC_conserving}
In studying out-of-equilibrium interacting problems within a diagrammatic approach, the self-consistent formulation of the Dyson equations -- i.e. a non-perturbative treatment where all GFs appearing in a self-energy are fully dressed, resulting in non-linear Dyson equations -- can become crucial for three main reasons. Firstly, the long-time behavior and in particular the steady state may not be accessible perturbatively. In fact, for a system to be able to forget about its initial state the memory terms appearing in the Dyson equation (see e.g. Eq.~\eqref{eq:Dyson_time}) must deviate from the initial state in an non-perturbative manner. Secondly, the integrals of motion of a problem are only correctly included within the so called conserving approximations, which themselves can be derived from an appropriate thermodynamic functional and always result in self-consistent theories. Thirdly, in the absence of a small coupling strength the expansion parameter for perturbative diagrammatics becomes of order one, as is for instance the case close to phase transitions.

In case of the driven-dissipative system described here, it is not a priori impossible to describe the steady-state perturbatively. This is because the relaxation happens also without interactions between light and matter.
On the other hand, it is unfortunately impossible to build a proper functional, since it would be incompatible with the approximate nonlinear Feynman rules introduced above. Having a conserving approximation in our case is however not crucial. This is a consequence of the incoherent, transversal drive and Markovian losses of the full, microscopic theory introduced later. These neither conserve energy nor quasi-momentum. Therefore, the only conserved quantity is the total number of atoms, which we approximately enforce, at least on average, by means of the nonlinear Feynman rules. While dropping these would allow to construct a conserving effective action, the resulting theory would not conserve the atom number either, since the approximate formulation of radiative decay in Eqs.~\eqref{eq:atomloss} explicitly breaks the corresponding symmetry of the atomic sector under the $U(1)$ transformation $\hat{\mathbf{a}}_a\to\hat{\mathbf{a}}_a e^{i\phi}$ and $\hat{\mathbf{a}}_a^\dagger\to\hat{\mathbf{a}}^\dagger_a e^{-i\phi}$.

In summary, the first two reasons requiring a self-consistent approach do not apply to our case. Still, when considering polariton interactions later in section \ref{sec:four_level}, we will largely make use of self-consistent solutions of the Dyson equations in order to include the important non-perturbative effects in regimes of single-atom cooperativity close to one: $C^{\rm sa}\sim 1$.

\subsection{Results: Linear susceptibility and slow light}\label{sec:EIT_results}

Following the non-linear Feynman rules and neglecting saturation effects the probe photon propagator is fully determined by the polarization bubble shown in Fig.~\ref{fig:polbubble}a). In this low excitation density limit the retarded photon propagator $G_P$ can then be directly obtained from Eqs.~\eqref{eq:GP_LO}, \eqref{eq:Sigma_LO} and \eqref{eq:Sigma_LO_K}, which can be simplified to (see also App.~\ref{app:loop_reduction})
\begin{align}
\begin{split}
\Sigma_P^R(\omega)&=g_P^2(1-n_V)G_e^R(\omega)\\
\delta\Sigma_P^K(\omega)&=g_P^2(2-n_V)\delta G_e^K(\omega)\,,
\end{split}
\end{align} 
where 
\begin{align}\label{eq:polarizability}
\begin{split}
G_{e}^R(\omega)=G_{e,0}^R(\omega)&=\frac{1}{\omega-\frac{\Omega^2}{\omega-\Delta_s+i\epsilon/2}+i\gamma_e/2}\\
G_{e}^K(\omega)=G_{e,0}^K(\omega)&=-2i\Im{G_{e,0}^R(\omega)}
\end{split}
\end{align}
are the components of the bare propagators of the excited state $|e\rangle$ obtained by inverting $\left[G_{a,0}^{R}\right]^{-1}$ in Eq.~\eqref{eq:matrixG0}. One should note that this solution involves no approximations beyond the linearization of the spin degree of freedom, which we showed in Sec.~\ref{sec:NLFR} to be fully compensated by simple nonlinear Feynman rules. As such, it is not surprising that upon identification of $G_{e}^R$ with the polarizability of the medium the present approach reproduces the exact linear polarizability
\begin{align}
\chi(\omega)=\frac{n\mu_{eg}^2}{\epsilon_0\hbar}G_e^R(\omega-\omega_e)\,,
\end{align}
where $\mu_{eg}$ denotes the dipole moment of the $g-e$ transition.
Hence, as pointed out earlier, $G_P^R$ no longer describes free photons, but the eigenmodes of the system, which are photons hybridized with the medium. The dispersion of these new degrees of freedom, the polaritons, has three branches resulting from the coupling of two atomic transitions and the photonic dispersive mode, which far away from the atomic resonance $\Delta_e$ is essentially that of the free photon. Due to the vanishing losses of state $|s\rangle$, however, the central branch -- the so called dark-state polariton, which is a combination of a photon and an atom in state $|s\rangle$ without any admixture of the lossy $|e\rangle$ -- is very long lived. Within the functional-integral description, the trivial calculation leading to Eq.~\eqref{eq:polarizability} thus fully captures the phenomenon of EIT. On a more pedagogical note, the destructive interference at the heart of EIT becomes particularly apparent upon inspection of the diagrammatic expression for $G_e$ shown in Fig.~\ref{fig:polbubble}b).

Since the dark-state polariton is a linear superposition of a localized atom and propagating photon, its group velocity can be tuned by adjusting the ratio $\Omega/g_P$. However, without losses in state $|s\rangle$, the linewidth is modified at the same rate, such that the penetration depth of photons into the waveguide is not affected. This can be easily verified by comparing the group velocity of the dark-state polariton with its linewidth. Linearizing the dispersion of the free photons, which on the energy scale of the susceptibility of the medium (set by $\gamma_e$) is typically well justified, the group velocity can be determined from the pole of the polariton Green's function $G_P^R(\omega,k)$ given by Eqs.~\eqref{eq:GP_LO}, \eqref{eq:SigmaP_LO} and \eqref{eq:polarizability}. In the limit of mostly atom-like dark-state polariton, where the ratio between atomic and photonic contributions $\theta=g_P^2(1-n_V)/\Omega^2$ becomes large, an expansion around the EIT window results in the condition
\begin{align}\label{eq:vgcond}
\begin{split}
\left[G_P^R(\omega,k)\right]^{-1}=&\theta(\omega-\Delta_s)-v_P(k-k_\text{EIT})\\&+i\xi(\omega-\Delta_s)^2+i\kappa_P/2\overset{!}{=}0\;,
\end{split}
\end{align}
where $v_P$ is the local group velocity of the bare photon near the resonance at $k=k_\text{EIT}$ with the laser acting on the $|s\rangle-|e\rangle$ transition.
\begin{figure*}[!ht]
\begin{center}
\includegraphics[width=\textwidth]{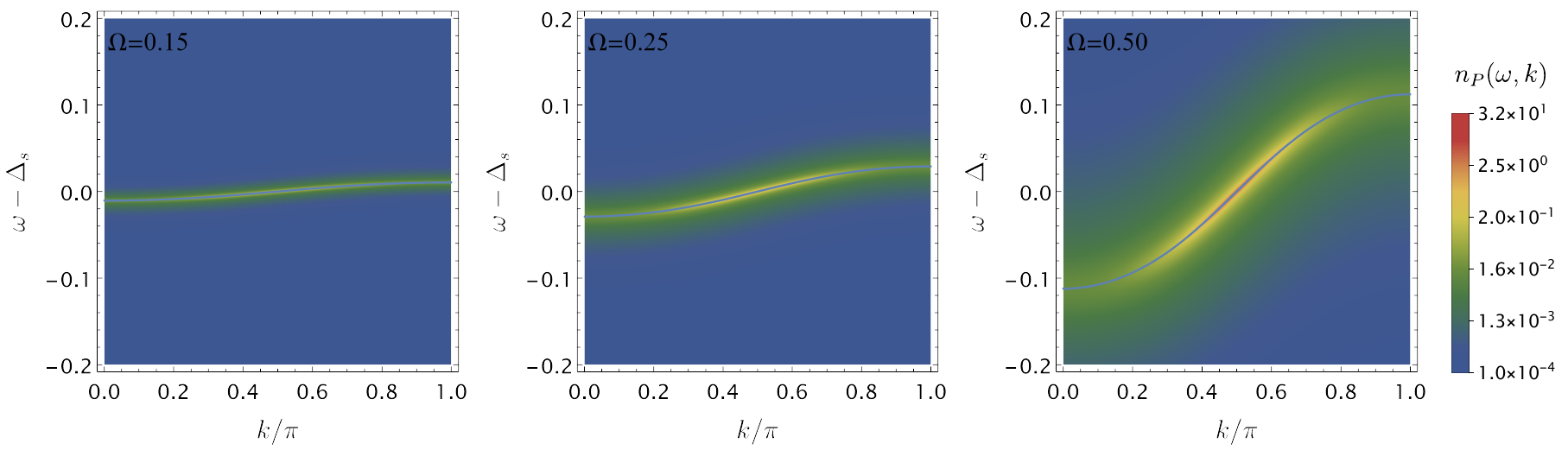}
\caption[Number densities for imperfect EIT conditions]{Frequency- and momentum-resolved number density in the vicinity of the EIT condition. The corresponding dispersion of the dark-state polariton has been added in the form of a blue line. The parameters used are $\kappa_0=2$, $\kappa_s=1$, $\omega_0=\Delta_s=n_V=0$, $g_P=10$, $\kappa_P=0.5$, $\gamma_e=1$ and $\Delta_P(k)=-50 \cos{k}$.}
\label{fig:EIT}
\end{center}
\end{figure*}
Furthermore, we have introduced the convenient abbreviation $\xi=\gamma_e\theta/(2\Omega^2)$. At the center of the EIT window the group velocity is given by
\begin{align}
v_g=\frac{d\omega_\text{res}}{dk}=\frac{v_P}{\sqrt{\theta^2+2\xi\kappa_P}}\sim\Omega^2\;,
\end{align}
where $\omega_\text{res}$ satisfies the condition \eqref{eq:vgcond}. On the other hand, at $k_\text{EIT}$ the linewidth of the dark-state polariton is given by
\begin{align}
\Delta\omega=\frac{\sqrt{-\theta^2-\xi \kappa_P+\sqrt{\theta^4+2\xi\theta^2\kappa_P+2\xi^2\kappa_P^2}}}{\sqrt{2}\xi}\sim\Omega^2\;.
\end{align}
\begin{figure*}[ht]
\begin{center}
\includegraphics[width=\textwidth]{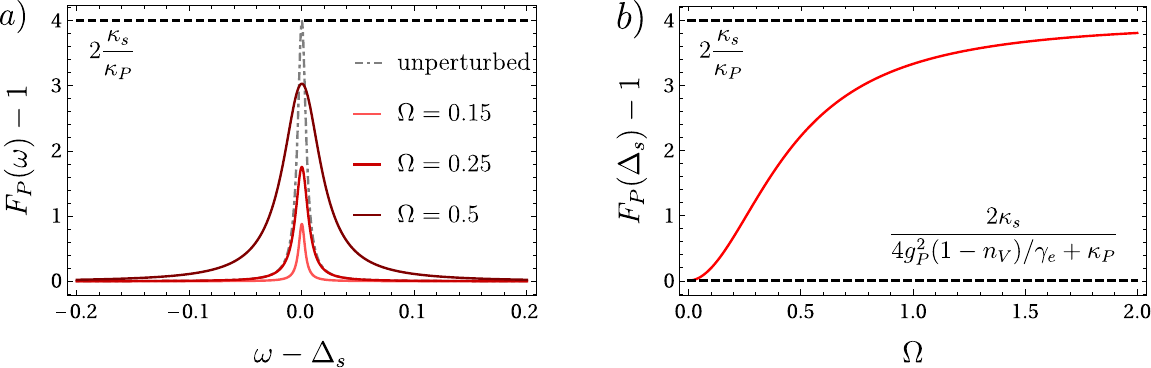}
\caption[Distribution function of EIT conditions with losses]{a) Distribution function of the perturbed dark-state polariton in the vicinity of the EIT condition at the same parameters as in Fig.~\ref{fig:EIT} with finite linewidth $\gamma_s=10^{-3}$. Clearly the occupation of the slower polaritons is more strongly suppressed by these losses. For comparison we also added (gray dash-dotted line) the distribution function of the unperturbed EIT polaritons (i.e.~$\gamma_s=\epsilon$) for $\Omega=0.25$. b) The EIT window is mostly destroyed for $\Omega\lesssim 0.2$, despite the very weak losses parametrized by $\gamma_s$. On the other hand polaritons with $\Omega\gtrsim 1$ are largely unaffected.}
\label{fig:FpEITd}
\end{center}
\end{figure*}
Expanding around large $\theta$, we find 
\begin{align}
\begin{split}
v_g&\approx v_p\frac{\Omega^2}{g_P^2(1-n_V)}=\frac{v_p}{\theta}\\
\Delta\omega&\approx\frac{\Omega^2\kappa_P}{2 g_P^2(1-n_V)}=\frac{\omega_p}{\theta}
\end{split}
\end{align}
and therefore
\begin{align}
\frac{v_g}{\Delta\omega}=\frac{2v_P}{\kappa_P}\;,
\end{align}
which agrees with the result for the free photon. Consequently, the effective probe photon propagation range
\begin{align}
L_P=v_P/\kappa_P
\end{align}
is unaffected by the formation of dark-state polaritons and the accompanying reduction of the group velocity. Independent of the mixing angle $\theta$ the inverse interaction range between atoms thus remains a small parameter suitable for a perturbative expansion unless the single atom cooperativity $C_P^{sa}$ becomes large, in which case all orders in $L_P^{-1}$ have to be included. Note that at fixed $g_P$ both the group velocity and linewidth of the dark-state polariton can be conveniently tuned by adjusting the Rabi amplitude $\Omega$. We illustrate this by showing a logarithmic density plot of the frequency- and momentum-resolved number density of polaritons $n_P(\omega,k)$ in Fig.~\ref{fig:EIT}, where the increase in group velocity and decay rate with growing $\Omega$ are clearly visible.

In the absence of the fluctuation-dissipation theorem the distribution function $F$ introduced in Sec.~\ref{sec:Keldysh} becomes an interesting quantity as it measures the strength of the drive that a given degree of freedom experiences, independent of its actual susceptibility. Since the atoms as well as drive and decay are assumed to be distributed homogeneously in space, $F_P$ is independent of momentum. In Fig.~\ref{fig:FpEITd}a) we illustrate that despite the broad drive by $\kappa_s(\omega)$, the distribution function of the dark-state polariton has a very sharp peak centered around the resonance with the laser on the $|e\rangle-|s\rangle$ transition, where it reaches the largest value possible $F_P(\Delta_s)=2\kappa_s/\kappa_P+1$. This narrow window of highly occupied polaritons is however very sensitive to losses in state $|s\rangle$. We illustrate this in Fig.~\ref{fig:FpEITd}a) by increasing the linewidth of the metastable state to $\gamma_s=10^{-3}$. With slow polaritons being mostly atomic it is clear that already a very small loss rates $\gamma_s$ drastically increases the opaqueness of the waveguide. This is captured by the suppression of the peak in the distribution function in Fig.~\ref{fig:FpEITd}a). Faster and therefore broader EIT polaritons are much less susceptible and thus the maximal value of $F_P(\omega)-1$ once again approaches $2\kappa_s/\kappa_P+1$ for $\Omega\to\infty$, whereas it drops to the typically much smaller value $2\kappa_s/(4g_P^2(1-n_V)/\gamma_e+\kappa_P)+1$ as $\Omega\to 0$ (see Fig.~\ref{fig:FpEITd}b)).\\
In the following we will induce similar losses through dissipative interactions between atoms in state $|s\rangle$. Doing so for highly sensitive slow polaritons will create a strong positive feedback that ultimately gives rise to a first order phase transition. As we shall argue in more detail below, this entirely dissipative feedback effect differs from the more conventional interaction-induced detuning commonly experienced in Rydberg gases.

\section{Application: strongly interacting photons using atoms near waveguides}\label{sec:four_level}

% \jlc{This could possibly be our best option to split the paper in two. This would also mean that we don't need to introduce subsubsubsections (all following sections until Rydberg atoms need an additional sub if we stick with only one paper).}

\begin{figure*}[htp]
\begin{center}
\includegraphics[width=\textwidth]{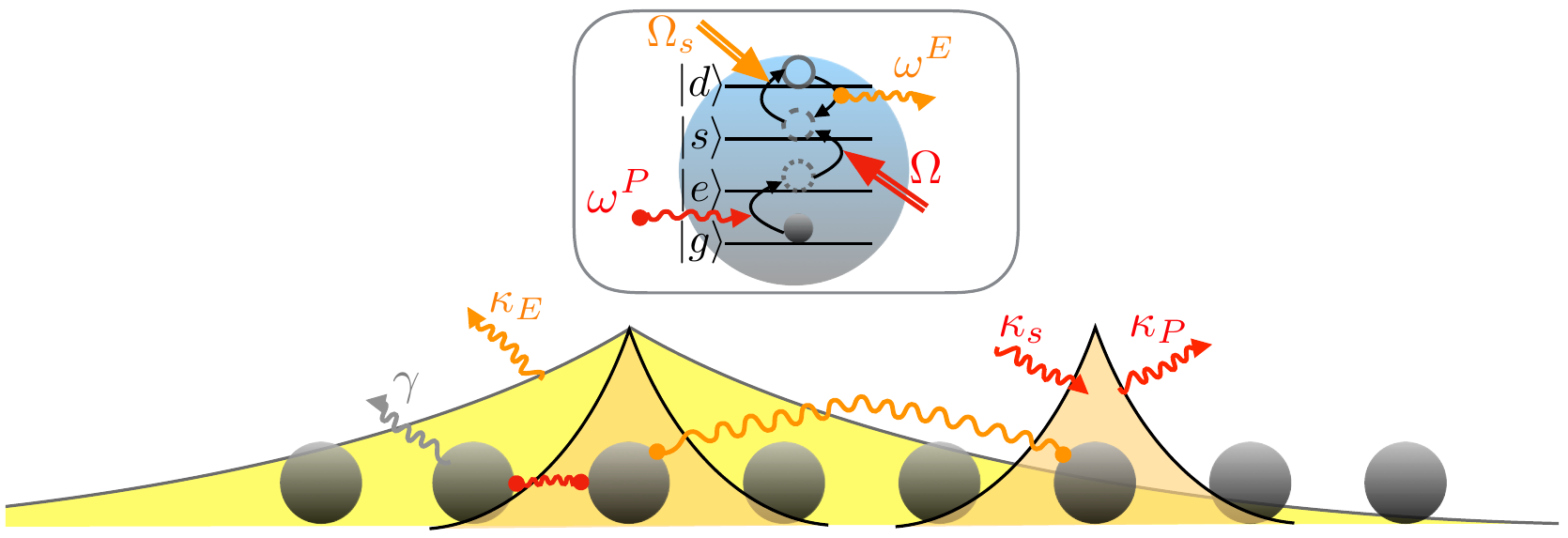}
\caption{Schematic representation of the setup implementing polariton interactions. Atoms are trapped in a regular chain and coupled to the photons propagating inside a waveguide. Inset: Internal level structure of the atoms. External lasers with Rabi amplitudes $\Omega$ and $\Omega_s$ drive transitions between the metastable state $|s\rangle$ and decaying, excited states $|e\rangle$ and $|d\rangle$. The two orthogonal polarizations of the photon modes $P$ and $E$ within the waveguide are (almost) resonant with the $|g\rangle-|e\rangle$ and $|s\rangle-|d\rangle$ transition respectively. Without the excited state $|d\rangle$ the system therefore reduces to the well known three-level scheme used above to discuss EIT. The state $|e\rangle$ dissipates excitations at a rate $\gamma$ and photons in both $P$ and $E$ photons are dissipated from the waveguide at a rate $\kappa_P$ and $\kappa_E$, respectively. The $P$ photons are pumped in at a rate $\kappa_s$.}
\label{fig:N_scheme}\end{center}
\end{figure*}

The direct photon-photon interaction arising from individual atom saturation is extremely weak \cite{bajcsy_2009}. Such an interaction can be made much stronger by introducing a mechanism for the atoms to interact with one another over a distance. Here, this is achieved via an additional set of exchange-photon modes with dispersion $\omega_k^E$. These are orthogonally polarized with respect to the $P$-modes introduced above. 
In fact, in PCWs the two transverse light polarizations do not mix and their band structures can be tuned independently. These engineered photon band-structures potentially allow to control not only
the photon dispersion but also both the strength and the range of
interactions \cite{chang_many_body_2015,douglas_molecules_2016,shi2015multiphoton}, as well
as the coupling with the environment \cite{asenjo_radiance_2017}. It is therefore possible to trap the atoms in a chain that is commensurate with the periodicity of the PCW, have them hybridize with the propagating probe photons and simultaneously make the resulting polaritons interact via localized exchange photons of the orthogonal polarization. Alternatively, the atoms could also be held in place in the evanescent field of a tapered fiber using tweezers \cite{Thompson1202, Endresaah3752}. In this case, the exchange photons could be associated with a higher-order guided mode, operating near cutoff. A schematic representation of the setup we consider is shown in Fig.~\ref{fig:N_scheme}.\\
Specifically, it is possible to use the exchange photons to couple a second excited state $|d\rangle$ to the state $|s\rangle$. To adjust the admixture of $|d\rangle$, we introduce a second driving laser of frequency $\omega_L^{(2)}$ and Rabi amplitude $\Omega_s$. 
In the actual calculations shown here we will for concreteness choose a quadratic dispersion for the $E$-photons $\omega^E(k)=\omega^E_0-\alpha_E(k-k_E)^2$ around the band edge $\omega^E(k_E)=\omega^E_0$, which is assumed to be slightly detuned against the $|s\rangle-|d\rangle$ transition. In general the parabolic approximation to $\omega_k^E$ is justified by tuning the laser frequency in the vicinity of a dispersion minimum or maximum. In particular, tuning to within the band gap creates a bound state, since the exchange photon cannot propagate and becomes localized around the atom that has emitted it \cite{Douglas2015a}. This bound state with a localization length
\begin{align}
L_E=\sqrt{\alpha_E/\kappa_E}
\end{align}
facilitates a strong interaction with other atoms within the region of localization, which takes the form $\sim \hat{a}^\dagger_s(z)\hat{a}^\dagger_s(z')\hat{a}_s(z')\hat{a}_s(z)$. 
On the other hand, since we are eventually interested in the interaction induced modifications to the dispersion of the propagating photons, we will require no approximations to the dispersion $\omega^P_k$. As already stressed above, the actual form of the photon dispersion does not play a qualitative role.\\

In summary, the EIT-Hamiltonian from the previous section is extended as follows 
\begin{align}
\begin{split}
\hat{H}=\hat{H}&_\text{EIT}\\+\hbar&\left[\sum_z\left\{\omega_d\hat{a}_d^\dagger(z)\hat{a}_d(z)+\frac{1}{L}\int_k \left(\omega_k^E\hat{b}_E^\dagger(k) \hat{b}_E(k)\right.\right.\right.\\&\left.\left.\left.+g_E \left(e^{i k z} u_k^E(z)\hat{b}_E(k)\hat{a}^\dagger_d(z) \hat{a}_s(z) + h.c.\right)\right)\right\}\right]\,.
\end{split}
\end{align}
Furthermore, the losses of state $|d\rangle$ are modeled in the already familiar linearized approximation
\begin{align}
\mathcal{L}_{\gamma_d}\rho&=-\hbar\sum_z \frac{\gamma_d}{2}\left(\left\{\hat{a}_d^\dagger(z)\hat{a}_d(z),\rho\right\}-2\hat{a}_d(z)\rho\hat{a}_d^\dagger(z)\right)\,.
\end{align}
In a functional-integral formulation the non-interacting part of the action still takes the same form as before, with the Green's functions replaced by the full expressions given in App.~\ref{app:matrix_GF}. The interactions between exchange photons and atoms give rise to an additional term in the interacting part, which now reads
\begin{widetext}
\begin{align}\label{eq:Sint_full}
\begin{split}
S_\text{int}=&\frac{1}{\sqrt{2}}\int\frac{d\omega}{2\pi}\int\frac{dk}{2\pi}\sum_z\\ & \left\{g_P e^{ikz} u_k^P(z)\left[b_P^\text{q}(k) \left(\bar{a}_e^\text{q}(z) a_g^\text{q}(z)+\bar{a}_e^\text{cl}(z)a_g^\text{cl}(z)\right)+b_P^\text{cl}(k)\left(\bar{a}_e^\text{cl}(z)a_g^\text{q}(z)+\bar{a}_e^\text{q}(z) a_g^\text{cl}(z)\right)\right]\right.\\&\left.+g_E e^{ikz} u_k^E(z)\left[b_E^\text{q}(k) \left(\bar{a}_d^\text{q}(z) a_s^\text{q}(z)+\bar{a}_d^\text{cl}(z)a_s^\text{cl}(z)\right)+b_E^\text{cl}(k)\left(\bar{a}_d^\text{cl}(z)a_s^\text{q}(z)+\bar{a}_d^\text{q}(z) a_s^\text{cl}(z)\right)\right]+c.c.\right\}.
\end{split}
\end{align}
\end{widetext}
With this being said, the topology of the Feynman diagrams in matrix notation is not affected by these extensions, only the associated values of Green's functions and vertices change.

We do however have to verify the validity of the non-linear Feynman rules. The dominant effect experienced by probe photons, i.e.~the hybridization with excited atoms, remains the same and is perfectly captured by the simplified non-linear Feynman rules discussed in the previous section. However, for higher order self-energy corrections to the probe photon propagator that involve the exchange photon, as well as for the polarization bubble of the exchange photon itself, the simplified Feynman rules do not work quite as well. The reason for this is that both states $|s\rangle$ and $|d\rangle$, necessarily appearing in the polarization bubble of an exchange photon, have non-vanishing self-energies. There might be then an interval in real time where these insertions into the bare propagators are incompatible with each other due to the atomic Hilbert space restriction. This means that the simplified non-linear Feynman rules, which act in a non-time-resolved fashion, no longer correctly capture the polarizability of the atoms. However, if the effective coupling rate between states $|s\rangle$ and $|d\rangle$ is small compared to $\gamma_d$, the excited atom will likely have decayed before it can be transferred into another state. To ensure this, we will exclusively work in a regime of small $\Omega_s/\gamma_d$. Note, however, that this condition will be significantly modified upon inclusion of strong interpolariton interactions, wherefore we will also require $\left(\Omega_s^\text{eff}\right)^2/(\gamma_d^\text{eff}\gamma_s^\text{eff})\ll 1$ for the fully dressed quantities.\\
In order to test that the choice of the specific implementation of the non-linear Feynman rules -- of which many different versions are available -- does not affect the results, we compare the two extreme options. One is the most strict implementation of the Feynman rules, where all diagrams that could at least partially be forbidden are entirely excluded. The other option corresponds to the opposite choice, where all at least partially allowed diagrams are fully included. In the following, we will refer to these two options as the ``strict'' and ``lenient'' implementation of the Feynman rules. If we observe only small differences between the results from both options, the ambiguity in the non-linear Feynman rules is of no quantitative significance and either version can be used to provide a lowest order approximation to the actual (time-dependent) selection rules.

Before we consider the effect of exchange photons, let us first see how the properties of the EIT-polaritons are affected by coupling the state $|s\rangle$ to $|d\rangle$ via the laser with Rabi frequency $\Omega_s$, but still in the absence of $E$ photons.
In this case, the Green's function $G_P^R$ remains exactly computable in the limit of vanishing polariton density, however now the polarizability is given by
\begin{align}
\label{eq:Gepol}
G_{e}^R(\omega)=\frac{1}{\omega+i\gamma_e/2-\frac{\Omega^2}{\omega-\Delta_s+i\epsilon/2-\frac{\Omega_s^2}{\omega-\Delta_s-\Delta_d+i\gamma_d/2}}}\,.
%G_{e}^R(\omega)=\frac{1}{\omega-\frac{\Omega^2}{\omega-\Delta_s-\frac{\Omega_s^2}{\omega-\Delta_s-\Delta_d+i\gamma_d/2}+i\epsilon/2}+i\gamma_e/2}\,.
\end{align}
Since the admixture of $|d\rangle$ to $|s\rangle$ introduces losses $\gamma^\text{eff}_s\approx \Omega_s^2\gamma_d/(\Delta_d^2+\gamma_d^2/4)$ to the metastable atomic state -- and therefore to the dark-state polariton -- without increasing its group velocity, the waveguide is no longer fully transparent. Given that already weak losses $\gamma_s^\text{eff}\gtrsim\kappa_P/\theta$ reduce the probe photon range to $L_P^\text{eff}=v_P/(\gamma_s^\text{eff}\theta)$, the delicate transparency window is easily destroyed by a small coherent coupling on the $s-d$ transition.

\subsection{Sorting Feynman diagrams}
\label{sec:1oL_all}

\begin{figure}[htp]
\begin{center}
\includegraphics[width=\columnwidth]{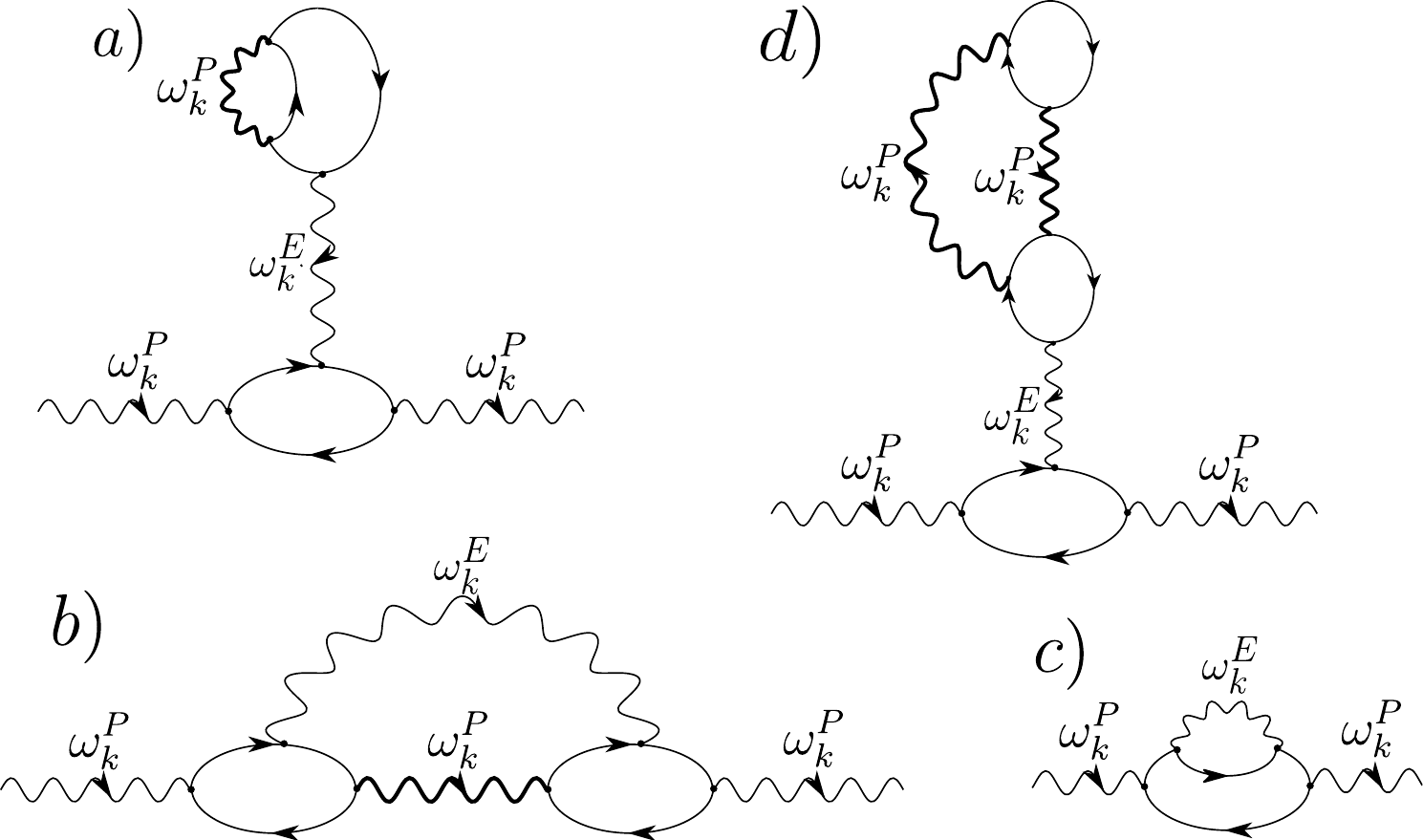}
\caption{All contributions to the polariton self-energy at next to leading order in $1/L$. The bold lines for the probe photon indicate that all powers of the leading order polarization bubble (see Fig.~\ref{fig:polbubble}a)) have to be inserted as well. For the purpose of clarity, we are not specifying the atomic states and also not using the loop-reduction simplification illustrated in App.~\ref{app:loop_reduction}.}
\label{fig:1overL}
\end{center}
\end{figure}

The strong dependence of EIT polaritons at large $\theta$ on the lifetime of the metastable state $|s\rangle$ can be exploited to enhance the effect of interactions. 
%an effect similar to that used in dissipative state preparation \cite{diehl2008quantum, Kraus2008}. 
However, one quickly realizes that to leading order in $1/L$, that is to say simultaneously in $1/L_E$ and $1/L_P$, the polaritons cannot interact. Indeed, to order $(1/L)^0$ the only interaction is a Hartree self-energy for the $s$-propagator of the type shown in Fig.~\ref{fig:perturbative}a) with $|g\rangle, |e\rangle$ and the $P$ photon replaced by $|s\rangle, |d\rangle$ and an $E$ photon, respectively. While one can include arbitrarily many Hartree insertions, as soon as a photon insertion of the type shown in Fig.~\ref{fig:perturbative}b) appears in an atomic line, it will necessarily induce a suppression by $1/L$. Avoiding such a suppression will exclude the appearance of any atomic $g$- or photonic $P$-propagators in self-energies for the $s$-propagator, and therefore prevent us from populating the $|s\rangle$ or the $|d\rangle$ level. The latter are not directly pumped and consequently, without $\mathcal{O}(1/L)$-insertions, empty. The distribution functions $F_{s,d}(\omega)$ are thus identical to one, which means that all particle-hole diagrams, i.e.~loops with counter-propagating atomic excitations, and in particular all Hartree diagrams involving $|s\rangle$ and $|d\rangle$ vanish. This is nothing else than the statement that there can be no interaction between atoms in state $|s\rangle$ if that level is not populated. 
\\
Therefore, in our expansion, interactions between polaritons only start to play a role at $\mathcal{O}(1/L)$ and the leading order investigated in the last section is indeed a theory of non-interacting polaritons. All the diagrams for the $P$-photon self-energy up to order $1/L$ are shown in Fig.~\ref{fig:1overL}. Note that, since in leading order in $1/L$ only the ground-state is occupied, the exchange photon propagator is bare. Furthermore, the version of diagram c) with the $E$-propagator
substituted by a $P$-propagator has to be excluded according to the Feynman rules discussed in section \ref{sec:NLFR}. 
In general, the order of a diagram is given by $(1/L)^{n}$, where $n$ is the number of total loops minus the number of atomic loops.\\
The fact that interactions take place at higher loop-order is a generic feature of polaritons formed by hybridizing probe photons with internal atomic excitations: If the atoms are initialized in the ground state and only probe photons are capable of exciting this initial configuration, then one will first need to populate the interacting atomic level, before atoms -- and thus polaritons -- can interact. 

%\subsection{\fs{The limit of long-ranged atom-atom interactions}}
%\label{sec:LEinf}

\subsection{Reduced theory for dissipatively-interacting polaritons}\label{sec:Gerry}

We will begin our discussion of interactions between polaritons with the limit of infinitely ranged exchange photons ($L_E\to\infty$), which implies infinitely ranged atom-atom interactions. In this case all diagrams can be resummed completely, resulting in a fully controlled field theory of a non-equilibrium system with strong light matter interactions. In particular, no further assumptions regarding $L_P$ are required and we are allowed to enter the regime of large single-atom cooperativities with respect to the propagating photons. We shall see that in this regime new many-body phases emerge and that the corresponding phase transitions can be described in a quantitative manner.

Before presenting the full theory in the $L_E\to\infty$ limit, in the present section we will consider only a particular subclass of the next-to-leading order interactions which corresponds to a spectrally-resolved mean-field approximation. We will see that already this simplified approach can have very interesting effects on the polariton transparency window and induce a phase transition in the steady state. Importantly, while this reduced set of diagrams will not typically yield quantitative results, it helps to illustrate many useful physical concepts and provides a simple application of the techniques outlined above. We therefore employ it as an instructive introduction into the theory of strongly interacting polaritons. By neglecting the information about spectral lineshapes, an even simpler and more physically transparent (though quantitatively uncontrolled) set of algebraic mean-field equations can be readily derived from the spectrally-resolved counterpart.

\subsubsection{Dyson equations}\label{sec:Dyson}

\begin{figure}[t]
\begin{center}
\includegraphics[width=\columnwidth]{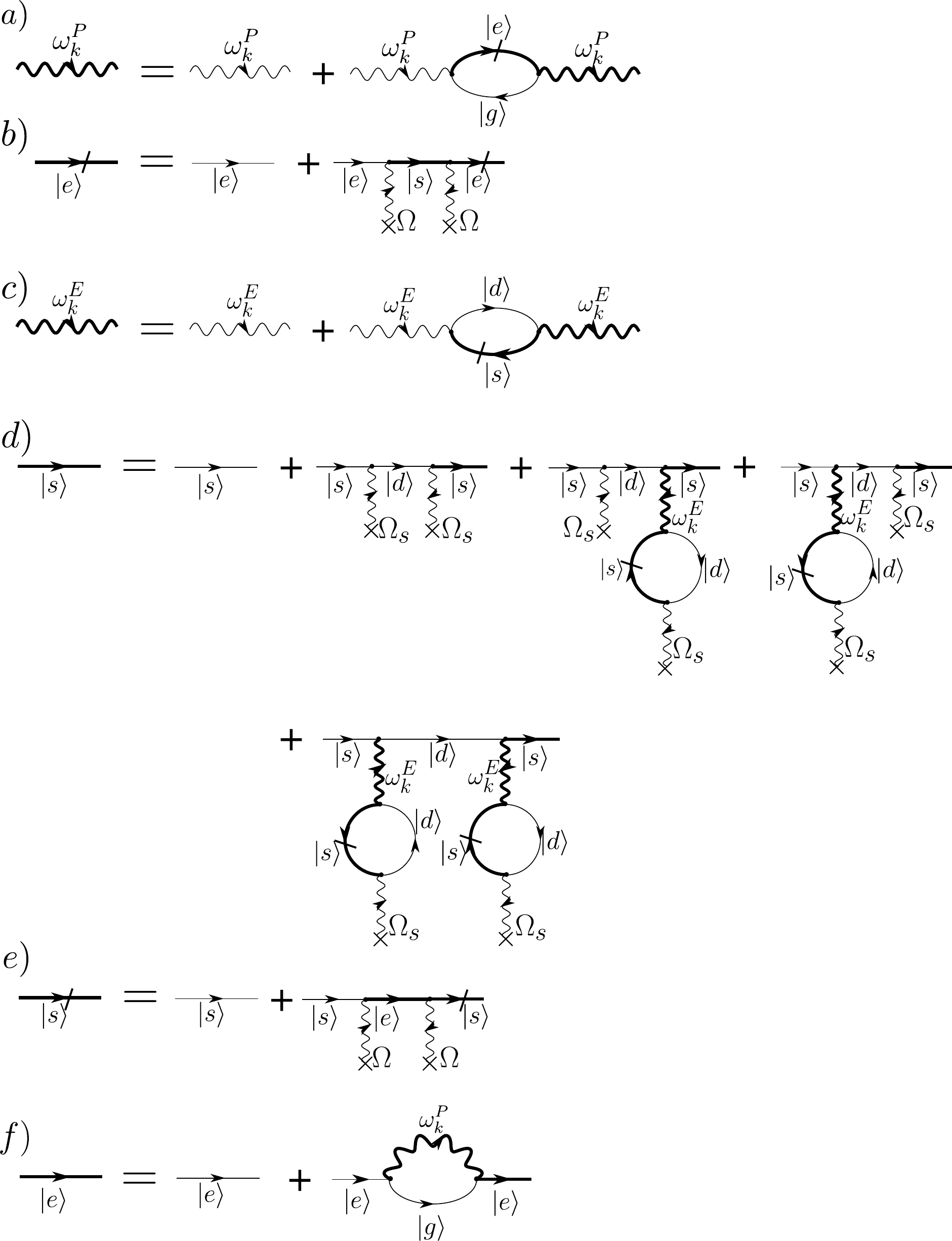}
\caption{Diagrammatic representation of the Dyson equations in the reduced Hartree-like approximation. The latter neglects all interaction diagrams at next-to-leading order except that in Fig.~\ref{fig:1overL}a). For the purpose of clarity, we refrain from using the loop-reduction simplification introduced in App.~\ref{app:loop_reduction}.}
\label{fig:Hartree}
\end{center}
\end{figure}

To distinguish between diagrams in $1/L$ we return to the simultaneous expansion in $1/L_E$ and $1/L_P$, which in next to leading order is shown in Fig.~\ref{fig:1overL}. Of these diagrams a) and d) are suppressed by $1/L_P$, c) is proportional to $1/L_E$ and b) depends on a combination of both lengths that approaches $1/\max{(L_E,L_P)}$ if both length scales differ a lot. Consequently, with $L_E\to\infty$ only diagrams \ref{fig:1overL}a) and d) need to be considered. In a perturbative expansion, that is if the single atom cooperativity $C_P^{\rm sa}=g_P^2/(\kappa_P\gamma_e L_P)\ll 1$, no self-consistent treatment beyond the resummation of all polarization bubbles that give rise to EIT (introduced already in Fig.~\ref{fig:polbubble} and indicated by the bold lines in Fig.~\ref{fig:1overL}) is required. At the same time, these weak interactions only sightly perturb the bare EIT and no qualitatively new effects are encountered. These would indeed require coupling strengths that are comparable with the bare $|s\rangle$-to-$|d\rangle$ coupling $\Omega_s$. This requirement breaks the strict confines of the $1/L_P$-expansion (see Sec.~\ref{sec:ana_gerry}). We therefore have to extend our analysis to strong single atom cooperativities, where all diagrams of the same class as \ref{fig:1overL}a) and d) have to be taken into account. Since the corresponding computations become somewhat involved, we will introduce the idea of the self-consistent resummation of a class of diagrams and the resulting physical consequences first by using only the diagram in \ref{fig:1overL}a), and emphasize connections to the random phase approximation in dynamical screening and mean-field theory. The full theory for $L_E\to\infty$ will be presented in section \ref{sec:LE_inf}. Clearly, every $|s\rangle-|d\rangle$ transition can either be directly driven by the laser acting on a single atom, or by the exchange of an $E$-photon with another atom that in turn couples to the laser. The interchangeability of the single- and multiple-atom processes gives rise to an infinite set of diagrams that is conveniently captured by a self-consistent treatment.\\
% \begin{figure}[htp]
% \begin{center}
% \includegraphics[width=0.8\textwidth]{Gerry.pdf}
% \caption{Same diagram as in Fig.~\ref{fig:1overL}a), but with all propagators and external fields labeled explicitly.}
% \label{fig:Gerry}
% \end{center}
% \end{figure}
The resulting approximation is depicted diagrammatically in Fig.~\ref{fig:Hartree}. In particular the direct and indirect absorption and emission of a laser photon coupled to the $|s\rangle-|d\rangle$ transition gives rise to the four last diagrams in \ref{fig:Hartree}d) which, involving bold lines themselves, actually correspond to an infinite number of self-energy diagrams if expressed in terms of bare propagators. Additionally, having extended the expansion from Fig.~\ref{fig:1overL} to large $C_P^{sa}$ the exchange photon is now dressed according to \ref{fig:Hartree}c). As we explain in the following, the corresponding self-consistent Dyson equations can be simplified such that they require finding only a single number $\eta$ as the solution of a nonlinear integral equation.\\
In close analogy to the formalism of Sec.~\ref{sec:EIT}, the probe photon Green's functions are dressed by excitations induced in the medium. The result
\begin{align}
G_P^R(\omega,k)&=\!\left[G_P^A(\omega)\right]^*\!=\!\frac{1}{\omega-\Delta_P(k)-\Sigma_P^R(\omega,k)+i\kappa_P/2}\nonumber\\
G_P^K(\omega,k)&=G_P^R(\omega,k)\left(\Sigma_P^K(\omega,k)-i\kappa_P-2i\kappa_s\right)G_P^A(\omega,k)
\label{eq:GMsym}
\end{align}
is therefore still fully determined by the polarization bubble, which using the Kramers-Kronig relations can again be put in the closed form
\begin{align}
\label{eq:SigmaMsym}
\Sigma_P^R(\omega,k)&=\frac{g_P^2(1-n_V) }{\omega-\Omega^2 G_s^R(\omega)+i\gamma_e/2}
\nonumber\\
\Sigma_P^K(\omega,k)&=2i\Im\Sigma_P^R(\omega,k)\,.
\end{align}
However now the propagator of state $|s\rangle$
\begin{align}\label{eq:GsR_sym}
G_s^R(\omega)&=\left[G_s^A(\omega)\right]^*=\frac{1}{\omega-\Delta_s-\Sigma_s^R(\omega)+i\epsilon/2}% \nonumber\\
% G_s^K(\omega)&=G_s^R(\omega)\left(\Sigma_s^K(\omega)-i\epsilon\right)G_s^A(\omega)\,,
\end{align}
has a modified coupling to state $|d\rangle$:
\begin{align}\label{eq:SigmasR}
% \Sigma_s^R(\omega)&=\frac{\Omega^2}{\omega-\omega_e+\omega_L^{(1)}-
%   \Sigma_{e}^R\left(\omega+\omega_L^{(1)}\right)+i\gamma_e/2}+\cancel{\Sigma}_s^R\nonumber\\
\Sigma_s^R(\omega)&=\frac{\left(\Omega_s^\text{eff}\right)^2}{\omega-\Delta_d-\Delta_s+i\gamma_d/2}\,,
\end{align}
where $\Omega_s^\text{eff}=\Omega_s\left|1+\eta\right|$ includes the effects of the direct coupling rate $\Omega_s$ as well as those due to the interactions.
Here $\eta$ is simply a complex number, which stems from the fact that the exchange photon mediating the interaction between different polaritons carries zero momentum and -- in the rotating frame -- zero frequency as well.\\
In the polarization bubbles of the exchange photon the non-linear Feynman rules forbid a dressing of $G_s$ by $|d\rangle$, which thus requires the definition of a second type of $s$-propagator
\begin{align}\label{eq:GsslashR_sym}
G_{\fsl{s}}^R(\omega)&=\left[G_{\fsl{s}}^A(\omega)\right]^*=\frac{1}{\omega-\Delta_s-\Sigma_{\fsl{s}}^R(\omega)+i\epsilon/2}% \nonumber\\
% \tilde{G}_s^K(\omega)&=\tilde{G}_s^R(\omega)\Sigma_{\fsl{s}}^K(\omega)-i\epsilon\right)\tilde{G}_s^A(\omega)\,,
\end{align}
that couples exclusively to $|e\rangle$, which in turn can emit and reabsorb a probe photon. This is accounted for by defining
\begin{align}
\Sigma_{\fsl{s}}^R(\omega)&=\frac{\Omega^2}{\omega-
  \Sigma_{e}^R\left(\omega\right)+i\gamma_e/2}
\end{align}
and
\begin{align}
\Sigma_{\fsl{s}}^K(\omega)&=2i\Im\Sigma_{\fsl{s}}^R(\omega)+\delta \Sigma_{\fsl{s}}^K(\omega)\nonumber\\
\delta \Sigma_{\fsl{s}}^K(\omega)&=\frac{\Omega^2\left(\Sigma^K_{e} \left(\omega\right)-2i\Im\Sigma^R_{e} \left(\omega\right)\right)}{\left(\omega-\Re\Sigma^R_{e} \left(\omega\right)\right)^2+\left(\gamma_e/2-\Im\Sigma^R_{e}
\left(\omega\right)\right)^2}\,.
\end{align}
It is at this point, that the self-consistency loop closes, meaning that the Dyson equations form a closed set of equations: The self-energy $\Sigma_e^R$ depends only on the probe photon propagator from Eq.~\eqref{eq:GMsym} and the bare ground-state Green's function via
\begin{widetext}
\begin{align}\label{eq:Sigma_e^R_Gerry}
\Sigma_e^R(\omega)=\frac{i}{2}g_P^2\int_{-\infty}^\infty\frac{d\omega'}{2\pi}\int_{-\pi}^{\pi}\frac{dk}{2\pi}G_P^R(\omega-\omega',k)G_g^K(\omega')+G_P^K(\omega-\omega',k)G_g^R(\omega')
\end{align}
\end{widetext}
and
\begin{align}\label{eq:Sigma_e^K_Gerry}
\begin{split}
\delta\Sigma_e^K(\omega)&=\Sigma_e^K(\omega)-2i\Im{\Sigma_e^K(\omega)}\\&=\!\frac{i}{2}g_P^2\!\!\int_{-\infty}^\infty\!\!\frac{d\omega'}{2\pi}\!\!\int_{-\pi}^{\pi}\!\frac{dk}{2\pi}\delta G_P^K(\omega-\omega',k)\delta G_g^K(\omega')\,,
\end{split}
\end{align}
As announced at the beginning of this section, the self-consistent functional equations $G_{P,\star}^R=G_P^R\left[G_{P,\star}^R,G_{P,\star}^K\right]$ and $G_{P,\star}^K=G_P^K\left[G_{P,\star}^R,G_{P,\star}^K\right]$ have been reduced to a single parameter satisfying a fixed point equation $\eta_\star=\eta\left(\eta_\star\right)$. As mentioned before, this is in part due to the Hartree nature of the interactions considered here, which implies that the functional form of $\Sigma_{s}^R$ is fixed and analytically known. On the other hand it is a consequence of the non-linear Feynman rules, which enforce an unoccupied propagator $G_s^K$ and therefore $\Sigma_s^K=2i\Im{\Sigma_s^R}$, which reduces the number of coupled equations.\\
The frequency integral in the first of the two expressions in Eq.~\eqref{eq:Sigma_e^R_Gerry} is trivial, as $G_g^K(\omega)\propto \delta(\omega)$. Since the poles of $G_P^R$ can be found analytically, also the frequency integral in the second term of $\Sigma_e(\omega)$ can be solved exactly via the residue theorem, such that only the momentum integration has to be evaluated numerically. After application of the residue theorem one obtains
\begin{widetext}
\begin{align}\label{eq:Sigmae}
\Sigma^R_{e}(\omega)=&-\sum_n \int\frac{dk}{2\pi}g_P^2\frac{\kappa_s}{2\Im(\omega_n(k))}\frac{1}{\omega-\omega_n(k)+i\epsilon/2}\frac{f(\omega_n(k)) f^*(\omega_n^*(k))}{\prod_{m\neq n}(\omega_n(k)-\omega_m(k)) (\omega_n(k)-\omega_m^*(k))} \notag \\
&+\int \frac{dk}{2\pi} \frac{1}{2}g_P^2(4-2n_V) G_P^R(\omega+i\epsilon/2,k)\nonumber\\
\Sigma^K_{e}(\omega)=&2i\Im\Sigma^R_{e}(\omega)-i\kappa_s \int \frac{dk}{2\pi} g_P^2(2-2n_V) G_P^R(\omega,k)G_P^A(\omega,k)\,,
\end{align}
where $n \in \{1,2,3,4\}$, $\omega_n(k)$ are the poles of $G_P^R(\omega,k)$ and
\begin{align}
f(\omega)=(\omega+i\gamma_e/2)(\omega-\Delta_s+i\epsilon/2)(\omega-\Delta_s-\Delta_d+i\gamma_d/2)- \Omega^2(\omega-\Delta_s-\Delta_d+i\gamma_d/2)-\left(\Omega_s^\text{eff}\right)^2(\omega+i\gamma_e/2)\,.
\end{align}
\end{widetext}
With all Green's functions depending solely on the parameter $\eta$, we are left with the task to solve for it self-consistently. The
corresponding equation can again be read off from Fig.~\ref{fig:Hartree} and states
\begin{align}\label{eq:Cconst}
\eta=\frac{\Sigma_E^R(0)}{-\Delta_E(0)-\Sigma_E^R(0)+i\kappa_E/2}\,.
\end{align}
So far, there is no ambiguity regarding the non-linear Feynman rules. In the polarization bubbles of the exchange photon however, these partially forbid dressing the propagator of state $|d\rangle$ via couplings to the metastable state. Employing the strict interpretation where $G_d^R$ remains undressed, the exchange photon self-energy reads
\begin{align}\label{eq:SigmaE}
\Sigma_E^R(\omega)=\frac{i}{2}\int\frac{d\omega'}{2\pi}g_E^2 G_{\fsl{s}}^R(\omega') G_{\fsl{s}}^A(\omega')\delta \Sigma_{\fsl{s}}^K(\omega')G_d^R(\omega+\omega')
\end{align}
with
\begin{align}
G_d^R(\omega)=G_{d_0}^R(\omega)=\frac{1}{\omega-\Delta_s-\Delta_d+i\gamma_d/2}\,.
\end{align}
If on the other hand the lenient rule is applied one is to use
\begin{align}\label{eq:lenient}
G_d^R(\omega)=\left(\left[G_{d_0}^R\right]^{-1}(\omega)-\frac{\left(\Omega_s^\text{eff}\right)^2}{\omega-\Delta_s-\frac{\Omega^2}{\omega+i\gamma_e/2}}\right)^{-1}\,,
\end{align}
which includes all possible admixtures of atomic states to $|d\rangle$, as the insertion of the ground-state can always be excluded by the methods introduced in Sec.~\ref{sec:NLFR}. Furthermore, $\Sigma_{\fsl{s}}^R$ is to be complemented by
\begin{align}\label{eq:Sigma_fsl_mod}
\Sigma_{\fsl{s}}^R\to\Sigma_{\fsl{s}}^R+\frac{\left(\Omega_s^\text{eff}\right)^2}{\omega-\Delta_s-\Delta_d+i\gamma_d/2}\,,
\end{align}
with the dependence of $\delta\Sigma_{\fsl{s}}^K$ on $\Sigma_e^R$ and $\delta\Sigma_e^K$ remaining unaffected.\\
Choosing among these two ways of applying Feynman rules affects the propagation of the exchange photons 
and hence the light-mediated atom-atom interactions. The photon propagator is ultimately given by
\begin{align}
G_E^R(\omega,k)&=\left[G_E^A(\omega)\right]^*=\frac{1}{\omega-\omega_E(k)-\Sigma_E^R(\omega,k)+i\kappa_E/2}\nonumber\\
G_E^K(\omega,k)&=G_E^R(\omega,k)\left(\Sigma_E^K(\omega,k)-i\kappa_E\right)G_E^A(\omega,k)\,,
\end{align}
with 
\begin{align}\label{eq:Gerry_last}
\Sigma_E^K(\omega,k)=\Sigma_E^{K_0}(\omega,k)=2i\Im\Sigma_E^R(\omega,k)\,.
\end{align}\\
Interestingly, the phase of $\eta$ can be adjusted via the detuning between the band-edge of the exchange photon and the laser $\Omega_s$. Its amplitude depends on the spectral density of atoms in the metastable state $n_s(\omega)$ and on the coupling constants, giving a great deal of control over the type and strength of backaction to be realized.\\
For numerical purposes, iterating equations~\eqref{eq:GMsym} through \eqref{eq:Gerry_last} having previously initialized the system with some $\Omega_s^\text{eff}=\Omega_s$ is very inefficient, as convergence will fail when approaching a phase-transition \cite{Lang2016}. We avoid this problem by instead fixing $\Omega_s^\text{eff}$ and determining $\Omega_s(\Omega_s^\text{eff},\eta)$, which requires no iterations at all. This actually means that the value of $\Omega_s$ corresponding to the solution is not known a priori. However, for the computation of the entire phase-diagram this does not matter as eventually a result for any value of $\Omega_s$ will have been produced.\\

%\fps{\subsubsection{Reformulation in terms of polarizabilities and connection to mean-field theory}}
\subsubsection{Qualitative picture from mean-field approximation}

Before we proceed to the numerical analysis, it is instructive to develop an intuitive understanding of the mechanisms in effect here. To do so, we neglect the $\omega$ and $k$ dependence of the indirect drive of state $|s\rangle$ and assume its linewidth to vanish. In this case equation \eqref{eq:Cconst} simplifies to
\begin{align}\label{eq:MF3}
\eta\approx\frac{2 g_E^2 n_s}{i\gamma_d(-\Delta_E(0)+i\kappa_E/2)-2g_E^2 n_s}\,.
\end{align}
Employing the previously discussed ratio between atomic and photonic admixtures $\theta\approx g_P^2(1-n_V)/\Omega^2$ and extending the arguments from section~\ref{sec:EIT_results}) to finite lifetime of the state $|s\rangle$ to approximate the density of excitations in the latter one finds
\begin{align}\label{eq:MF2}
n_s\approx\frac{\theta}{1+\theta}\frac{2\kappa_s}{\kappa_P+\theta\gamma_s^\text{eff}}\,.
\end{align}
This depends on the effective lifetime of that state induced by coupling to $|d\rangle$, which for large losses $\gamma_d\gg|\Delta_s+\Delta_d|$ is given by
\begin{align}\label{eq:MF1}
\gamma_s^\text{eff}\approx\frac{4\Omega_s^2|1+\eta|^2}{\gamma_d}\,.
\end{align}

These simple algebraic equations are able to describe qualitatively the relatively complex feedback mechanism resulting from the screening of laser induced losses. This type of feedback is different from the more direct one obtained from density-density interactions in Rydberg gases \cite{weath_rybist_exp_2013,diehl_rydbist_2014,weath_rybist_exp_2016, maghrebi_bistability_2016}, where the real interaction-shift of the polariton energy detunes the latter with respect to the transparency window (see section \ref{sec:Rydberg}).
  
\begin{figure}[htp]
\begin{center}
\includegraphics[width=\columnwidth]{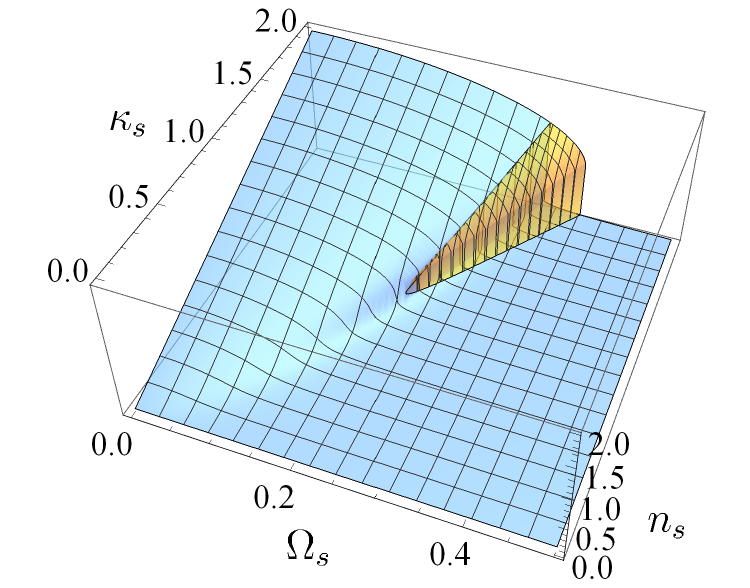}
\caption{The mean-field approximation to the polariton density $n_s$ shows a bistability for strong driving $\kappa_s\gtrsim 1$ and otherwise the same parameters as in Fig.~\ref{fig:EIT} ($\kappa_0=2$, $\omega_0=\Delta_s=n_V=0$, $g_P=10$, $\kappa_P=0.5$, $\gamma_e=1$, $\Delta_P(k)=-50 \cos{k}$) together with $\Delta_E(k=0)=-1$, $\Delta_d=0$, $\gamma_d=10$, $\kappa_E=5$ and $g_E=10$.}
\label{fig:MF}
\end{center}
\end{figure}

The numerical solution of Eqs.~\eqref{eq:MF2} to \eqref{eq:MF3} is shown in Fig.~\ref{fig:MF}. It exhibits a strongly nonlinear dependence of $n_s$ on $\Omega_s$, which for sufficient drive strength $\kappa_s$ even gives rise to a bistability, i.e.~increasing $\Omega_s$ the system will evolve along the yellow surface where it exists, whereas in the opposite direction $n_s$ always stays on the blue surface. While this demonstrates that the mean-field approach can qualitatively capture strong interactions, equation~\eqref{eq:MF2} inaccurately approximates the line-shape of the interacting EIT polaritons resulting in an unphysically large fraction $n_s>1$ of atoms in state $|s\rangle$ (see Fig.~\ref{fig:MF_vs_Gerry}).
The inability to provide accurate and quantitative predictions within mean field physically arises because of the approximation of a constant mixing angle, which is only valid to lowest order in an expansion around the unperturbed EIT window. Furthermore, the mean field theory is not easily extended to finite interaction ranges. This motivates the field theoretic method introduced in the next sections.

\subsubsection{Results: Non-equilibrium phase transition of the transparency window}
\label{sec:IIT}

\begin{figure*}[htp]
\begin{center}
\includegraphics[width=\textwidth]{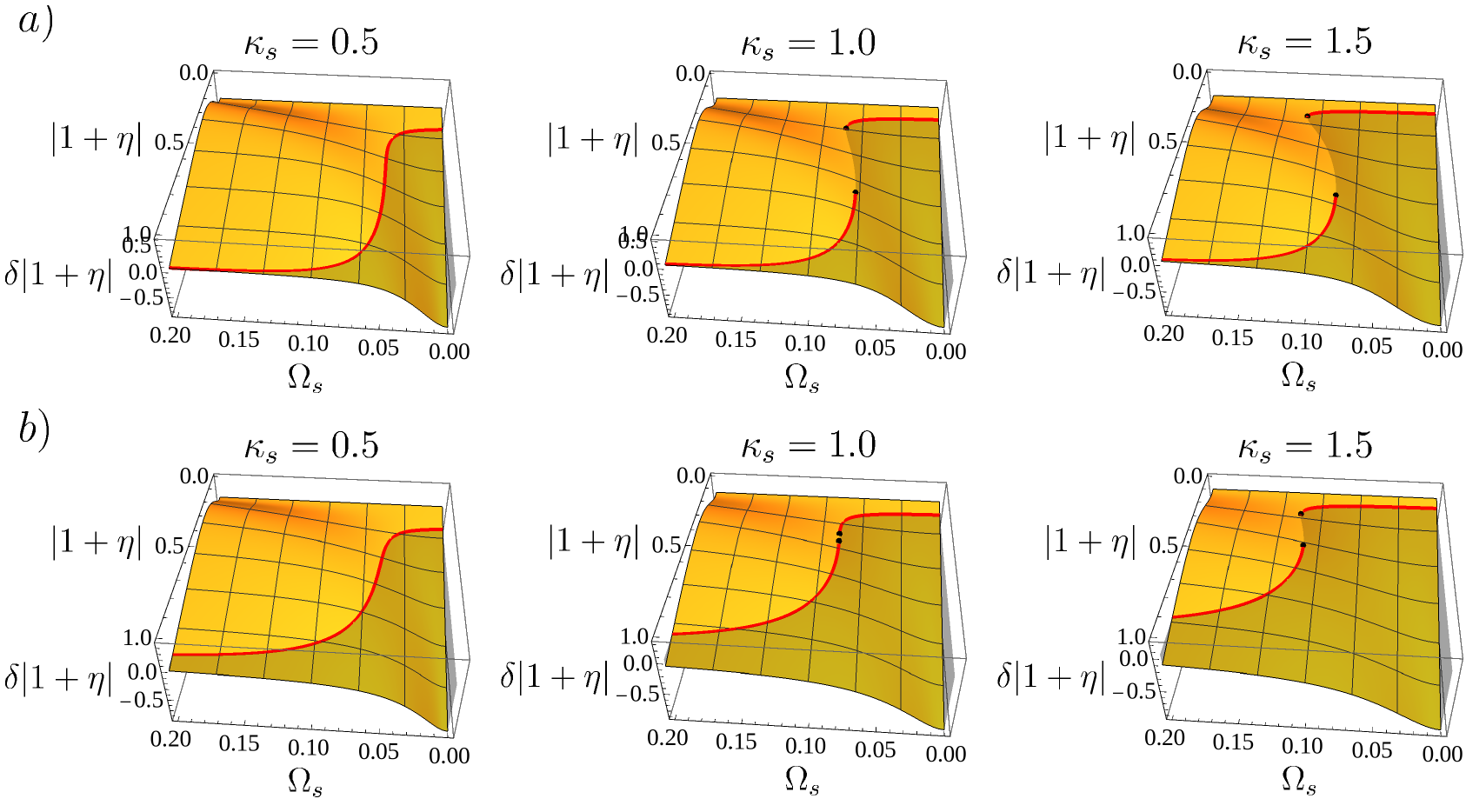}
\caption{Flow diagram of the effective relative coupling strength $|1+\eta|$ as a function of the externally adjustable parameter $\Omega_s$, where otherwise the same parameters as in Fig.~\ref{fig:EIT} ($\kappa_0=2$, $\kappa_s=1$, $\omega_0=\Delta_s=n_V=0$, $g_P=10$, $\kappa_P=0.5$, $\gamma_e=1$, $\Delta_P(k)=-50 \cos{k}$) are used together with $\Delta_E(k=0)=-1$, $\Delta_d=0$, $\gamma_d=1$, $\kappa_E=5$ and $g_E=10$ and lenient(strict) Feynman rules in the first(second) row.}
\label{fig:Flow0}
\end{center}
\end{figure*}
We now show how a non-perturbative field theoretic treatment not only gives rise to a quantitatively accurate description of a phase transition and bistability, but predicts other features that are not possible to capture within mean field theory. As can be gleaned from the mean-field equations~\eqref{eq:MF2} to \eqref{eq:MF1}, the restoration of the transparency window arises from a destructive interference between the laser and the exchange photon that drastically reduces the coupling to state $|d\rangle$, which is predicted by our approach and involves the last four diagrams in Fig.~\ref{fig:Hartree}d\\
This many-body phenomenon, which can be named ``interaction-induced transparency'' as opposed to the standard single-particle ``electromagnetically-induced transparency'', is analyzed elsewhere \cite{lang_EIT_short}, also in relation to its observability for realistic experimental parameters in the context of PCW and tapered fibers. In the remainder of this section, we provide a complementary analysis focusing on the nature of the underlying non-equilibrium phase transition and discuss the fundamental mechanism from a more formal perspective as an application of our diagrammatic approach.\\
The reconstruction of the transparency window can be attributed to the positive feedback brought about by the dependence of $\eta$ on the excitation density: $\eta \propto n_s$, which stabilizes both a low density i.e. opaque phase and a high density i.e. transparent phase, separated by a first-order phase transition. The mechanism behind this can be understood by studying Fig.~\ref{fig:Flow0}, which shows the amplitude and sign of the variation in the flow of the quantity $|1+\eta|$ during the evaluation of the self-consistency equation \eqref{eq:Cconst}. If the system is initialized with a certain value of $\eta$ such that $\delta|1+\eta|$ is positive, the system will flow towards the opaque phase and vice versa, if $\delta|1+\eta|<0$, the system is unstable towards the transparent phase. Consequently, only those parameter combinations with $\delta|1+\eta|=0$ and a negative slope in $\delta|1+\eta|$ as a function of $|1+\eta|$ are stable and therefore marked with a red line in Fig.~\ref{fig:Flow0}. In sufficiently strongly driven systems we witness the emergence of a bistability: for a given Rabi amplitude $\Omega_s$ two stable solutions exist. They differ significantly in the effective coupling $\Omega_s^\text{eff}$ and in the occupation of dark-state polaritons. Quite surprisingly we find a stable transparent solution with $\Omega_s^\text{eff}\ll\Omega_s$, which entails significantly reduced losses compared to the non-interacting case with $g_E=0$. Remarkably, the stable ratio $\Omega_s^\text{eff}/\Omega_s$ is smallest for purely dissipative interactions, that is, when $\Sigma_E^R(0)$ is purely imaginary. In this case, the phase shift between the E-photon-mediated driving of the $s-d$ transition and the direct driving via $\Omega_s$ is the most destructive. This results in small losses for the dark-state polaritons, at least if there are enough to create a sufficiently large backaction in the form of $\Sigma_E^R(0)$. A comparison between Fig.~\ref{fig:Flow0}a) and \ref{fig:Flow0}b) demonstrates that for these rather small values of $\Omega_s$ the choice of the non-linear Feynman rules does not affect the results appreciably. For the remainder of this section, we will therefore focus on the strict implementation of the Feynman rules.\\

\begin{figure*}[htp]
\begin{center}
\includegraphics[width=\textwidth]{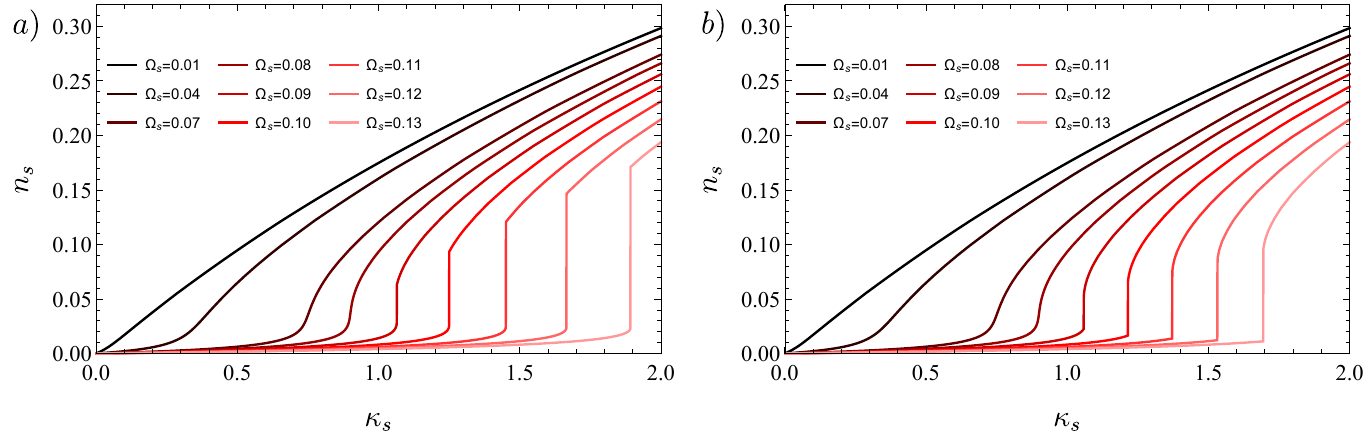}
\caption{Hysteresis of the polariton density, evidenced in $n_s$ for scans at the incoherent drive strength $\kappa_s$ for different values of $\Omega_s$. Brighter colors corresponds to larger values of $\Omega_s$. In panel a) the system is initialized in the opaque phase with $\kappa_s=0$, whereas panel b) uses $\kappa_s=2$ in the transparent phase as a starting point. Below the critical Rabi amplitude $\Omega_{s_c}\approx0.0851$ both scans are identical. However above $\Omega_{s_c}$ the initial phase is stabilized against fluctuations induced by slow scans and a hysteresis curve becomes observable. The parameters used are the same as in Fig.~\ref{fig:Flow0}.}
\label{fig:ScanUpDown}
\end{center}
\end{figure*}

\begin{figure}[htp]
\begin{center}
\includegraphics[width=\columnwidth]{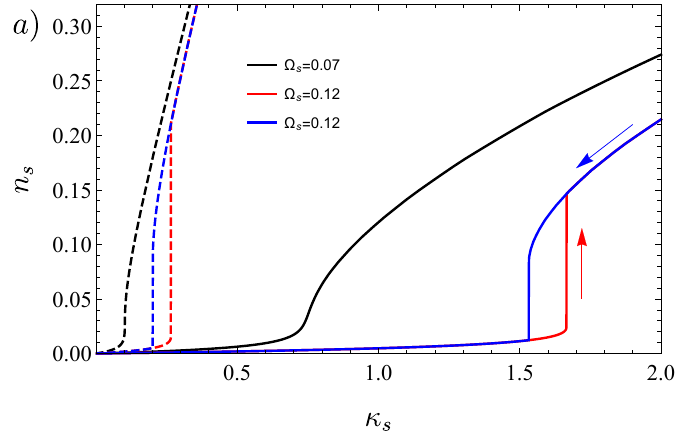}
\caption{A comparison between the spectrally resolved solution to Eqs.~\eqref{eq:GMsym} through \eqref{eq:Gerry_last} (full lines) with the mean-field theory of Eqs.~\eqref{eq:MF2} to \eqref{eq:MF3} (dashed lines) highlights the quantitative inadequacy of mean-field theory. In black we show a scan of the drive strength $\kappa_s$ at $\Omega_s=0.07$, which is below the critical Rabi amplitude in both cases. In red and blue a scan through the bistable regime is performed with the former (latter) increasing (lowering) the drive. The parameters have been chosen as in Fig.~\ref{fig:Flow0}.}
\label{fig:MF_vs_Gerry}
\end{center}
\end{figure}

\begin{figure}[htp]
\begin{center}
\includegraphics[width=\columnwidth]{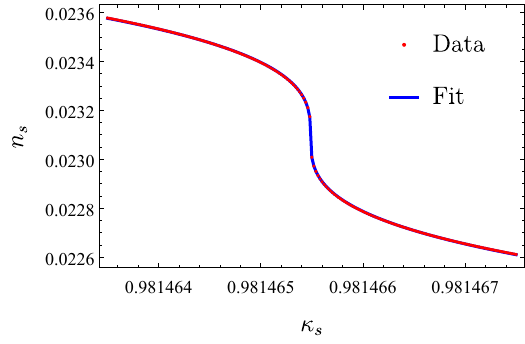}
\caption{Power law fit to the density of excited atoms at the critical coupling strength $\Omega_{s_c}\approx 0.0851$ for the same parameters as in Fig.~\ref{fig:Flow0}. The critical exponent of the order parameter as a function of the drive strength is determined to be $\delta=3\pm0.01$.}
\label{fig:critExp}
\end{center}
\end{figure}

\begin{figure}[htp]
\begin{center}
\includegraphics[width=\columnwidth]{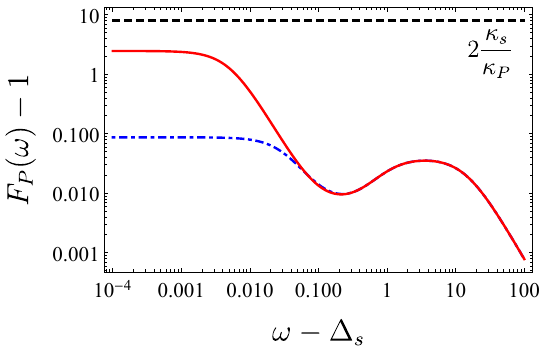}
\caption{Double logarithmic plot of the distribution functions $F_P(\omega)$ of the transparent phase in red and the opaque phase dash-dotted and in blue for the same parameters as in Fig.~\ref{fig:Flow0}. Near the EIT condition $F_P-1$ differs by more than an order of magnitude.}
\label{fig:FpGerry}
\end{center}
\end{figure}

\begin{figure*}[htp]
\begin{center}
\includegraphics[width=\textwidth]{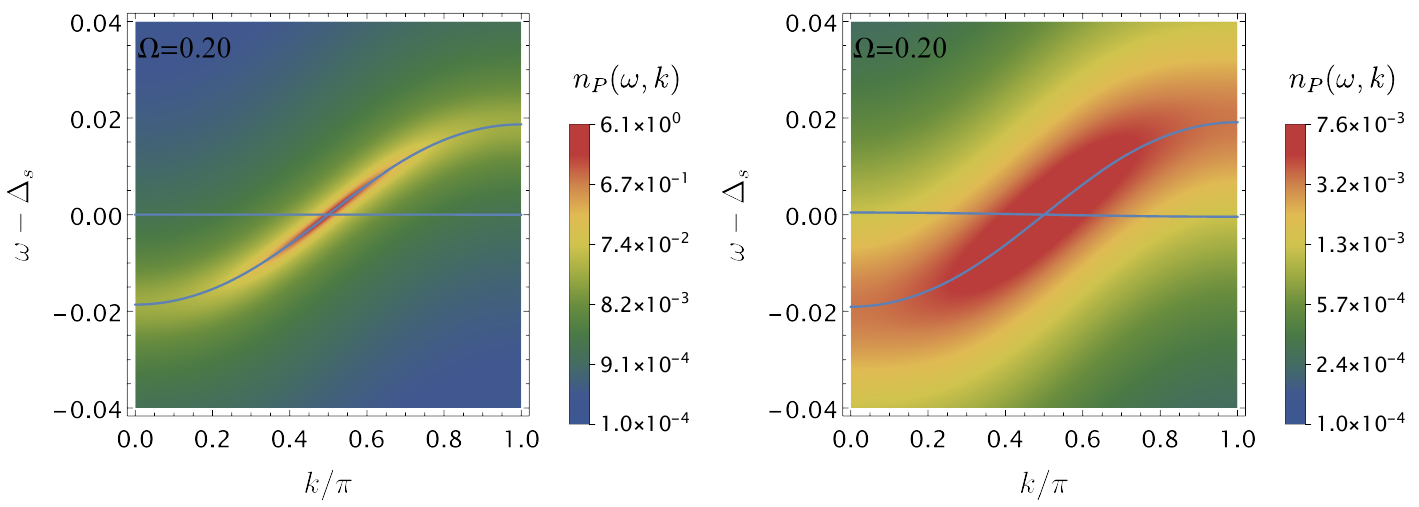}
\caption{The frequency and momentum resolved photon number density $n_P(\omega,k)$ in the transparent phase (left) exhibits an almost perfect transparency window, whereas the opaque phase (right) with the same parameters shows strongly dissipative polaritons (note the difference by almost three orders of magnitude in the maximal spectral density). Results are shown for $\Omega_s=0.14$, where for the parameters of Fig.~\ref{fig:Flow0} combined with $\kappa_s=2$ the system is bistable. The almost flat blue line corresponds to the atomic level $|d\rangle$, that hybridizes with the probe photon to form a fourth, bright polariton branch. For the selected small value of $\Omega_s$ this hybridization remains weak and the new polariton branch consequently is essentially unoccupied.}
\label{fig:EITGerry}
\end{center}
\end{figure*}

In combination with the possibility of the simultaneous stability of an opaque and a transparent phase, a first order phase transition similar to that between a gaseous and a liquid phase emerges: above a critical bare laser strength $\Omega_{s_c}$ an increasingly strong hysteresis is observed as the source intensity $\kappa_s$ is increased. This is shown in Fig.~\ref{fig:ScanUpDown} (see also Fig.~\ref{fig:MF_vs_Gerry} for a comparison with the mean-field approximation). However, at exactly the critical laser strength, the first order phase transition ends in a critical point, where the phase transition is continuous and of mean-field type. This is to be expected by a Hartree-type theory with infinitely ranged interactions and we verify this by fitting the numerical data for $n_s(\kappa_s,\Omega_{s_c})$ with a power law and extracting the critical exponent $\delta=3\pm0.01$, consistent with the Ising universality class\cite{Ma_book}, (see Fig.~\ref{fig:critExp}).\\
We note that in the regime of the first order phase transition, the difference in polariton density between the opaque and transparent solution is typically large. This can be seen from the distribution function (see Fig.~\ref{fig:FpGerry}) as well as from the frequency- and momentum-resolved photonic number density of Fig.~\ref{fig:EITGerry}. One thus concludes that, far away from the critical point in the opaque phase the system behaves essentially as a non-interacting theory: the occupation numbers are so small that interactions via exchange photons play no role and the bare -- but due to $\Omega_s$, lossy -- EIT is recovered. \\
In the transparent phase on the other hand an only weakly perturbed three-level-scheme is restored, which seems to imply that the effective degrees of freedom are again only weakly interacting.
Correspondingly, many simple correlation functions can be described by an effective free theory. However, except for the limit of vanishing $\Omega_s$, the response of the system to external perturbations will be very different compared to the free theory discussed in Sec.~\ref{sec:EIT}.

\subsubsection{Analytic estimates and requirements of the bistable regime}
\label{sec:ana_gerry}

Due to the simplicity of the reduced theory presented in this section, we can actually give some analytic estimates for the conditions necessary for a phase transition.
Due to the typically large atomic admixture $\theta$ to the dark-state polaritons, even for relatively strong driving $\kappa_s \sim \kappa_P$, a slow group velocity gives rise to only a small photon number density
\begin{align}
n_P=i\int\frac{d\omega}{4\pi}\int\frac{dk}{2\pi}\delta G^K_P(\omega,k)<\theta n_s\ll 1\,.
\end{align}
Here the first inequality results from the fact that only photons in a narrow frequency interval actually form dark-state polaritons. Most photons instead hybridize into bright polaritons, that involve the decaying excited atomic states, resulting in even smaller occupations.\\
Of the two contributions to $\Sigma_e^R$ in \eqref{eq:Sigmae}, the second one thus dominates. Typically, in PCW or tapered fibers, the photonic bandwidth is several orders of magnitude larger than the inverse life times of all atomic states. It is therefore well justified to approximate the photon spectrum as linear. We do so by writing their retarded Green's function as a sum of left- and right-movers
\begin{align}
\begin{split}
G_P^R(\omega)=&\frac{1}{\omega-\Delta_P^{(0)}-v_P k-\Sigma_P^R(\omega)+i\kappa_P}\\&+\frac{1}{\omega-\Delta_P^{(0)}+v_P k-\Sigma_P^R(\omega)+i\kappa_P}.
\end{split}
\end{align}
For $L_P=\kappa_P/v_P\gg 1$ the EIT window in momentum space is much narrower than the inverse lattice constant $1/a$ and thus far away from the band edge a linearized spectrum suffices to reproduce the results obtained from any Bloch wave with the same group velocity in the EIT window.\\
Together with the observation that, since the atoms are fixed in space, $\Sigma_P^R(\omega)$ is momentum independent, this allows to find
\begin{align}
\begin{split}
\Sigma_e^R&\approx-i g_P^2(2-n_V)|u_k^P(0)|^2/v_P\\
\delta\Sigma_e^K&\approx-g_P^2(1-n_V)\frac{2}{v_P}\frac{i\kappa_s}{\kappa_P/2-\Im{\Sigma_P^R(\omega)}}\,,
\end{split}
\end{align}
where the momentum integral has been approximated by an integral along the entire real axis.
This result can be used to approximate the number density of atoms in the metastable state by
\begin{widetext}
\begin{align}\label{eq:ns_approx}
\begin{split}
n_s&=i\int_{-\infty}^{\infty}\frac{d\omega}{4\pi}|G_s^R(\omega)|^2\delta\Sigma_{\fsl{s}}^K(\omega)=i\int_{-\infty}^{\infty}\frac{d\omega}{4\pi}|G_{\fsl{s}}^R(\omega)|^2|G_e^R(\omega)|^2\delta\Sigma_{e}^K(\omega)
%\\&=\int_{-\infty}^{\infty}\frac{d\omega}{2\pi}\kappa_s(\omega)g_P^2\left(\int_{-\pi}^{\pi}\frac{dk}{2\pi}|G_P^R(\omega,k)|^2\right)\left|\frac{\Omega}{(\omega-\Delta_s)\left(\omega-2 g_P^2\left(\int_{-\pi}^{\pi}\frac{dk}{2\pi}G_P^R(\omega,k)\right)+i\gamma_e/2\right)-\Omega^2}\right|^2
\\ &\approx \int_{-\infty}^{\infty}\frac{d\omega}{2\pi}\kappa_s(\omega)\left|\frac{1}{\frac{(\omega-\Delta_s)}{\Omega^2}\left(\omega+(2-n_V)i\frac{g_P^2}{v_P}+i\gamma_e/2\right)-1}\right|^2\frac{g_P^2}{\Omega^2}\frac{1-n_V}{v_P\left(\kappa_P/2-\Im{\Sigma_P^R(\omega)}\right)}\,.
\end{split}
\end{align}
\end{widetext}
As can be extracted from Fig.~\ref{fig:Flow0} the system becomes bistable once
\begin{align}
0\overset{!}{>}\frac{d\Omega_s}{d\Omega_s^\text{eff}}=\frac{d}{d\Omega_s^\text{eff}}\frac{\Omega_s^\text{eff}}{|1+\eta(\Omega_s^\text{eff})|}\,,
\end{align}
which, using the explicit form \eqref{eq:Cconst} can be rewritten as
\begin{align}
\frac{d\eta}{d\Omega_s^\text{eff}}\overset{!}{>}\left|\frac{-\Delta_E(0)+i\kappa_E/2}{\Omega_s^\text{eff}\left(-\Delta_E(0)-\Sigma_E^R(0)+i\kappa_E/2\right)}\right|\,.
\end{align}
In the ideal case of a resonance between the exchange photon and the corresponding laser ($\Delta_E(0)=0$) as well as strong coupling $g_E$, such that $|\Sigma_E^R(0)|\gg\kappa_E$, this still requires
\begin{align}
\frac{d n_s \Omega_s^\text{eff}}{d\Omega_s^\text{eff}}<0\,.
\end{align}
A condition that can be satisfied only if
\begin{align}\label{eq:Bi_cond}
\Im{\Sigma_P^R(0)}-\Omega_s^\text{eff}\frac{d}{d\Omega_s^\text{eff}}\Im{\Sigma_P^R(0)}>\kappa_P/2\,,
\end{align}
where we used \eqref{eq:ns_approx} with the absolute value approximated by unity as an upper bound. Since the minimum of the frequency dependent loss rate
\begin{align}
-\Im{\Sigma_P^R(\omega)}\approx \sigma + \xi (\omega-\Delta_s)^2
\end{align}
with
\begin{align}
\sigma=\frac{2 (\Omega_s^\text{eff})^2 g_P^2(1-n_V)|u_P^k(0)|^2}{\gamma_e (\Omega_s^\text{eff})^2+\tilde{\gamma}_d\Omega^2}
\end{align}
and
\begin{align}
\tilde{\gamma}_d=\frac{\gamma_d^2+4\Delta_d^2}{\gamma_d}\,,
\end{align}
for slow polaritons is tightly focused around $\omega=\Delta_s$, this is a reasonably good approximation. Using the just stated expansion of the probe photon self-energy around $\Delta_s$, one finds the left hand side of Eq.~\eqref{eq:Bi_cond} to be maximized for 
\begin{align}
\Omega_s^\text{eff}=\sqrt{\frac{\gamma_d \Omega^2}{3\gamma_e}}\,,
\end{align}
where one finds a strong collective coupling satisfying
\begin{align}
g_P^2>2\gamma_e\kappa_P
\end{align}
or equivalently a large collective cooperativity $C_P>2$ to be a necessary condition for the emergence of a bistability. While, due to the rough approximations used here, this is only a lower bound on the collective cooperativity, it clearly shows that the type of phase transition discussed here is not amenable to a purely perturbative approach.\\
Instead of calculating a lower bound for the collective cooperativity $C_P$ we can also search for a rough estimate that includes all relevant scales. To do so, we approximate $n_s\approx p  C_P^{\rm sa}\Omega_s^2/\Omega^2$, where $p=2\kappa_s(0)/\kappa_P$ is the pump ratio, indicating how strongly the probe photons near the EIT condition are driven compared to their losses. Inserting this expression for $n_s$ into $|\Im\Sigma_E^R(0)|\gtrsim \kappa_E$, which is necessary for a highly non-perturbative regime, yields the final strong coupling condition 
\begin{align}\label{eq:strong_coupling}
p C_E C_P^{\rm sa} \frac{\Omega_s^2}{\Omega^2}\gtrsim 1\,.
\end{align}
In agreement with the superficial analysis of Sec.~\ref{sec:Dyson}, we find that a phase transition needs a single atom cooperativity for the probe photon of order one, but is nevertheless accessible to an expansion in $1/L_E$ as only the collective cooperativity $C_E$ has to be of comparable size.
An actual bistability additionally requires an efficient backaction of the losses in the dressed state $|s\rangle$ onto the dark-state polariton density. Therefore, typical systems that exhibit a phase transition satisfy Eq.~\eqref{eq:strong_coupling} by more than one order of magnitude. For example, for the parameters of the critical point in Fig.~\ref{fig:critExp}, one has $p C_E C_P^{\rm sa} \Omega_s^2/\Omega^2\approx 58$.

\subsection{Quantitative theory in the infinite-range limit}
\label{sec:LE_inf}

\begin{figure*}[htp]
\begin{center}
\includegraphics[width=0.9\textwidth]{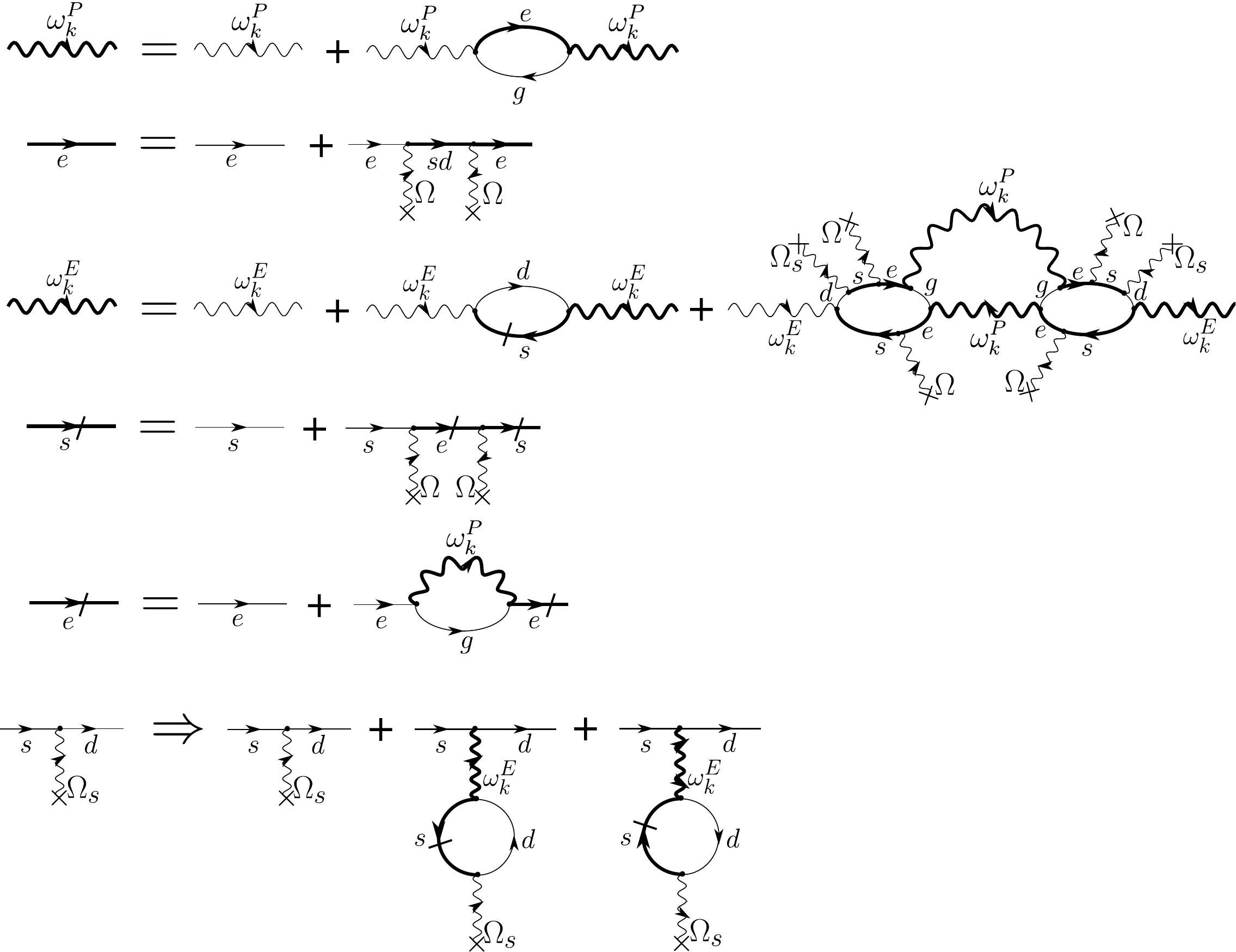}
\caption{Complete diagrammatics in the limit $L_E\to\infty$. Note that we do not show the anomalous components of the self-energy for the E-photon. Those are however obtained from the last diagram in the second line upon an exchange of laser and E-photons acting on the same transition. We included the anomalous components in the calculation (see App.~\ref{app:anomalousGF}).}
\label{fig:Gerry+ECar}
\end{center}
\end{figure*}

The reduced class of diagrams discussed in the previous section is helpful to obtain a general idea about the emergence of a phase transition between the two limits of a perfectly restored transparency window deep within the transparent phase on the one hand, and an empty system in the opaque phase on the other hand. Our main goal, however, is the quantitative description that extends all the way to the critical point and the bistable region. In order to achieve this, one has to include all diagrams that can be created self-consistently from the two diagrams in Fig.~\ref{fig:1overL}a) and d). The resulting theory is illustrated in terms of Feynman diagrams in Fig.~\ref{fig:Gerry+ECar}, which differs from the reduced theory of the previous section by the addition of the Fock diagram to the Dyson equation of the exchange photon (see last diagram in the third line of Fig.~\ref{fig:Gerry+ECar}).
Note that at this level of the theory the exchange photon obtains a Nambu structure, which requires us to extend the Kramers-Kronig relations of App.~\ref{app:KK} to anomalous Green's functions, which we do in App.~\ref{app:anomalousGF}.

\subsubsection{Dyson equations}\label{sec:Gerry+ECar}
Having introduced the anomalous non-equilibrium Green's functions we can now solve the self-consistent Dyson equations shown in Fig.~\ref{fig:Gerry+ECar}, where in order to simplify the notation we have introduced the matrix Green's function $\mathcal{G}_{sd}$ for the states $|s\rangle$ and $|d\rangle$. In absence of any diagrams of order $1/L_E$, this propagator is fully determined by the corresponding submatrix of $\left[G_{a,0}^R\right]^{-1}$ (see Eq.~\eqref{eq:matrixG0}), but with the effective Rabi amplitude $\Omega_s^\text{eff}=\Omega_s|1+\eta|$:
\begin{align}
\begin{split}
\mathcal{G}^{R/K}_{sd}=&\begin{pmatrix}
G^{R/K}_{ss}&&G^{R/K}_{sd}\\
G^{R/K}_{ds}&&G^{R/K}_{dd}
\end{pmatrix}\\=&\begin{pmatrix}
\omega-\Delta_s+i\epsilon/2&&-\Omega_s^\text{eff}\\
-\Omega_s^\text{eff}&&\omega-\Delta_d-\Delta_s+i\gamma_d/2
\end{pmatrix}^{-1}\,.
\end{split}
\end{align}
In fact, as indicated by the last line in Fig.~\ref{fig:Gerry+ECar}, and in analogy to Sec.~\ref{sec:Gerry}, $\Omega_s$ has to be replaced everywhere by $\Omega_s^\text{eff}$ and $G^{R/K}_{ss}$ supersedes the otherwise identical expression $G^{R/K}_s$ used in Sec.~\ref{sec:Gerry}.
Apart from these notational remarks, the only physical difference between the present theory and the one discussed in section \ref{sec:Gerry} is in the propagator of the exchange photon, which acquires a new self-energy contribution $\Sigma_{E}^{R_2}$:
\begin{align}\label{eq:GERtwo}
G^R_E=\left(\left[G^{R_0}_E\right]^{-1}-\Sigma^{R_1}_{E}-\Sigma^{R_2}_{E}\right)^{-1}\,.
\end{align}
While the first term $\Sigma^{R_1}_E(\omega)$ remains exactly the same as Eq.~\eqref{eq:SigmaE}, the second, due to the Nambu structure takes the lengthy form
\begin{widetext}
\begin{align}\label{eq:ECarR}
\begin{split}
\Sigma^{R_2}_E(\omega,k)&=\frac{i}{2}g_P^4 g_E^2\Omega^4\Omega_s^2|1+\eta|^2(1-n_V)^2\int\frac{d\omega'}{2\pi}\frac{dp}{2\pi}\left|G^R_e(\omega')G_{ss}^R(\omega')\right|^2\delta G^K_P(\omega',p)\\\times&\left[\left[G^R_{ss}(\omega+\omega')G^R_e(\omega+\omega')\right]^2 G^R_P(\omega+\omega',p+k)\begin{pmatrix}
\left[G^R_d(\omega+\omega')\right]^2&& G^A_d(\omega')G^R_d(\omega+\omega')\\
G^R_d(\omega')G^R_d(\omega+\omega')&&\left|G^R_d(\omega')\right|^2
\end{pmatrix}\right.\\&\left.+\left[G^A_e(\omega'-\omega)G^A_{ss}(\omega'-\omega)\right]^2G^A_P(\omega'-\omega,p-k)\begin{pmatrix}
\left|G^R_d(\omega')\right|^2&&G^A_d(\omega'-\omega)G^A_d(\omega')\\
G^A_d(\omega'-\omega)G^R_d(\omega')&&\left[G^A_d(\omega'-\omega)\right]^2
\end{pmatrix}\right]\,.
\end{split}
\end{align}
\end{widetext}
Some care has to be taken when it comes to determining $\eta$: $\Sigma_E^{R_2}$ is actually indistinguishable from $\Sigma_E^{R_1}$ once one of their external legs is substituted with the laser field $\Omega_s$. Consequently, coupling to the coherent field with $\Sigma_E^{R_2}$ would overcount the diagrams in the last line of Fig.~\ref{fig:Gerry+ECar}. Therefore, $\eta$ is given by
\begin{align}\label{eq:chi}
\eta=\sum_{j}\Sigma^{R_1}_{E_{jj}} G^R_{E_{j1}}\bigg|_{k,\omega=0}\,.
\end{align}
Note that strictly speaking using the real and positive definition $\Omega_s^\text{eff}=\Omega_s|1+\eta|$ in the anomalous components of the exchange photon Green's function is wrong, since it leads to an incorrect behavior of $G_E$ under a global $U(1)$ transformation. This does not matter, however, since all observables depend only on the gauge invariant $|1+\eta|^2$, which allows us to simplify our calculations. By fixing the real value $\Omega_s^\text{eff}$ one can then directly determine the corresponding experimentally relevant parameter $\Omega_s$. From a computational point of view, this makes for a very cheap calculation, as the two-dimensional convolution in Eq.~\eqref{eq:ECarR} -- which has to be calculated only once -- only has to be evaluated at $k=\omega=0$.\\
Similar to the previous section, the simplified non-linear Feynman rules are not uniquely defined and we thus again have to choose between the strict and lenient way of implementing the rules in order to estimate the error bounds of the simplified diagrammatics. We do so in the same fashion as before, i.e. for the strict rule we use $G_{d_0}^{A/R}$ in Eqs.~\eqref{eq:ECarR}, and \eqref{eq:SigmaE}. For the lenient version we employ $G_d^{A/R}$ according to Eq.~\eqref{eq:lenient} together with the replacement \eqref{eq:Sigma_fsl_mod} for the very same equations.

\begin{figure*}[htp]
\begin{center}
\includegraphics[width=0.8\textwidth]{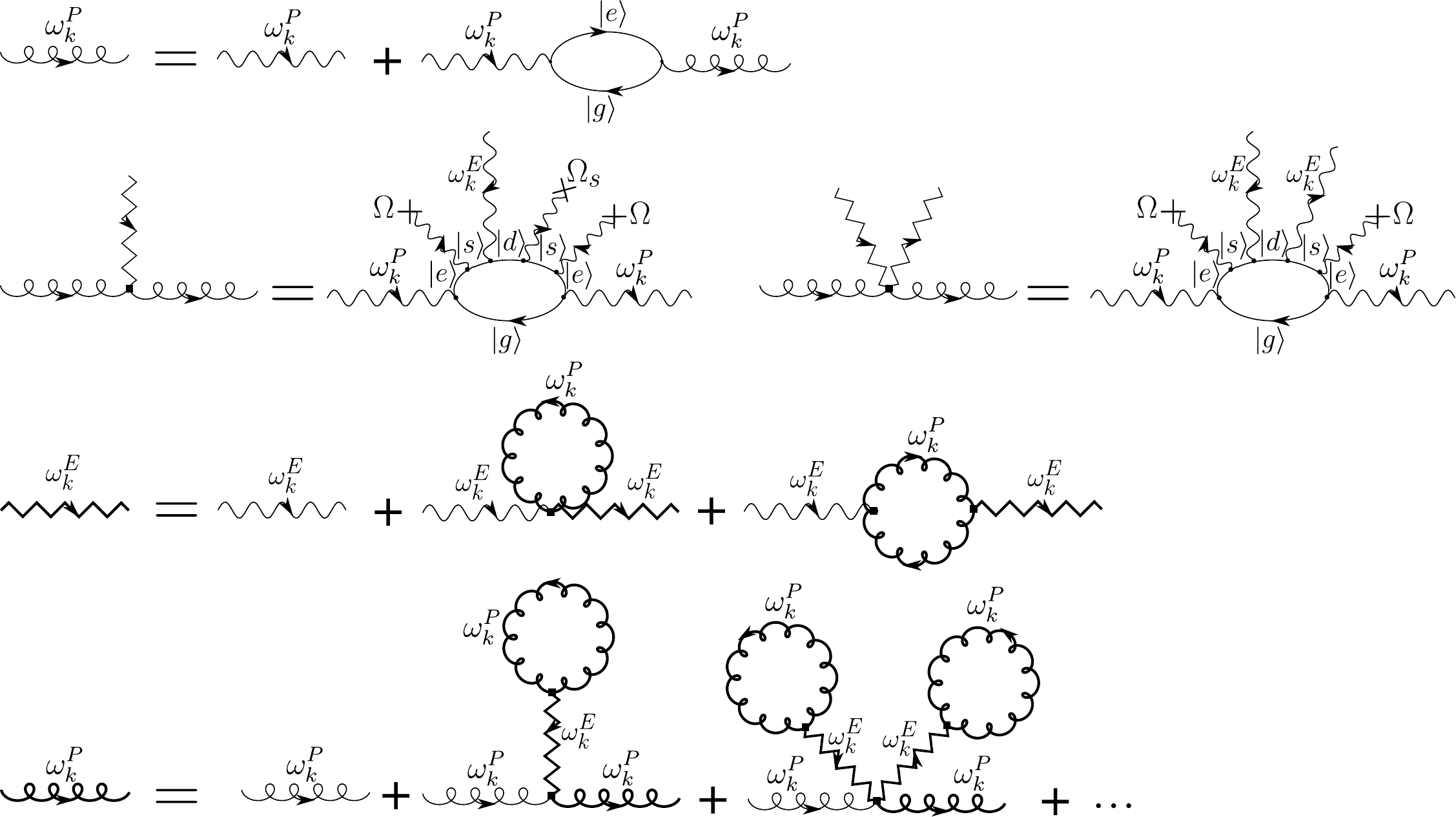}
\caption{Effective theory of dark-state polaritons in the limit of infinite interaction range, i.e. $L_E\to\infty$.  While there is an infinite set of vertices coupling a single probe photon to an arbitrary number of exchange photons, these are conveniently summed up in the geometric series embedded in $\Sigma_P^R$. To make this apparent, in the first and third line we therefore introduce propagators for the dressed probe and exchange photons, which are nothing else than the polaritons, that form the relevant collective excitations of the system. The second line then defines the first two vertices in perturbation theory, with higher orders generated in the same fashion. With these new, effective degrees of freedom and corresponding interactions, the theory shown in Fig.~\ref{fig:Gerry+ECar} becomes completely equivalent to the geometric series in the last line.}
\label{fig:effTheory}
\end{center}
\end{figure*}

Before we proceed to discuss the results obtained from the set of coupled Dyson equations introduced in this section, it is instructive to view these calculations from a more conceptual point of view: despite the potentially large single atom cooperativity experienced by the probe photons, their density is assumed to be small, such that dark-state polaritons in the absence of exchange photons are non-interacting quasi-particles. This is correctly captured by the non-linear Feynman rules, which allow for an exact diagrammatic solution of the interacting theory in the $g-e-s-P$ sector. If we now consider the additional coupling to level $d$ and include the $E$-photons, we can eliminate the atomic degrees of freedom to obtain an effective theory for the dressed propagating and exchange photons. Indeed, on the one hand the atomic level structure contains the microscopic details necessary for the formation of polaritons, which within the effective theory is incorporated in the dressed $P$-photons, and on the other hand the atoms serve as interaction vertices between one probe photon and an arbitrary number of exchange photons. While the latter may be strongly dressed with probe photons themselves, there are only two processes for this that are allowed by the atomic vertices, namely those in the third line of Fig.~\ref{fig:Gerry+ECar}. 
The diagrammatic representation of the effective theory is shown in Fig.~\ref{fig:effTheory}. We stress that this is completely equivalent to the theory presented in Fig.~\ref{fig:Gerry+ECar}. In the first line of Fig.~\ref{fig:effTheory}, the free polariton propagator is defined and indicated as a curly-line. In the second line the interaction vertices between the polariton and the $E$-photons are illustrated. Out of these, only the first two are shown but actually an infinite number of $E$-lines is allowed in the vertex indicated by the dots in the last line of Fig.~\ref{fig:effTheory}, where all possible interaction-corrections to the polariton self-energy are shown. Luckily all of these vertices can be conveniently resummed as a geometric series, as we have demonstrated earlier in the derivation of the self-consistent equations. Similarly, all possible contributions to the $E$-photon self-energy are shown in the third line. However, as every vertex has to involve exactly one probe photon, the number of diagrams here is limited to two.

\subsubsection{Quantitative results and validity}
\begin{figure*}[htp]
\begin{center}
\includegraphics[width=\textwidth]{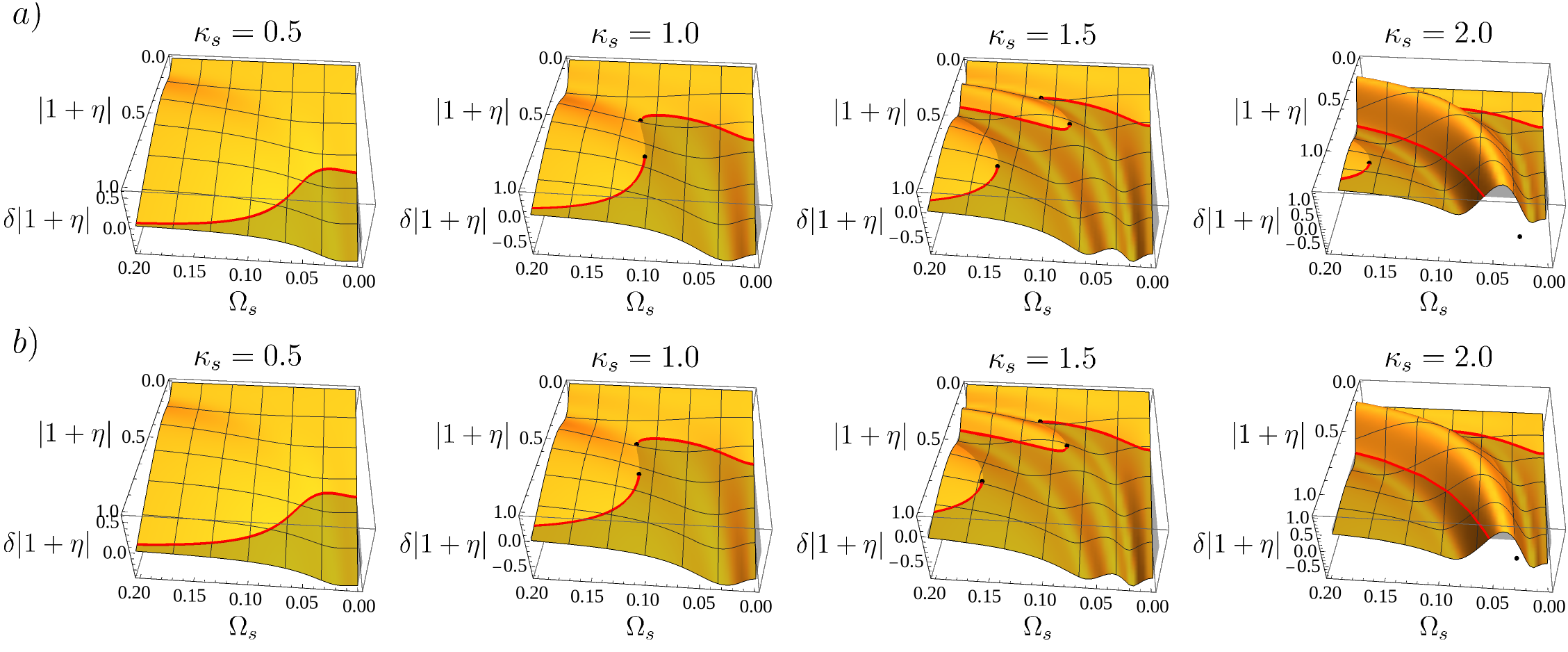}
\caption{Flow diagram of the effective relative coupling strength $|1+\eta|$ as a function of the externally adjustable parameter $\Omega_s$, where apart from $\gamma_d=10$ the same parameters as in Fig.~\ref{fig:Flow0} ($\kappa_0=2$, $\kappa_s=1$, $\omega_0=\Delta_s=n_V=0$, $g_P=10$, $\kappa_P=0.5$, $\gamma_e=1$, $\Delta_P(k)=-50 \cos{k}$, $\Delta_E(k=0)=-1$, $\Delta_d=0$, $\kappa_E=5$, $g_E=10$) are used together with the lenient and strict interpretation of the non-linear Feynman rules in the first row and second row, respectively. Note the emergence of a tristable region, where in addition to the opaque and transparent phases a new, strongly interacting semi-transparent phase appears.}
\label{fig:Flow1}
\end{center}
\end{figure*}

With the inclusion of all effects at leading order in $1/L_E$, $\eta$ is no longer bounded from below by $-1$. In fact, it can achieve arbitrarily small values, which can be understood by a closer examination of the effects of $\Sigma_E^{R_2}$ in terms of the effective theory in Fig.~\ref{fig:effTheory}, where it is represented as the last diagram of the third line. Within this framework, one immediately realizes that $\Sigma_E^{R_2}$ describes in fact a particle-hole excitation of a probe photon. Since, however, this photon itself is strongly dressed by the medium inducing EIT, its distribution $F_P(\omega)$ is sharply peaked. This allows for a resonant reallocation of photons from highly occupied frequencies and momenta towards low-occupation regions, by means of the particle-hole excitations in $\Sigma_E^{R_2}$. Where this is possible, it will act as a locally inverted environment for the exchange photons, thereby effectively driving them. Since there is no other diagram to counter this effect, the exchange photon propagator can develop a divergence, resulting in $\eta\to-\infty$, which is unphysical. While in general there is nothing wrong with the inverted bath experienced by the exchange photons, one has to pay attention to its effect on the cooperativity $C_E$.% \jlc{No, actually $C_E\propto 1/\kappa_E$ grows, but in the limit $L_E\to\infty$ this has no impact on $C_E^{\text{sa}}=0$, only at the transition $C_E^\text{sa}$ is a problem, since the order of the limits $L_E\to\infty$ and $\kappa_E\to 0$ matters. However I believe that the combination of these two limits is somewhat pathological, since there is no effect that prevents the condensation if $L_E\to\infty$ and only at the transition everything breaks down at once. So yes, the argument I give here is incomplete, because I use the divergence of $C_E$ to argue about the need to include corrections in $1/L_E$, which is only true for a finite $L_E$, which is not subject to this section. But it is hard to find a way around this, given that I cannot identify any stabilizing effect in the Hartree diagrams.}
The latter namely grows as the divergence in $G_E^R$ is approached. Consequently, diagrams at higher order in $1/L_E$ have to be included and these will in turn prevent the unphysical instability in the exchange photon propagator. We will outline the underlying processes in the next section. Nevertheless, as long as $L_E$ remains large enough, $-\eta$ can still become large without forcing us to include subleading orders in $L_E$. This can happen to such an extent that it actually overcompensates the bare coupling $\Omega_s$ up to the point where a new, strongly interacting phase emerges. This new phase, which will be referred to as ``intermediate phase'', is stable, as evidenced by the flow diagrams \ref{fig:Flow1}, which we show again both for the lenient and the strict implementation of the Feynman rules. As there is hardly any quantitative differences between the two versions, we will in the following focus on the strict rules.\\
Previously, we presented an argument for the emergence of the bistability, whereby an increase in $\Omega_s^\text{eff}$ was met with a sufficiently fast decrease of $n_s$ (and of $\eta$), so that $\Omega_s$ itself was reduced, resulting in a non-unique identification $\Omega_s(\Omega_s^\text{eff})$, i.e. a bistability. It is exactly the opposite effect that stabilizes the intermediate phase, whereby for small $\Omega_s$ an increase of $\Omega_s^\text{eff}$ increases the efficiency of the drive experienced by $G_E^R$, such that $|\eta|$ grows until this effect is exactly balanced by the effects of increased losses discussed in Sec.~\ref{sec:Gerry}. If this happens at $\eta\lesssim -1$ a stable intermediate phase exists.

\begin{figure}[htp]
\begin{center}
\includegraphics[width=\columnwidth]{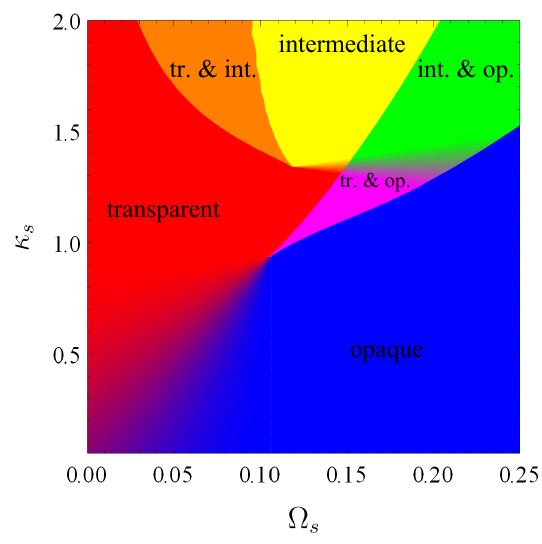}
\caption{The quantitative phase diagram in the limit $L_E\to\infty$ and with the parameters of Fig.~\ref{fig:Flow1}, shows three distinct phases. While the transparent and opaque phase can be adiabatically connected to free theories far away from the multistable regime, the same cannot be said for the strongly interacting intermediate phase. The region of coexistence between opaque and transparent phase is indicated in magenta, that between transparent and intermediate phase in orange, and the remaining bistable area in green. All multistable regions are labeled by the initial characters of the coexisting phases.}
\label{fig:PD1b}
\end{center}
\end{figure}

As can be observed in Fig.~\ref{fig:PD1b}, where the losses $\gamma_d$ have been increased tenfold compared to Fig.~\ref{fig:Flow0}, the stability of the transparent phase is strongly enhanced in comparison with the results of Sec.~\ref{sec:Gerry}. This is a consequence of the slow dark-state polaritons, which require that each probe photon during its lifetime excites on average multiple atoms. As such, while the field content of the two contributions $\Sigma_E^{R_1}$ and $\Sigma_E^{R_2}$, as well as the relative detunings between atoms, lasers and guided photons, allow no distinction between them, $\Sigma_E^{R_2}$ is favored combinatorically by a factor $\sim C_P^{\rm sa}$. In essence, on can think of the last diagram in the third line of Fig.~\ref{fig:Gerry+ECar} or equivalently Fig.~\ref{fig:effTheory} as an antenna increasing the amplitude of the indirect coupling beyond that of the direct laser driven transition between $|s\rangle$ and $|d\rangle$. Consequently, slow polaritons with infinitely ranged interactions are typically dominated by these diagrams. If the gain of the antenna $\sim C_P^\text{sa}$ is large and increases sufficiently with $\Omega_s^\text{eff}$, as suggested by the leading dependence $\Sigma_E^{R_2}\sim(\Omega_s^\text{eff})^2$, it can counteract the reduction in polariton density, thereby stabilizing the intermediate phase.

\begin{figure}[htp]
\begin{center}
\includegraphics[width=\columnwidth]{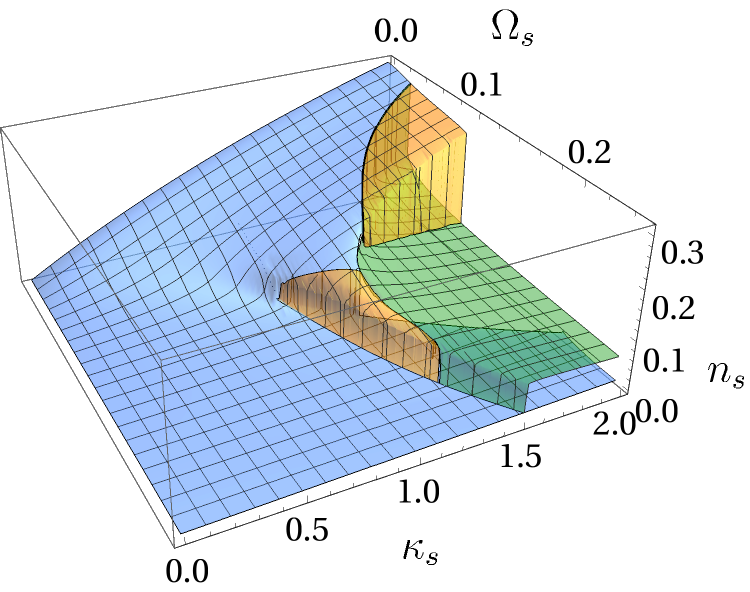}
\caption{The number density of atoms in the metastable state $|s\rangle$ can be used to characterize the three distinct phases. For the slow polaritons obtained for the parameters of Fig.~\ref{fig:Flow1} that are also used here, $n_s$ is a good estimate of the dark-state polariton density. The density of the intermediate phase is highlighted in green and those of the adiabatically connected transparent and opaque phases in blue. If these coexist the transparent solution is shown in yellow. As a testament to the overcompensation of $\Omega_s$ by $\eta$ the density of the strong coupling intermediate phase decreases as drive intensity $\kappa_s$ is increased.}
\label{fig:ns1b}
\end{center}
\end{figure}

\begin{figure*}[htp]
\begin{center}
\includegraphics[width=\textwidth]{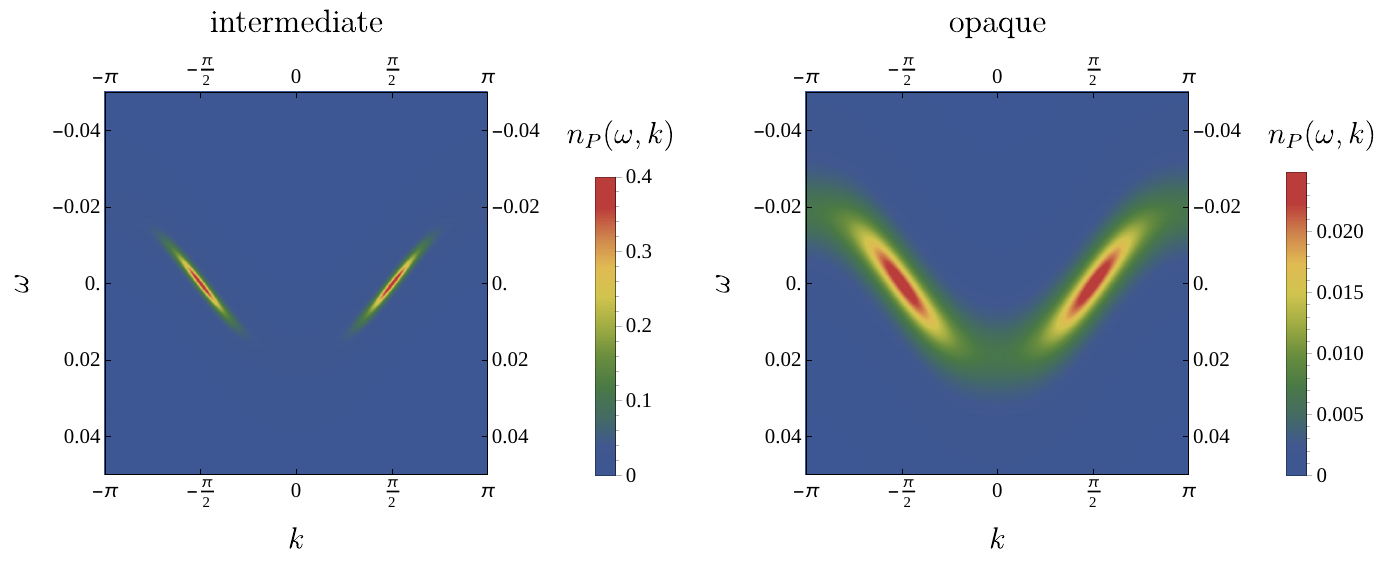}
\caption{Comparison of the transparency window for the two different stable phases shows a distinct ordering in the brightness of the dark-state polaritons. Except for $\kappa_s=2$ and $\Omega_s=0.21$ the parameters of Fig.~\ref{fig:Flow1} were used.}
\label{fig:GPK_1b}
\end{center}
\end{figure*}

As is indicated by the color gradients in Fig.~\ref{fig:PD1b}, the transparent and opaque phase are adiabatically connected. The same is true for the transparent and intermediate phase as the latter emerges from the former at large drive strengths $\kappa_s$. 
In order to more closely investigate the properties of each phase, we provide a plot of the number density of atoms in the state $|s\rangle$ (Fig.~\ref{fig:ns1b}), which shows that in every phase the polariton density and therefore their lifetime decreases as $\Omega_s$ is increased. However, in case of the intermediate phase $n_s$ and the polariton lifetime decrease also with increasing $\kappa_s$, which implies that the interaction strength is increased. This demonstrates that the intermediate phase is indeed stabilized by the overcompensation of $\Omega_s$ via strong interactions and its properties are not directly linked to either the weakly interacting limit $\Omega_s/\kappa_s\to\infty$ or the unperturbed polaritons at $\kappa_s/\Omega_s\to\infty$. We therefore use the strong backaction condition of a negative slope in the polariton density $d n_s/d\kappa_s$ as the defining property to distinguish between the transparent and intermediate phase in Fig.~\ref{fig:ns1b}. The relatively low density and the increased linewidth of the dark-state polaritons (see Fig.~\ref{fig:GPK_1b}) in the intermediate phase actually helps with the numerical investigation, as the discretization of momentum and frequency space can be performed at a lower resolution and saturation effects can more readily be discarded.


% \subsection{Verification: Controlled expansion to finite $L_E$}
\subsection{Verification: Corrections at order $1/L_E$}
\label{sec:LE_finite}
\begin{figure}[htp]
\begin{center}
\includegraphics[width=\columnwidth]{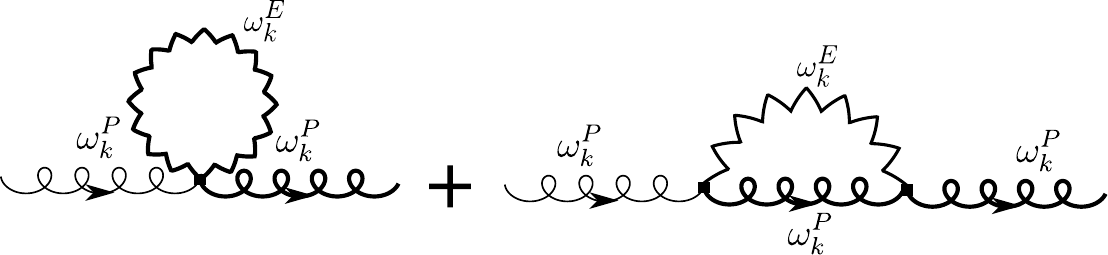}
\caption{Addition to the effective Theory in Fig.~\ref{fig:effTheory} at next-to-leading order in $1/L_E$.}
\label{fig:effTheoryAdd}
\end{center}
\end{figure}
\begin{figure*}[htp]
\begin{center}
\includegraphics[width=0.85\textwidth]{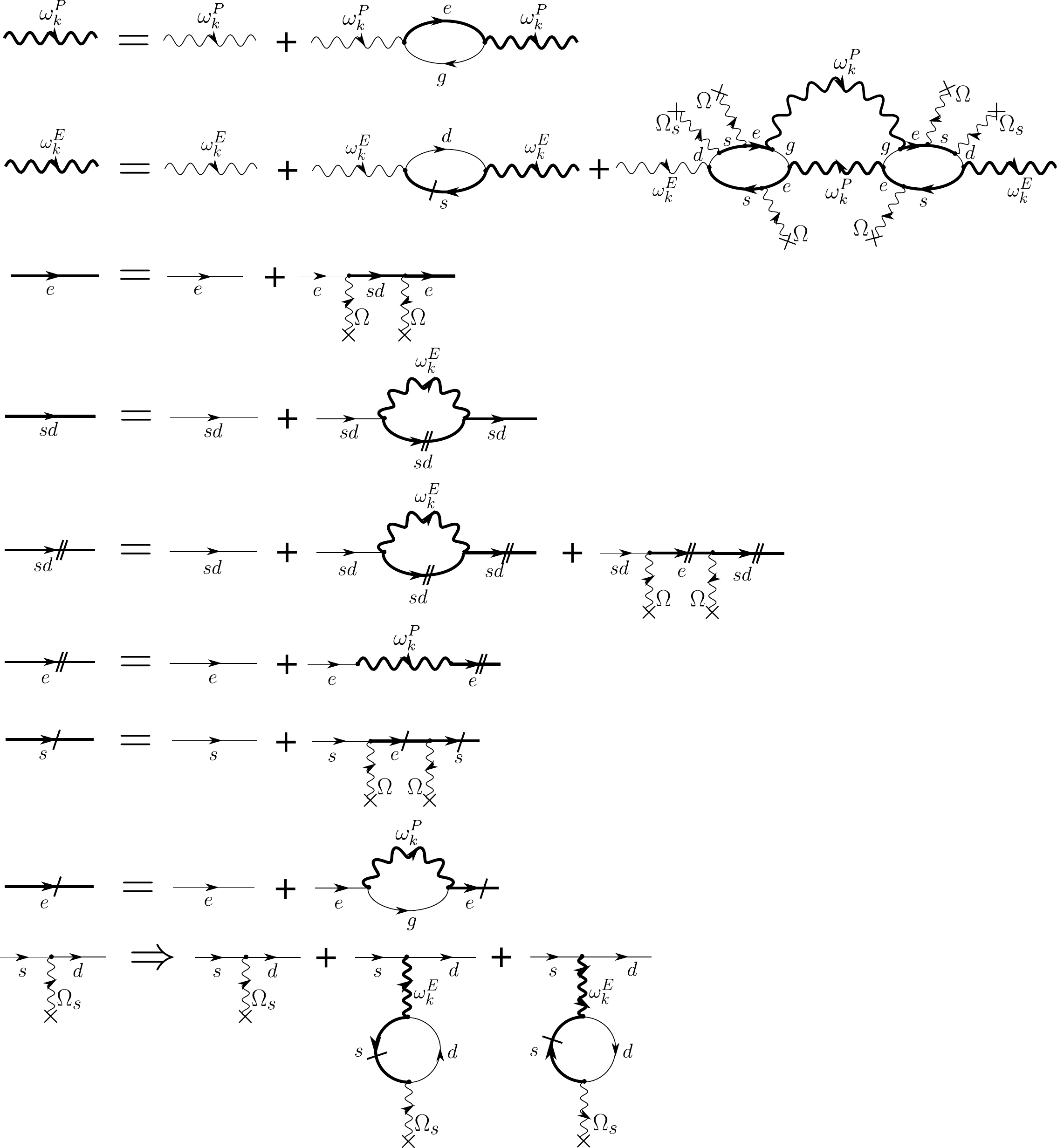}
\caption{Complete set of coupled Dyson equations at next-to-leading order. The loop-reduction procedure of App.~\ref{app:loop_reduction} is employed here and, depending on the Feynman rules at use, $d$-propagators are either bare or given as part of $G_{sd}$, of which the $s$-propagator in the last diagram of the third line is just the $(11)$-component.}
\label{fig:Everything}
\end{center}
\end{figure*}

As was summarized at the end of the last section, the restriction to a theory that exclusively resums all diagrams of the effective theory in Fig.~\ref{fig:effTheory} is not always quantitatively justified. In particular, for current experiments with PCWs \cite{kimble_2014_crystal} the range of the exchange photons is limited due to imperfections in the fabrication that cause rather large losses $\kappa_E$. 
Therefore, in this section we will go one step further and include all diagrams in next-to-leading order. In the absence of any other method applicable to the required long chains of atoms far from equilibrium, the purpose of this section will be twofold: On the one hand, it will serve as a verification of the validity of the expansion in $1/L_E$. On the other hand, it will improve the accuracy of our predictions as it will in particular allow us to include scattering between polaritons, that is, processes involving momentum transfer. In terms of the effective theory in Fig.~\ref{fig:effTheory}, the only modification is the inclusion of the two diagrams in Fig.~\ref{fig:effTheoryAdd}. Equivalently, in terms of the original theory including the atomic degrees of freedom, we obtain the Dyson equations shown diagrammatically in Fig.~\ref{fig:Everything}. One can identify these self-energies with the full set of self-consistently generated diagrams from the next-to-leading order corrections in $1/L_E$ and $1/L_P$ to the probe-photon propagator shown in Fig.~\ref{fig:1overL}.
Since at the present order of expansion the form of the Dyson equations as well as the approach to their solution becomes rather involved, we move the corresponding coupled set of equations to Appendix \ref{sec:app_1overLE}, and only present the results next.

\subsubsection{Results}
When including the effects of a finite interaction range, care has to be taken as not to break any of the assumptions underlying the quantitative validity of the approximations at use. In particular, if the interaction becomes too short-ranged, the losses in state $|d\rangle$ caused by emission of exchange photons and described by the second diagram of the fourth and fifth line of Fig.~\ref{fig:Everything} -- or equivalently the (22)-component of Eq.~\eqref{eq:SigmasdR} -- become large as a result of the narrow linewidth of state $|s\rangle$ for long lived dark-state polaritons. These effects are included in $G_d^R$ in the lenient interpretation of the non-linear Feynman rules, but not for the strict rules. As these atomic Green's functions form the vertex of the effective theory, the differences will grow upon iteration of the self-consistency equations. The uncertainty regarding the results of the exact Feynman rules for four-level atoms thus grows with decreasing $L_E$.
Since the additional scattering effects that arise from the inclusion of $1/L_E$ effects into the description cannot themselves create any new instabilities and instead remedy those that could otherwise exist in $G_E^R$, relatively large values of $C_E^\text{sa}$ can be treated without much more than quantitative corrections to the previously discussed results. The main limitation for an extension to even smaller values of $L_E$ or larger values of $C_E^\text{sa}$ lies in the discrepancy between the different interpretations of the Feynman rules, which eventually will have to be specified in more detail.
\begin{figure}[htp]
\begin{center}
\includegraphics[width=\columnwidth]{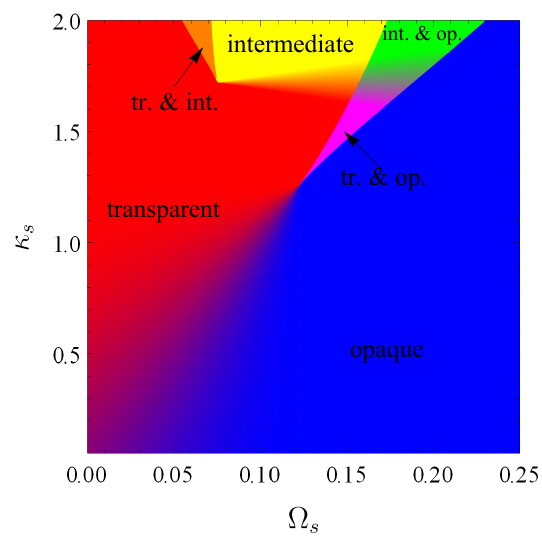}
\caption{Phase diagram including corrections due to the finite interaction range. The color coding is the same as in Fig.~\ref{fig:PD1b}. The bistability between transparent and intermediate phase is less pronounced and for large $\Omega_s$ the opaque phase is more prevalent. The parameters are identical to those in Fig.~\ref{fig:Flow1}, except for $\alpha_E=1000$ and $k_E=0$ (setting the interaction range $L_E\approx 14$ and profile) and the use of lenient Feynman rules.}
\label{fig:PD3}
\end{center}
\end{figure}
This time, for a change, we discuss our results using numerical data obtained from the lenient Feynman rules, which requires exactly the same amount of numerical effort as the strict rules. The resulting phase diagram depicted in Fig.~\ref{fig:PD3} is similar to that in Fig.~\ref{fig:PD1b}. However, the quantitative corrections due to the finite interaction range reduce the extent of the intermediate phase. This is to be expected as the corrections in $1/L_E$ counteract the previously discussed effective drive of the exchange photons. This comparison shows that all the corrections to the next-to-leading order are of the same order as the small parameter of our expansion, that is, our expansion is indeed controlled.\\
\begin{figure*}[htp]
\begin{center}
\includegraphics[width=\textwidth]{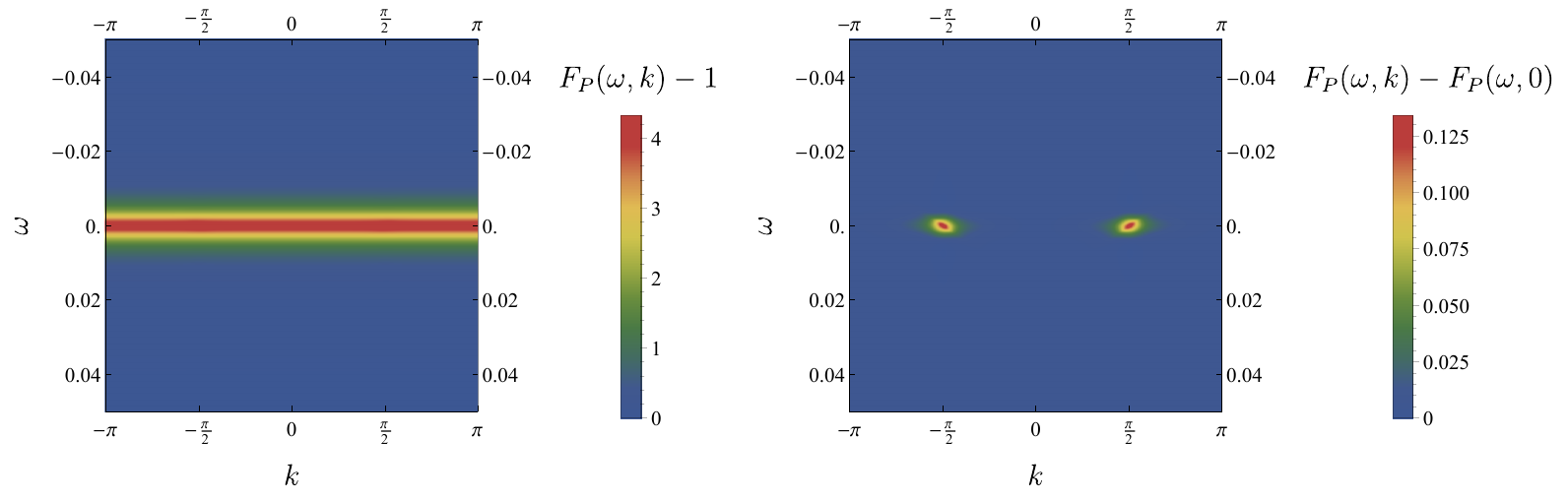}
\caption{For parameters where the expansion remains quantitatively controlled, the distribution function $F_P(\omega,k)$ shows hardly any visible momentum dependence and thus only weak signatures of scattering. To make the weak momentum dependence visible, we subtracted the momentum independent background $F_P(\omega,k=0)$. Here the transparent solution is depicted for the same parameters as in Fig.~\ref{fig:PD3} ($\kappa_0=2$, $\kappa_s=1$, $\omega_0=\Delta_s=n_V=0$, $g_P=10$, $\kappa_P=0.5$, $\gamma_e=1$, $\gamma_d=10$ $\Delta_P(k)=-50 \cos{k}$, $\Delta_E(k=0)=-1$, $\alpha_E=1000$, $k_E=0$, $\Delta_d=0$, $\kappa_E=5$, $g_E=10$), with $\kappa_s=1.8$, $\Omega_s=0.07$.}
\label{fig:FP_1_3}
\end{center}
\end{figure*}

The restriction to large interaction ranges imposed by the discrepancy between the approximate implementations of the Feynman rules, together with the fact that scattering between dark-state polaritons is dominated by forward scattering -- the exchange photons are most efficiently coupled to at $k=0$ -- renders the effects of scattering on the probe photons actually negligible in this regime. As a demonstration of the smallness of the redistribution due to scattering, one can examine the distribution function $F_P(\omega,k)$, which, even in the transparent phase where resonant scattering is strongest, is almost entirely momentum independent (see Fig.~\ref{fig:FP_1_3}). Only upon subtraction of the momentum independent background a slight increase in $F_P(\omega,k)$ near the EIT window can be observed.

For the present case of scattering with small momentum transfers, the most significant effect of the inclusion of $1/L_E$ corrections is the avoidance of the divergence in $G_E^R$ appearing as an artifact of the $L_E\to\infty$ theory: while the exchange photons can still experience an effective drive due to the redistribution of energy between dark-state polaritons, this effect is significantly weakened by the increasing dissipative nature of the atomic vertex brought about by the aforementioned losses in $|d\rangle$. As the exchange photon experiences fewer and fewer losses, those of $|d\rangle$ namely increase, thereby weakening the coupling between probe and exchange photons enough to stabilize the system.

\section{Application: photons in an atomic medium with Rydberg interactions}\label{sec:Rydberg}
% \jlc{There are many different effects that we could discuss here, for example scattering between polaritons or a single polariton that scatters off a localized defect or just the self-energy in a hand wavy explanation, bound states, crystallization etc. Here is my attempt at minimal effort:}
We consider now a different laser scheme with respect to section \ref{sec:four_level}. Starting again from the three-level scheme of Fig.~\ref{fig:setup}b), we do not add a laser exciting to a fourth level but consider the s-level to be a Rydberg state. We will mostly use the same notation as above to illustrate how interactions between Rydberg polaritons fit into a $1/L$ expansion. Instead of driven exchange photons with significant losses and a tunable dispersion, unguided photons with low energies mediate the interactions between Rydberg atoms. In fact the quadratic Stark shift that gives rise to the interatomic van-der-Waals potential $V(\mathbf{x})=-C/|\mathbf{x}|^6$ requires the exchange of two photons. Their dynamics however happens on timescales much shorter than those experimentally relevant and can therefore be neglected. With this knowledge it is well justified to replace the two-photon interaction by the effective potential $V(\mathbf{x-x'})n_s(\mathbf{x})n_s(\mathbf{x}')$.

\begin{figure}[h!]
\begin{center}
\includegraphics[width=\columnwidth]{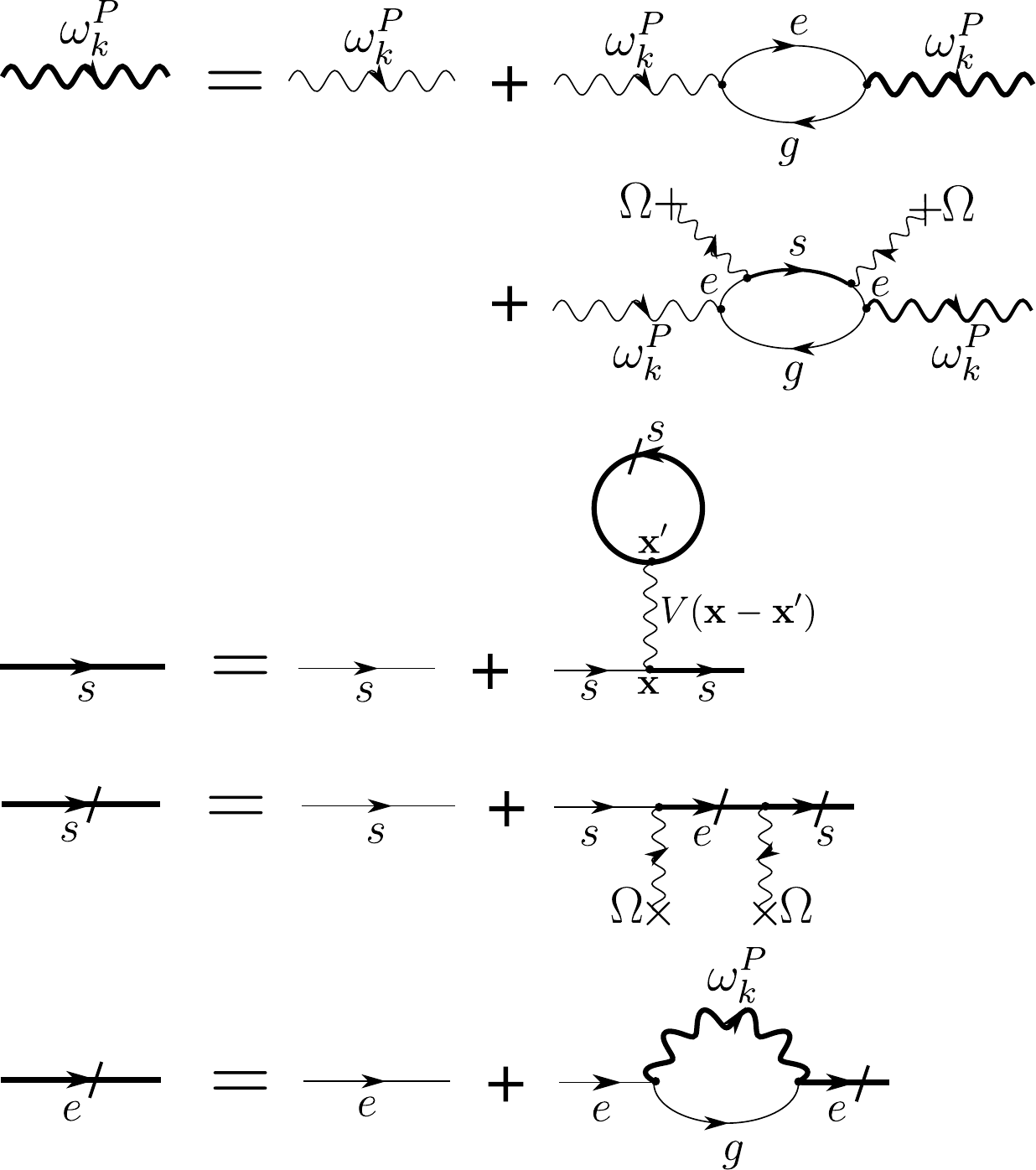}
\caption[Expansion in the interaction range for Rydberg atoms]{Leading order of the expansion in Feynman diagrams around the limit of infinitely ranged interactions between Rydberg polaritons. Note that, due to the use of a fixed potential, only a single interaction diagram has to be considered. Otherwise the self-consistent treatment is similar to that in Sec.~\ref{sec:Gerry}.}
\label{fig:RydbergDiagrams}
\end{center}
\end{figure}
Diagrammatically, the resulting theory looks very similar to the one discussed in the previous sections, the only modification being the replacement of $G_E^R$ coupling between states $|s\rangle$ and $|d\rangle$ by $V(\mathbf{x})$ acting directly on $|s\rangle$. The non-interacting Rydberg polariton theory is illustrated in the first line of Fig.~\ref{fig:RydbergDiagrams} with the $|s\rangle$-propagator considered as bare. Interactions are then taken into account by dressing this state with density-density interactions that take a similar form as those considered in Sec.~\ref{sec:Gerry}. The resulting Feynman diagram in the second line of Fig.~\ref{fig:RydbergDiagrams} has to be treated self-consistently following the same procedure as in Sec.~\ref{sec:Gerry}, with the main difference compared to Fig.~\ref{fig:1overL}a) being the absence of state $|d\rangle$ and the external source $\Omega_s$. It is readily evaluated as
\begin{align}\label{eq:SigmaRydberg}
\Sigma_s^R(\mathbf{x},t)=\int d^3xV(\mathbf{x}-\mathbf{x}')n_s(\mathbf{x}',t)
\end{align}
with the functional dependencies $G_s^R\left[\Sigma_s^R\right]$, $G_e^R\left[G_s^R\right]$ and $G_P^R\left[G_e^R\right]$ identical to Sec.~\ref{sec:Gerry}.

Interestingly, the leading diagrammatic contributions for the setups discussed in the previous sections actually disappear in the context of Rydberg polaritons. Self-interactions of a Rydberg atom by emission and absorption of a photon induce a Lamb shift that is already included in the bare energy of the atomic state. A repeated interaction between two Rydberg atoms on the other hand has to be treated with the nonlinear Feynman rules. By arguments identical in spirit to those of section \ref{sec:Feynman} it reduces to terms already included in \eqref{eq:SigmaRydberg}. Last but not least, a self-interaction of a Rydberg polariton through the interaction of two distinct atoms, similar to Fig.~\ref{fig:1overL}b), is excluded by the instantaneous nature of interactions. As such, the limit of a low Rydberg polariton density results in a less complicated, but conceptually similar expansion to that derived for PCWs and TNWs. However, the absence of an external coupling similar to $\Omega_s$, with which the interaction can interfere destructively, prevents the emergence of phase transitions of the type discussed before.
\begin{figure}[h!]
\begin{center}
\includegraphics[width=\columnwidth]{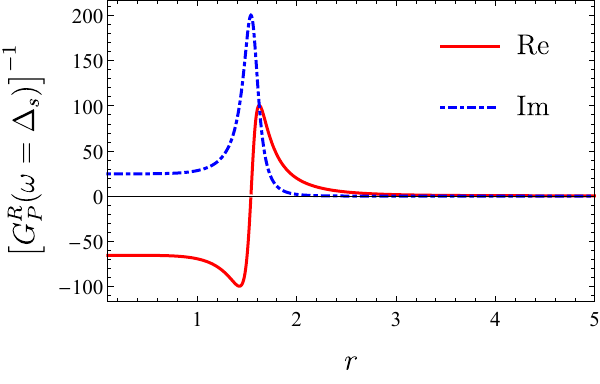}
\caption[Rydberg blockade in the Green's function of the Rydberg polariton]{Rydberg blockade experienced by a Rydberg polariton as a function of the distance from a stationary Rydberg atom at $r=0$. The imaginary part of the inverse propagator indicating the losses becomes very large at a distance set by the blockade radius $R_b$. At the same time the real part that gives rise to deflection also grows. Parameters used are $\kappa_P=1/5$, $\gamma_e=1/4$, $\Delta_s=1/3$, $\Omega=1/2$, $n_V=0$ and $g_P=5$.}
\label{fig:RydbergBlockade}
\end{center}
\end{figure}

Instead, interesting questions include the scattering of Rydberg polaritons and the stability of regular structures (i.e.~n-particle bound states) or even crystals. Here we only want to give a brief idea of how these questions can be approached in terms of a $1/L$ expansion and therefore discuss the simple case of a Rydberg polariton scattering off a fixed Rydberg atom at the origin. In this case the polariton Green's function is given by
\begin{align}
\begin{split}
G_P^R&(\omega,\mathbf{k},\mathbf{x})=\\ &\left(\!\omega-\omega_P(k)-\frac{g_P^2(1-n_V)}{\omega-\frac{\Omega^2}{\omega-\Delta_s-V(\mathbf{x})}+i\gamma_e/2}\!+ i\kappa_P/2\!\right)^{\!\!-1}\!\!,
\end{split}
\end{align}
where the inversion in momentum space first requires a Fourier transform from $\mathbf{x}$ to the momentum difference between incoming and outgoing polariton and is then to be understood as the inverse with respect to the convolution and thus a non-trivial operation. Nevertheless, assuming a slow, and thus well-localized, incoming Rydberg polariton of fixed frequency $\omega_\text{EIT}=\Delta_s$ corresponding to the EIT window at $|\mathbf{x}|\to\infty$, we can calculate its losses as a function of $r=|\mathbf{x}|$ and determining the blockade radius. The losses are given by the imaginary part of the inverse propagator
\begin{align}
\kappa_P^\text{eff}(\omega,r)=\kappa_P+\frac{g_P^2(1-n_V)\gamma_e}{\left(\omega-\frac{\Omega^2}{\omega-\Delta_s-V(\mathbf{x})}\right)^2+\gamma_e^2/4}\,,
\end{align}
and illustrated in Fig.~\ref{fig:RydbergBlockade}. The pronounced maximum that forms at the blockade radius $R_b$ is determined by equating the frequency shift due to $\Sigma_e^R$ with the bare losses $\gamma_e/2$. As a result one finds
\begin{align}
R_b= \left(\frac{C(\gamma_e+2\Delta_s)}{2\Omega^2}\right)^{1/6}\,,
\end{align}
%which differs from the commonly derived result $R_b=\left(C/\Omega\right)^{1/6}$ for lossless excited states \cite{Saffman2010}
which agrees with existing work \cite{pohl_ryd_old}.
Finally, an expansion of $\kappa_P^\text{eff}(\Delta_s,r)$ around $r\to\infty$ reproduces the related result $\kappa_P^\text{eff}(\Delta_s,r)\propto r^{-12}$.

\section{Conclusions and outlook}

We have developed a non-equilibrium diagrammatic approach to strongly interacting photons in optically dense atomic media. It provides quantitative results in the regime of low atom-excitation densities where saturation effects are negligible, provided the atom-atom interactions -- which the photons inherit -- have a large effective range compared to the atomic interspacing. Such a regime can still feature single-atom cooperativities of order or larger than one and thus show strong nonlinearities and emergent many-body phenomena, like the non-equilibrium phase transitions we describe here.

The formalism applies to a broad class of hybrid (nano)photonic devices coupled to arrays of interacting atoms and constitutes a novel theoretical approach to such driven-dissipative many-body systems. Important future applications involve in first place the description of experiments where photon wavepackets propagate through the medium, the description of which requires to solve the Dyson's equations presented above without relying on space- and time-translation invariance. A second interesting avenue is the study of possible crystalline phases of photons appearing in the steady-state \cite{Otterbach2013,ostermann2016spontaneous,hartmann_photon_solid}, which can be also efficiently described within our formalism.

\section{Acknowledgments}
DC acknowledges support from ERC Starting Grant FOQAL, MINECO Severo Ochoa Grant No. SEV-2015-0522, CERCA Programme/Generalitat de Catalunya, Fundacio Privada Cellex, Fundacin Ramn Areces, and Plan Nacional Grant ALIQS, funded by MCIU, AEI, and FEDER.

\appendix

\section{Kramers-Kronig relations}\label{app:KK}
Due to causality, each vertex involves either one or three quantum fields. There are thus four copies of each vertex, differing only in the Keldysh index while otherwise being identical. As every retarded Green's function has to connect a quantum and a classical field, while the Keldysh component connects only classical fields, the distinction in the number of quantum indices in a vertex results in it being connected to different components of the matrix Green's function. In general, this gives rise to a large amount of Feynman diagrams to be calculated. Luckily, most of these can be related to one another using Kramers-Kronig relations. In the following we will illustrate how these relations can be exploited.

To do so it is convenient to use the distribution function $F(\omega)$ defined in Eq.~\eqref{eq:F}, which encodes how strongly the effective environment tries to populate a certain degree of freedom. In the absence of particles one finds $F(\omega)=\mathbb{1}$, which allows to express the Keldysh component through the retarded Green's function. This is nothing else than the formal statement, that vacuum field theory requires no Keldysh formalism in the first place, since no violation of detailed balance is possible. We will denote this special case of an unoccupied degree of freedom by a subscript $0$ attached to the causality index ($K$) as in $G^{K_0}$. For these empty modes, invoking Kramers-Kronig relations for the retarded Green's function
\begin{align}\label{eq:KKR}
\begin{split}
\Re{G^R(\omega)}&=\mathcal{P}\int_{-\infty}^{\infty}\frac{d\omega'}{\pi}\frac{\Im{G^R(\omega')}}{\omega'-\omega}\\
\Im{G^R(\omega)}&=-\mathcal{P}\int_{-\infty}^{\infty}\frac{d\omega'}{\pi}\frac{\Re{G^R(\omega')}}{\omega'-\omega}
\end{split}
\end{align}
allows to find some simplifications. Indeed, any two Green's functions $G_1(\omega)$ and $G_2(\omega)$ will obey
\begin{align}\label{eq:KKSigmaR}
\begin{split}
\int_{-\infty}^{\infty}&\frac{d\omega'}{2 \pi}G_1^{K_0}(\omega')G_2^R(\omega-\omega') \\&=\int_{-\infty}^{\infty}\frac{d\omega'}{2 \pi}G_1^R(\omega')G_2^{K_0}(\omega-\omega') \\
\int_{-\infty}^{\infty}&\frac{d\omega'}{2 \pi}G_1^{K_0}(\omega')G_2^R(\omega+\omega') \\&=-\int_{-\infty}^{\infty}\frac{d\omega'}{2 \pi}G_1^A(\omega')G_2^{K_0}(\omega+\omega'),
\end{split}
\end{align}
which follow immediately from \eqref{eq:KKR} by splitting the retarded Green's function into real and imaginary part. Additionally, one has
\begin{align}\label{eq:KKSigmaK}
\begin{split}
&\int_{-\infty}^{\infty}\frac{d\omega'}{2 \pi}G_1^{K_0}(\omega')G_2^{K_0}(\omega-\omega') \\&=\!-\!\!\int_{-\infty}^{\infty}\!\frac{d\omega'}{2 \pi}\!\left(G_1^A(\omega')G_2^A(\omega-\omega')+G_1^R(\omega')G_2^R(\omega-\omega')\right)\\
&\int_{-\infty}^{\infty}\frac{d\omega'}{2 \pi}G_1^{K_0}(\omega')G_2^{K_0}(\omega+\omega') \\&=\!-\!\!\int_{-\infty}^{\infty}\!\frac{d\omega'}{2 \pi}\!\left(G_1^A(\omega')G_2^R(\omega+\omega')+G_1^A(\omega')G_2^R(\omega+\omega')\right),
\end{split}
\end{align}
which are easily proven by applying the Fourier transforms to the convolution and using that
\begin{align}
\mathcal{F}(\mathcal{H}(f))=-i \, sgn(t) \mathcal{F}(f),
\end{align}
where $f$ is any function and $\mathcal{H}$ and $\mathcal{F}$ are Hilbert and Fourier transform, respectively.\\
Similar identities, using the same method, can also be proven for more complicated products and higher order convolutions %see p.72 in my notes
of Green's functions. Since the bare atomic propagators for all but the ground state, as well as that of the bare exchange photon, involve no driving, these states can only be occupied by coupling to the ground state. Without these interactions, they remain unoccupied, their distribution function is the identity matrix and we can use the Kramers-Kronig identities to great effect. In particular one recovers the intuitive result that interactions cannot generate excitations from the vacuum, which formally takes the simple form $\Sigma^K_\nu\left[G^{K_0}_\mu,G^R_\mu\right]=2i\Im\Sigma^R_\nu\left[G^{K_0}_\mu,G^R_\mu\right]$.

\section{Derivation of a coloured bath}\label{app:Bath}

A coloured bath in the form of a frequency dependent coupling rate $\kappa_s(\omega)$ can be constructed by an indirect coupling to a Markovian bath. First an additional Gaussian mode with a coherence time and length much shorter than the relevant time and length scales for dark-state polaritons is itself driven by a Markovian bath:
\begin{subequations}
\begin{align}
\hat{H}_\text{aux}&=\omega_0 \hat{b}_\text{aux}^\dagger \hat{b}_\text{aux}\\
\mathcal{L}_\text{loss}\rho_\text{aux}&=-\hbar\frac{\kappa_1}{2}\left(\left\{\hat{b}^\dagger_\text{aux}\hat{b}_\text{aux},\rho_\text{aux}\right\}-2\hat{b}_\text{aux}\rho_\text{aux}\hat{b}^\dagger_\text{aux}\right)\\
\mathcal{L}_\text{drive}\rho_\text{aux}&=-\hbar\frac{\kappa_2}{2}\left(\left\{\hat{b}_\text{aux}\hat{b}^\dagger_\text{aux},\rho_\text{aux}\right\}-2\hat{b}^\dagger_\text{aux}\rho_\text{aux}\hat{b}_\text{aux}\right)\,,
\end{align}
\end{subequations}
where $\hat{b}^\dagger_\text{aux}$ and $\hat{b}_\text{aux}$ are bosonic creation and annihilation operators for the auxiliary mode with density matrix $\rho_\text{aux}$.
In a next step this mode then couples bilinearly to the propagating photons:
\begin{align}
\hat{H}_{ab}&=g_b \int_k(\hat{b}^\dagger_\text{aux} \hat{b}_P(k)+\hat{b}^\dagger_P(k) \hat{b}_\text{aux})\,.
\end{align}
In the limit of strong driving, where $\kappa_1/\kappa_2$ approaches unity from below, this construction, that effectively mimics a frequency dependent coupling of the system to a highly occupied incoherent bath, is described by the following addition to the Liouvillian
\begin{widetext}
\begin{align}
\begin{split}
\mathcal{L}_{\kappa_s}\rho(t)=-\frac{\hbar}{2}\int_k\int_{-\infty}^t dt'&\left(\kappa_s(t-t')\left[\left[\hat{b}_{P,I}(k,t)\right]^\dagger,\left[\hat{b}_{P,I}(k,t'),\rho_I(t')\right]\right]\right.\\ &\left. -\kappa_s(t'-t)\left[\hat{b}_{P,I}(k,t),\left[\left[\hat{b}_{P,I}(k,t')\right]^\dagger ,\rho(t')\right]\right]\right)\,,
\end{split}
\end{align}
\end{widetext}
where the additional index $I$ indicates that operators are to be evaluated in the interaction picture, where $H$ as well as all Lindblad operators (except $\mathcal{L}_{\kappa_s}$) contribute to the time-evolution.
Here $\kappa_s(t)\sim\exp{(-\kappa_0 |t|)}$ decays exponentially, which due to $\kappa_0=2(\kappa_1-\kappa_2)$ can be tuned by the properties of the Gaussian mode. While this construction is rather cumbersome if written as Liouvillian, in the functional-integral description however, the Gaussian mode can be integrated out immediately giving rise to a simple, closed expression for a colored bath.\\
Note that, in order to control the occupation of the propagating modes, it is sufficient to choose $\kappa_1=\kappa_2$ such that Stokes and anti-Stokes processes are equal in amplitude and then to adjust the loss rate $\kappa_P$ accordingly. This construction implies that the anti-Stokes processes are at least equally likely as the Stokes processes, such that the bath cannot be inverted and consequently lasing or condensation are excluded \cite{Byrnes2014}.\\

\section{Matrix Green's functions}\label{app:matrix_GF}
In this appendix we give the expression of the bare matrix Green's functions used in Sec.~\ref{sec:four_level}. Those of section \ref{sec:EIT} are then simply obtained by removal of state $|d\rangle$ from the Green's functions of the atoms and the exchange photon from the photon propagators.\\
For the non-interacting atoms coupled to the coherent laser fields, the inverse retarded Green's function in the basis
\begin{align}
\mathbf{a}^\text{(q,cl)}(\omega,z)=\begin{pmatrix}
a_d(\omega,z)\\
a_s(\omega,z)\\
a_e(\omega,z)\\
a_g(\omega,z)\\
\end{pmatrix}^\text{(q,cl)}
\end{align}
reads
\begin{widetext}
\begin{align}
\begin{split}
&\left[\bar{G}_{a,0}^R\right]^{-1}(\omega,\omega')=\\&\begin{pmatrix}
\left(\omega-\omega_d+i\frac{\gamma_d}{2}\right)\delta(\omega-\omega')&&  -\Omega_s\delta\left(\omega-\omega'-\omega_L^{(2)}\right)&&0&&0\\
-\Omega_s \delta(\omega-\omega'+\omega_L^{(2)})&&\left(\omega-\omega_s+i\frac{\epsilon}{2}\right)\delta(\omega-\omega')&&
-\Omega \delta(\omega-\omega'+\omega_L^{(1)})&&0\\
0&&-\Omega \delta(\omega-\omega'-\omega_L^{(1)})&& \left(\omega-\omega_e+i\frac{\gamma_e}{2}\right)\delta(\omega-\omega')&&0\\
0&&0&&0&&(\omega+i\frac{\epsilon}{2})\delta(\omega-\omega')\\
\end{pmatrix}\,.
\end{split}
\end{align}
However as it turns out, it is far more convenient to transform into a rotating frame, where the states $|e\rangle$, $|s\rangle$ and $|d\rangle$ rotate at frequencies $\omega_e$, $\omega_e-\omega_L^{(1)}$ and $\omega_e-\omega_L^{(1)}+\omega_L^{(2)}$ respectively.
Within this frame the atomic Green's function becomes time translationally invariant, that is $G_{a,0}^{-1}(\omega,\omega')=G_{a,0}^{-1}(\omega)\delta(\omega-\omega')$ with
\begin{align}\label{eq:matrixG0}
\left[G_{a,0}^R\right]^{-1}(\omega)=\begin{pmatrix}
\omega-\Delta_d-\Delta_s+i\frac{\gamma_d}{2}&&  -\Omega_s&&0&&0\\
-\Omega_s&&\omega-\Delta_s+i\frac{\epsilon}{2}&&
-\Omega &&0\\
0&&-\Omega && \omega+i\frac{\gamma_e}{2}&&0\\
0&&0&&0&&\omega+i\frac{\epsilon}{2}\\
\end{pmatrix}\,,
\end{align}
\end{widetext}
in the frequency shifted basis
\begin{align}
\mathbf{a}^\text{(q,cl)}(\omega,z)=\begin{pmatrix}
a_d(\omega+\omega_e-\omega_L^{(1)}+\omega_L^{(2)},z)\\
a_s(\omega+\omega_e-\omega_L^{(1)},z)\\
a_e(\omega+\omega_e,z)\\
a_g(\omega,z)\\
\end{pmatrix}^\text{(q,cl)}\,.
\end{align}
Here the detunings $\Delta_s=\omega_e-\omega_L^{(1)}-\omega_s$ and $\Delta_d=\omega_d-\omega_s-\omega_L^{(2)}$ between laser frequencies and atomic transitions have been introduced.
The corresponding Keldysh component of the inverse Green's function within the same frame of reference is given by
\begin{align}
D_{a,0}^K(\omega)=\begin{pmatrix}
i\gamma_d&&0&&0&&0\\
0&&i\epsilon&&0&&0\\
0&&0&&i\gamma_e&&0\\
0&&0&&0&&(3-2n_V)i\epsilon\\
\end{pmatrix}\,.
\end{align}
In direct analogy to Sec.~\ref{sec:appl_twolevel} $n_V\in[0,1]$ is the homogeneous number density of vacant lattice sites.

An equivalent rotation can also be performed for the retarded and Keldysh component of the inverse photon Green's function, which then are given by
\begin{align}
\left[G_{p,0}^R\right]^{-1}(\omega,k)=
\begin{pmatrix}
\omega-\Delta_E(k)+i\frac{\kappa_E}{2}&&0\\
0&&\omega-\Delta_P(k)+i\frac{\kappa_P}{2}\\
\end{pmatrix}
\end{align}
and
\begin{align}
D_{p,0}^K(\omega,k)=\begin{pmatrix}
i\kappa_E&&0\\
0&&i\kappa_P +2 i \kappa_s(\omega)\\
\end{pmatrix}
\end{align}
respectively. Here we expressed both functions in the basis
\begin{align}
\mathbf{b}^\text{(q,cl)}(\omega,k)=\begin{pmatrix}
b_E(\omega+\omega_L^{(2)},k)\\
b_P(\omega+\omega_e,k)\\
\end{pmatrix}^\text{(q,cl)}\,,
\end{align}
and introduced the detunings $\Delta_P(k)=\omega_P(k)-\omega_e$ and $\Delta_E(k)=\omega_E(k)-\omega_L^{(2)}$. Note that here we have already performed the Gaussian integration over the auxiliary field $b_\text{aux}$ used in App.~\ref{app:Bath} to model a coloured bath, after which the inverse probe photon propagator in general is modified by the subtraction of $g_b^2 G_b(\omega)$. Assuming very strong coupling to the incoherent source, however, $\kappa_1$ and $\kappa_2$ diverge, while $\kappa_0=2(\kappa_1-\kappa_2)$ and $\kappa_s=2g_b^2\kappa_1$ are kept finite. In this limit $g_b^2 G_b^R(\omega)$ vanishes, while $\kappa_s(\omega)=-ig_b^2 G_b^K(\omega)/2=\kappa_s/((\omega-\omega_0)^2+\kappa_0^2)$ remains finite.\\
Introducing the field vectors $\mathbf{a}=\left\{\mathbf{a}^\text{cl},\mathbf{a}^\text{q}\right\}$ and $\mathbf{b}=\left\{\mathbf{b}^\text{cl},\mathbf{b}^\text{q}\right\}$ the non-interacting part of the action $S=S_0+S_\text{int}$ is conveniently expressed in terms of the bare Green's functions (indicated by the lower index $0$):
\begin{align}
\begin{split}
S_0=\hbar\!\int\!\frac{d\omega}{2\pi}&\left(\!\sum_{z}\mathbf{a}^*\!(\omega,z)\mathcal{G}_{a,0}^{-1}(\omega)\mathbf{a}(\omega,z)\right.\\&\left.+\!\!\int\!\frac{dk}{2\pi}\mathbf{b}^*\!(\omega,k)\mathcal{G}_{p,0}^{-1}(\omega,k)\mathbf{b}(\omega,k)\!\!\right).
\end{split}
\end{align}
Here, as before in Sec.~\ref{sec:appl_twolevel}, the corresponding inverse Keldysh matrix Green's functions are given by
\begin{align}
\mathcal{G}_{\mu,0}^{-1}=
\begin{pmatrix}
0&&\left[G^A_{\mu,0}\right]^{-1}\\
\left[G^R_{\mu,0}\right]^{-1}&&D^K_{\mu,0}\\
\end{pmatrix}\,.
\end{align}
Lastly, the light-matter interactions are described by
\begin{widetext}
\begin{align}\label{eq:Sint}
\begin{split}
S_\text{int}=&\int\frac{d\omega}{2\pi}\int\frac{dk}{2\pi}\sum_z\\ & \left(\frac{1}{\sqrt{2}}g_P e^{ikz} u_k^P(z)\left[b_P^\text{q}(k) \left(\bar{a}_e^\text{q}(z) a_g^\text{q}(z)+\bar{a}_e^\text{cl}(z)a_g^\text{cl}(z)\right)+b_P^\text{cl}(k)\left(\bar{a}_e^\text{cl}(z)a_g^\text{q}(z)+\bar{a}_e^\text{q}(z) a_g^\text{cl}(z)\right)\right]+h.c.\right.\\
&\left.+\frac{1}{\sqrt{2}}g_E e^{ikz} u_k^E(z)\left[b_E^\text{q}(k) \left(\bar{a}_d^\text{q}(z) a_s^\text{q}(z)+\bar{a}_d^\text{cl}(z)a_s^\text{cl}(z)\right)+b_E^\text{cl}(k)\left(\bar{a}_d^\text{cl}(z)a_s^\text{q}(z)+\bar{a}_d^\text{q}(z) a_s^\text{cl}(z)\right)\right]+h.c.\,\right).
\end{split}
\end{align}
\end{widetext}
The same discussion used in Sec.~\ref{sec:Keldysh} for a single mode applies also to the Bloch functions $u_k^{P,E}(z)$, for which the $u_k^{P,E}(0)$ is assumed to be independent of k, so that it can absorb into the coupling constant via $g_{P,E}|u_k^{P,E}(0)|\to g_{P,E}$.

\section{Loop reduction}\label{app:loop_reduction}
Beyond the Kramers-Kronig relations, a further significant simplification can be achieved by noting that the atomic ground-state has no dynamics of its own. Hence any loop involving the bare Keldysh component of the ground-state propagator can be computed trivially\footnote{The retarded component instead contains an Heaviside theta and is thus still time-dependent.}. In and of itself this is not a particularly useful observation. In combination with the Kramers-Kronig relations and the non-linear Feynman rules introduced in section \ref{sec:NLFR}, however, several loop integrals can be computed exactly.\\ 
\begin{figure}[htp]
\begin{center}
\includegraphics[width=\columnwidth]{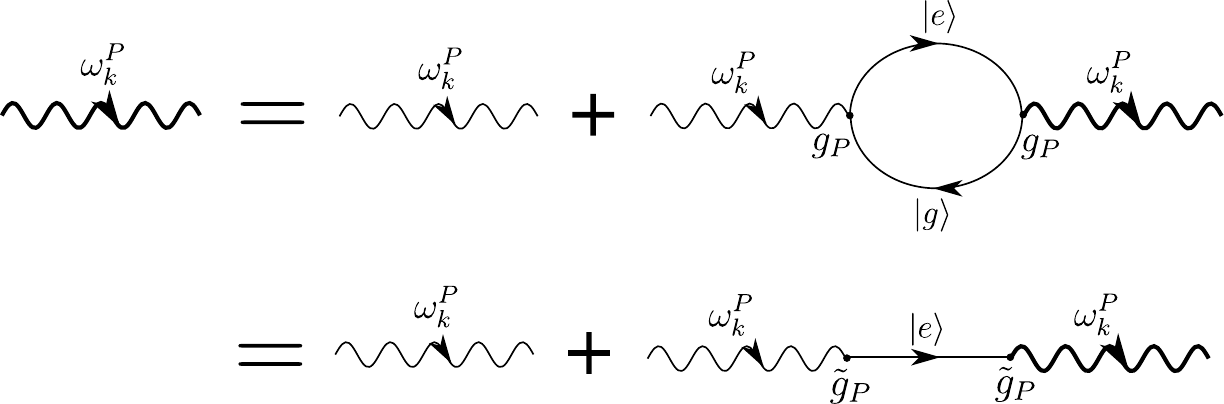}
\caption{The Probe photon propagator to leading order in $1/L_P$ can be simplified significantly using the loop reduction procedure.}
\label{fig:loop_reduction}
\end{center}
\end{figure}
To better understand how all of these properties come together, let us consider the case of a probe photon propagating through any polarizable medium. This process to leading order in $1/L$ is described by the diagrammatic equation in Fig.~\ref{fig:loop_reduction}. Here the propagator of the excited state cannot be specified further, since interactions with other excited atoms can and will dress it. The ground-state propagator on the other hand only couples to other states via the absorption of a probe photon. Employing the Feynman rules of section \ref{sec:NLFR} to the polarization loop in Fig.~\ref{fig:loop_reduction}, it will thus always be described by the bare Green's function. Consequently, the diagram of Fig.~\ref{fig:loop_reduction} for the retarded probe photon propagator is solved by \eqref{eq:GP_LO} and \eqref{eq:Sigma_LO}. Since furthermore $G_{g,0}^K(\omega)=-2\pi i(3-2n_V)\delta(\omega)$, the remaining integral can then be solved immediately, such that one finally obtains
\begin{align}\label{eq:SigmaP_LO}
\Sigma_P^R(\omega)=g_P^2(1-n_V)G_e^R(\omega)\,,
\end{align}
completely independent of the form of the interactions between excited states. A similar calculation gives the equivalent result
\begin{align}
\delta\Sigma_P^K(\omega)=g_P^2(2-n_V)\delta G_e^K(\omega)\,
\end{align}
for the Keldysh component of the self-energy. Hence, the result for the polarization bubble is the same as for a bilinear coupling converting a photon into an excited atomic state, albeit with the modified coupling constant $\tilde{g}_P=g_P\sqrt{2-n_V}$.\\
This identification changes the topology of diagrams. However, quite importantly the ordering in powers of the inverse interaction range remains unaffected.\\
Note that, despite the extremely long life time of the meta-stable state $|s\rangle$, due to the static laser field coupling to state $|e\rangle$ (and in Sec.~\ref{sec:four_level} also to $|d\rangle$) the corresponding Keldysh component $G_s^K(t-t')$ explicitly depends on time and a similar identity for particle hole loops involving $|s\rangle$ is not quite as useful.

\section{Alternative derivation of non-linear Feynman rules}\label{app:Feynman}
In section~\ref{sec:NLFR} we argued that in order to capture the properties of non-interacting polaritons giving rise to electromagnetically induced transparency, it is not actually necessary to implement non-linear Feynman rules in real time. Instead, it suffices to simply exclude all self-energy insertions into the ground-state propagator and to forbid any of the excited states to repeatedly couple to the ground-state. By these means we then derived the (exact) polarizability of the atomic medium in \ref{sec:EIT}. It is instructive to rederive this result directly from the coupled Lindblad equations of the spin operators $\sigma_{\mu\nu}$ introduced in Sec.~\ref{sec:appl_twolevel} and \ref{sec:EIT}.\\
The retarded polarizability of each atom in the ground-state takes the form
\begin{align}
P(t)=\theta(t)\left[\text{Tr}(\sigma_{ge}(t)\sigma_{eg}\rho)-\text{Tr}(\sigma_{eg}\sigma_{ge}(t)\rho)\right]\,.
\end{align}
The latter of these contributions vanishes identically as $\rho=\sigma_{gg}$. The time-evolution of $\sigma_{ge}(t)$ is given by
\begin{align}
\begin{split}
-i\dot{\sigma}_{ge}(t)=&\left[H,\sigma_{ge}(t)\right]\\&+i\gamma_e\left(\sigma_{ge}\sigma_{ge}(t)\sigma_{eg}-\frac{1}{2}\left\{\sigma_{eg}\sigma_{ge},\sigma_{ge}(t)\right\}\right)\;,
\end{split}
\end{align}
which very nicely simplifies, if one uses $\sigma_{eg}\sigma_{ge}=\sigma_{ee}$ as well as the observation that $H$ in the limit of low polariton densities acts trivially on the ground-state, which implies $\sigma_{ge}\sigma_{ge}(t)=0$ and $ H\sigma_{ge}(t)=0$. One thus ends up with
\begin{align}
i\dot{\sigma}_{ge}(t)=\sigma_{ge}(t)H-i\frac{\gamma_e}{2}\sigma_{ge}(t)\sigma_{ee}\;,
\end{align}
which has the solution
\begin{align}
\sigma_{ge}(t)=\sigma_{ge}e^{-i\tilde{H}t}\;,
\end{align}
where $\tilde{H}=H+i\frac{\gamma_e}{2}\sigma_{ee}$ is the non-hermitian effective Hamiltonian governing the time-evolution of the three level system in the presence of losses. Inserting this result back into the polarizability, we obtain
\begin{align}
P(t)=\theta(t)\left(e^{i\tilde{H}t}\right)_{22}\;,
\end{align}
which upon Fourier transformation turns into
\begin{align}
P(\omega)=\frac{i}{\omega-\frac{\Omega^2}{\omega-\Delta_s}+i\gamma_d/2}\;,
\end{align}
which coincides with the result obtained by means of the simplified non-linear Feynman rules in \eqref{eq:polarizability}. We thus have seen that, due to the absence of laser coupling between the atomic ground-state and the excited state, non-linear Feynman rules are easily implemented. As is mentioned in Sec.~\ref{sec:four_level}, the dynamics of the other external state of the four-level-scheme is not as simple. Thus a similar derivation for the susceptibility of the medium from the perspective of the exchange photon fails.\\
We can now go beyond the limit of low polariton densities and consider the effect of a finite density of excited atoms.  In this case, there is a finite fraction of atoms for which $\rho\neq\sigma_{gg}$. In particular due to EIT, excited atoms will predominantly occupy the metastable state ($\rho=\sigma_{ss}$). This however implies that $\sigma_{eg}\rho$ and $\rho\sigma_{eg}$ both vanish. The main effect of a finite density of polaritons will thus simply result in a reduced susceptibility of the atoms. This is easily included via a finite density of defects $n_V$ in the chain of atoms. If the number density of excited atoms remains small, this effect can be neglected all together, as it will have no effect on the stability of phases reported here.

\subsection{Anomalous Green's functions}
\label{app:anomalousGF}
The appearance of anomalous Green's functions for the exchange photons is caused by the presence of a background coherent contribution to the $E$-photon field. This has to be expected from the theory presented in Sec.~\ref{sec:Gerry}, as an E-photon can be resonantly excited by the external laser field with strength $\Omega_s$. The net effect is that a coherent component i.e. a condensate arises in the connected part of the two-point function of the exchange photon as an addition to the background coherent laser field. This effect is described by the last diagram in the third line of Fig.~\ref{fig:Gerry+ECar}, where at each atom the order in which the $E$-photon and the laser are coupled to a given transition can be chosen arbitrarily. While this formally creates anomalous contributions to the self-energy, it underlines the physical argument that from the perspective of fixed atoms exchange photons and laser photons are indistinguishable if their frequency is identical.\\
It is important to understand that this condensation of $E$-photons will not be inherited by the probe photons. Consequently, no polariton condensate can be generated unless the probe photons themselves are directly coupled to an inverted bath (which we excluded from the outset).
Similar to what is done in equilibrium, each component of the non-equilibrium Green's function can be augmented by a Nambu structure. For the retarded part of the Green's function one typically defines the following $2\times 2$ matrix
\begin{widetext}
\begin{align}\label{eq:GRnambu}
\begin{split}
G^R(\omega,p)&=-i\begin{pmatrix}
\left\langle b^\text{cl}(\omega,p)\bar{b}^\text{q}(\omega,p)\right\rangle&&
\left\langle b^\text{cl}(\omega,p)b^\text{q}(-\omega,-p)\right\rangle\\
\left\langle \bar{b}^\text{cl}(-\omega,-p)\bar{b}^\text{q}(\omega,p)\right\rangle&&\left\langle \bar{b}^\text{cl}(-\omega,-p)b^\text{q}(-\omega,-p)\right\rangle
\end{pmatrix}\\&=-i\begin{pmatrix}
\begin{tikzpicture}
\draw[-{Stealth[length=2.3mm]},snake it,densely dashed](0,0)--node[above]{$(\omega,p)$}++(1.95,0);
\draw[snake it] (1.95,0) --node[above]{$(\omega,p)$}++(2,0);
\end{tikzpicture}&&
\begin{tikzpicture}
  \draw[-{Stealth[length=2.3mm]},snake it,densely dashed]
 (2,0) --node[above]{$(-\omega,-p)\qquad\;\;$}++(-1,0);;
   \draw[snake it,densely dashed]
 (1,0) --(0,0);;
  \draw[-{Stealth[length=2.3mm]},snake it]
 (2,0) --node[above]{$\qquad\;\;(\omega,p)$}++(1,0);;
   \draw[snake it]
 (3,0) --(4,0);;
\end{tikzpicture}\\
\begin{tikzpicture}
  \draw[-{Stealth[length=2.3mm]},snake it,densely dashed]
 (2,0) --node[above]{$(\omega,p)\qquad\;\;$}++(-1,0);;
   \draw[snake it,densely dashed]
 (1,0) --(0,0);;
  \draw[-{Stealth[length=2.3mm]},snake it]
 (4,0) --(2.95,0);;
   \draw[snake it]
 (2,0) --node[above]{$\qquad\;\;(-\omega,-p)$}++(.95,0);;
\end{tikzpicture}&&\begin{tikzpicture}
\draw[{Stealth[length=2.3mm]}-,snake it](2,0)--node[above]{$(-\omega,-p)$}++(2,0);
\draw[snake it,densely dashed] (0,0) --node[above]{$(-\omega,-p)$}++(2,0);
\end{tikzpicture}
\end{pmatrix}
\end{split}
\end{align}
\end{widetext}
and the same for $G^K(\omega,p)$ with all quantum fields (index q) replaced by classical fields (index cl).\\%
The diagonal entries then describe ordinary Green's functions, while the off-diagonal, so-called anomalous components, are non-zero only in the presence of a condensate.\\
For the retarded Green's function the Dyson equation takes the same form as in equilibrium
\begin{align}
G^R=G^R_0+G^R_0 \cdot \Sigma^R \cdot G^R\,,
\end{align}
where $G^R_0$ is purely diagonal.
Retarded and advanced Green's functions are once again not independent and one finds the relation $G^R_{\sigma\rho}(\omega,p)=G^A_{\bar{\sigma}\bar{\rho}}(-\omega,-p)$, where $\sigma,\rho\in\{1,2\}$ and $\bar{\sigma}$ is the complement of $\sigma$. The proof follows from a direct inspection of the respective Feynman diagrams: exchanging the external legs reverses the direction of the momentum and energy flow and simultaneously reverses causality, thereby equating the off-diagonal entries of $G^R$ and $G^A$. Additionally $\left[G^A\right]^\dagger=G^R$ follows immediately once the definition \eqref{eq:GRnambu} is evaluated on the Keldysh contour. Therefore, the advanced Green's function never has to be calculated and the retarded Green's function can be restricted to only two independent functions:
\begin{align}
G^R(\omega,p)=\begin{pmatrix}
G^R_{11}(\omega,p)&&G^R_{12}(\omega,p)\\
\left[G^R_{12}\right]^*(-\omega,-p)&&\left[G^R_{11}\right]^*(-\omega,-p)
\end{pmatrix}\,.
\end{align}
Furthermore, considering all self-energy diagrams order-by-order, one can prove that $G^R_{\sigma\bar{\sigma}}(\omega,p)=G^R_{\bar{\sigma}\sigma}(\omega,p)$, which is true for any uniform bosonic system in and out of equilibrium \cite{Fetter_book}.
%page 215%
\\
Taking the conjugate transpose of the Keldysh component can be done immediately in frequency space and a comparison of elements reveals the anti-hermitian structure, i.e. $\left[G^K\right]^\dagger=-G^K$. We can thus still use the parametrization $G^K=G^R\cdot F-F \cdot G^A$, with a hermitian matrix $F$. The fact that $F=\sigma_z$ for an empty system accounts for the reversed order of operators between the components of the first and the second row of $G^K$. Exchanging incoming and outgoing particles in $G^K$ furthermore allows to identify $G^K_{\sigma\rho}(\omega,p)=G^K_{\bar{\sigma}\bar{\rho}}(-\omega,-p)$. By considering again all possible self-energy diagrams one finds the additional symmetry $G^K_{\sigma\bar{\sigma}}(\omega,p)=-\left[G^K_{\bar{\sigma}\sigma}\right]^*(-\omega,-p)$. The Dyson equation for the Keldysh component directly generalizes to the Nambu structure: $G^K=G^R\cdot\left(\Sigma^K-D_0^K\right)\cdot G^A$, which together with the other symmetries implies
\begin{align}
\Sigma^{K_0}(\omega,p)=2
\begin{pmatrix}
i\Im\Sigma^R_{11}(\omega,p)&&\Re\Sigma^R_{12}(\omega,p)\\
-\Re\Sigma^R_{12}(-\omega,-p)&&i\Im\Sigma^R_{11}(-\omega,-p)
\end{pmatrix}
\end{align}
for the empty system and
\begin{align}
\begin{split}
\delta\Sigma^K(\omega,p)&=\Sigma^K(\omega,p)-\Sigma^{K_0}(\omega,p)\\&=\begin{pmatrix}
\delta\Sigma^K_{11}(\omega,p)&&\delta\Sigma^K_{12}(\omega,p)\\
-\left[\delta\Sigma^K_{12}\right]^*(\omega,p)&&\delta\Sigma^K_{11}(-\omega,-p)
\end{pmatrix}
\end{split}
\end{align}
for excitations above the vacuum.  
With these definitions the Kramers-Kronig relations of App.~\ref{app:KK} remain valid without limitation. For simplicity we only provide the Kramers-Kronig relations for one-dimensional convolutions with some normal Green's function labeled $G_n$, keeping in mind that higher dimensional generalizations take exactly the same form:
\begin{align}
\begin{split}
\int\frac{d\omega'}{2\pi}\left(G^K_{\sigma\rho}(\omega')G^R_n(\omega-\omega')-G^R_{\sigma\rho}(\omega')G^K_n(\omega-\omega')\right)\\=-2\delta_{\rho2}\int\frac{d\omega'}{2\pi}G^R_{\sigma\rho}(\omega')G^R_n(\omega-\omega')\,,
\end{split}
\end{align}
as well as
\begin{widetext}
\begin{align}
\begin{split}
&\int\frac{d\omega'}{2\pi}\left(G^K_{\sigma\rho}(\omega')G^K_n(\omega-\omega')-G^R_{\sigma\rho}(\omega')G^R_n(\omega-\omega')-G^A_{\sigma\rho}(\omega')G^A_n(\omega-\omega')\right)\\&=-\int\frac{d\omega'}{\pi}\left(\delta_{\rho2}G^R_{\sigma\rho}(\omega')G^R_n(\omega-\omega')+\delta_{\sigma2}G^A_{\sigma\rho}(\omega')G^A_n(\omega-\omega')\right)\,.
\end{split}
\end{align}
\end{widetext}
Note that due to the symmetries of the diagonal entries of the Green's functions, these two relations already fully incorporate the four equations derived in App.~\ref{app:KK}.

\section{Dyson equations at order $1/L_E$}
\label{sec:app_1overLE}

As can be seen in Fig.~\ref{fig:Everything}, a fully self-consistent theory involving all effects at next-to-leading order in the inverse interaction range requires to solve an even larger number of coupled integral equations than in the formulation at lower order. This task might seem daunting at first sight, however, using the Kramers-Kronig relations (see App.~\ref{app:KK}) and the loop reduction procedure (see App.~\ref{app:loop_reduction}), every single diagram can once again be broken down into a combination of independent one-loop effects. Due to the non-linear Feynman rules (section \ref{sec:NLFR}), great care has to be taken in determining which of these single loop effects can be combined. We do so by introducing two different matrix Green's functions $\mathcal{G}_{sd}$ and $\tilde{\mathcal{G}}_{sd}$ for the states $|s\rangle$ and $|d\rangle$. To help distinguish these propagators in Feynman diagrams, we slash the propagator of $\tilde{\mathcal{G}}_{sd}$ twice. When appearing as an insertion inside the probe-photon propagator, $\mathcal{G}_{sd}$ cannot itself involve a self-energy that would return the atom to its ground state. There is thus only one contribution to the self-energy and the matrix propagator takes the fairly simple form
\begin{align}\label{eq:loop_begin}
\mathcal{G}^{R}_{sd}=\left(\left[\mathcal{G}^R_{sd}\right]^{-1}-\Sigma^R_{sd}\right)^{-1}\,,
\end{align}
where
\begin{widetext}
\begin{align}\label{eq:SigmasdR}
\Sigma^R_{sd}=\frac{i}{2}g_E^2\begin{pmatrix}
\delta G^K_{E_{22}}\star\tilde{G}^R_{dd}+G^R_{E_{22}}\star\delta\tilde{G}^K_{dd}&&G^R_{E_{21}}\star\delta\tilde{G}^K_{ds}+\left(\delta G^K_{E_{21}}+2 G^{K_0}_{E_{21}}\right)\star\tilde{G}^R_{ds}\\
\delta G^K_{E_{12}}\star\tilde{G}^R_{sd}+G^R_{E_{12}}\star\delta\tilde{G}^K_{ds}&&G^R_{E_{11}}\star\delta\tilde{G}^K_{ss}+\left(\delta G^K_{E_{11}}+2 G^{K_0}_{E_{11}}\right)\star\tilde{G}^R_{ss}
\end{pmatrix}
\end{align}
uses $\star$ to denote the convolution in $\omega$ and $k$.
The corresponding Keldysh component reads
\begin{align}
\delta\mathcal{G}^K_{sd}=\mathcal{G}^R_{sd}\cdot\delta\Sigma^K_{sd}\cdot\mathcal{G}^A_{sd}\,,
\end{align}
with
\begin{align}
\delta\Sigma^K_{sd}=\frac{i}{2}g_E^2\begin{pmatrix}
\left(\delta G^K_{E_{22}}+2G^{K_0}_{E_{22}}\right)\star\delta\tilde{G}^K_{dd}&&\left(\delta G^K_{E_{21}}+2\left[G^R_{E_{12}}\right]^*\right)\star\delta\tilde{G}^K_{ds}\\
\left(\delta G^K_{E_{12}}-2G^R_{E_{12}}\right)\star\delta\tilde{G}^K_{sd}&&\delta G^K_{E_{11}}\star\delta\tilde{G}^K_{ss}
\end{pmatrix}\,.
\end{align}
\end{widetext}
\begin{figure}[t]
\begin{center}
\includegraphics[width=\columnwidth]{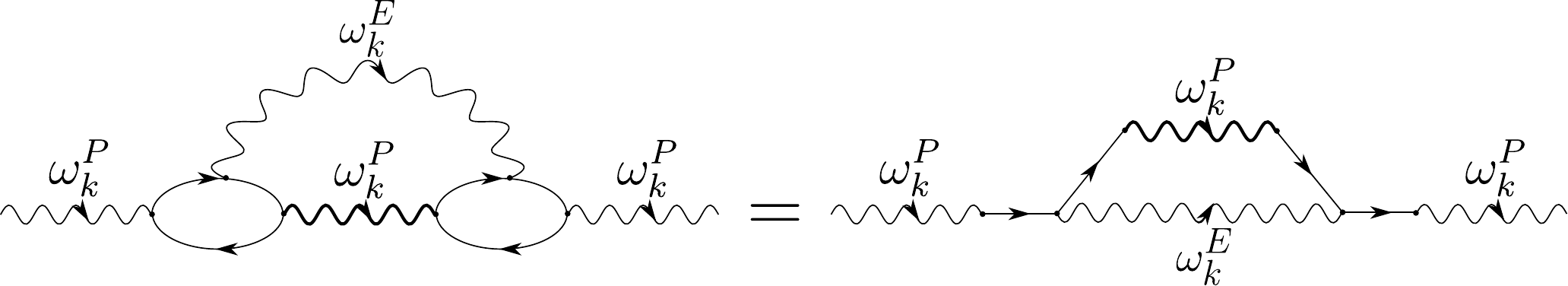}
\caption{Loop reduction procedure for the Feynman diagram in Fig.~\ref{fig:1overL}b), which formally delocalizes excited atoms. Note that during this procedure the probe photon coupling strength has to be modified by $g_P\to g_P\sqrt{2-2 n_V}$ to properly reflect the atom number density.}
\label{fig:loop_red_Mcar}
\end{center}
\end{figure}
Note that $\Sigma^{R/K}_{sd}$ are allowed to depend on the quasi-momentum $k$, since, due to the photon admixture with momentum transfer, they effectively no longer describe a completely stationary atom. This effective delocalization of the atoms is not an actual physical process, but rather a mathematical trick to accommodate the loop reduction procedure shown in Fig.~\ref{fig:loop_red_Mcar}. In fact, we construct $\tilde{\mathcal{G}}_{sd}$ via
\begin{align}
\tilde{\mathcal{G}}^R_{sd}=\left(\left[\mathcal{G}^R_{sd}\right]^{-1}-\begin{pmatrix}\Sigma^R_s&&0\\0&&0\end{pmatrix}\right)^{-1}
\end{align}
with
\begin{align}
\Sigma^R_s=\frac{\Omega^2}{\omega-g_P^2\left[(1-n_V)G_P^R+\frac{1}{2}G^R_g\star_\omega \delta G^K_P\right]+i\gamma_e/2}\,,
\end{align}
where $\star_\omega$ indicates a particle-hole convolution in frequency only, i.e.
\begin{align}
f\star_\omega h =\int_{-\infty}^\infty\frac{d\omega'}{2\pi}f(\omega')h(\omega'+\omega)\,.
\end{align} 
Furthermore, the related
\begin{align}
\delta\tilde{\mathcal{G}}^K_{sd}=\tilde{\mathcal{G}}^R_{sd}\cdot\left(\delta\Sigma^K_{sd}+\begin{pmatrix}\delta\Sigma^K_s&&0\\0&&0\end{pmatrix}\right)\cdot\tilde{\mathcal{G}}^R_{sd}\,,
\end{align}
where
\begin{align}
\delta\Sigma^K_s=\frac{g_P^2}{\Omega^2}(1-n_V)\delta G^K_P|\Sigma^R_s|^2\,
\end{align}
correctly includes all repeated scattering processes of a probe photon into a probe and an exchange photon. The loop reduction procedure therefore allows for a very cost-efficient inclusion of pairing effects between probe and exchange photons, that would otherwise require a self-consistent treatment of the corresponding T-matrix.\\
The Dyson equation for the probe photon propagator takes almost exactly the same form as it does for the non-interacting EIT:
\begin{align}
\begin{split}
G^R_P&=\left(\omega-\Delta_P(k)-g_P^2(1-n_V)G^R_e+i\kappa_P/2\right)^{-1}\\
\delta G^K_P&=\left(\Omega^2 g_P^2(2-n_V)\delta G^K_{ss}|G^R_e|^2-2i\kappa_s(\omega)\right)|G^R_P|^2\,,
\end{split}
\end{align}
with the only difference being hidden in the more elaborate form of $G_{ss}$.\\
In order to close the set of coupled equations one has to find the full exchange photon propagator, which again has two self-energy contributions as in Eq.~\eqref{eq:GERtwo}.
The first one
\begin{align}
\Sigma^{R_1}_E(\omega)=\frac{i}{2}g_E^2 \begin{pmatrix}
\left[\delta G^K_{\fsl{s}}\star_\omega G^R_{d}\right](\omega)&&0\\
0&&\left(\delta G^K_{\fsl{s}}\star_\omega G^R_{d}\right)(-\omega)
\end{pmatrix}
\end{align}
is already known from Sec.~\ref{sec:Gerry} and in the present notation involves the propagator $\delta G^K_{\fsl{s}}$ given by
\begin{widetext}
\begin{align}\label{eq:GKslash}
\delta G^K_{\fsl{s}}(\omega)=\frac{g_P^2}{\Omega^2}(1-n_V)\kappa_s(\omega)\left|\frac{\Sigma^R_{\fsl{s}}(\omega)}{\omega-\Delta_s-\Sigma^R_{\fsl{s}}(\omega)+i\epsilon/2}\right|^2\int\frac{dk}{2\pi}\delta G^K_P(\omega,k).
\end{align}
For the strict interpretation of the non-linear Feynman rules we use
\begin{align}
\Sigma^R_{\fsl{s}}(\omega)=\frac{\Omega^2}{\omega+i\gamma_e/2-g_P^2\int\frac{dk}{2\pi}\left[(1-n_V)G^R_P+\kappa_s(\omega)G^R_g\star_\omega \delta G^K_P\right]}
\end{align}
and $G_d^R=G_{d_0}^R$. One therefore recovers exactly the same expression for $\Sigma^R_{E_1}$ as in the formulation at leading order. In case of the lenient Feynman rules the denominator in the absolute value in Eq.~\eqref{eq:GKslash} is to be replaced by
\begin{align}
\omega-\Delta_s-\Sigma^R_{\fsl{s}}(\omega)+i\epsilon/2+\frac{\left(\Omega_s^\text{eff}\right)^2}{\omega-\Delta_d-\Delta_s+i\gamma_d/2}+\int_{-\pi}^{\pi}\frac{dk}{2\pi}\Sigma_{ss}^R(\omega,k)\,,
\end{align}
where $\Sigma_{ss}^R$ is the $11$-component of $\Sigma_{sd}^R$.
At the same time $G_d^R$ is given by $\tilde{G}_{dd}^R$.
This leaves $\Sigma_E^{R_2}$, which takes the same form as in Sec.~\ref{sec:Gerry+ECar}:
\begin{align}
\begin{split}
\Sigma^{R_2}_E(\omega,k)&=\frac{i}{2}g_P^4 g_E^2\Omega^4\left(\Omega_s^\text{eff}\right)^2(1-n_V)^2\int\frac{d\omega'}{2\pi}\frac{dp}{2\pi}\left|G^R_e(\omega')G_s^R(\omega')\right|^2\delta G^K_P(\omega',p)\\\times&\left[\left[G^R_s(\omega+\omega')G^R_e(\omega+\omega')\right]^2 G^R_P(\omega+\omega',p+k)\begin{pmatrix}
\left[G^R_d(\omega+\omega')\right]^2&& G^A_d(\omega')G^R_d(\omega+\omega')\\
G^R_d(\omega')G^R_d(\omega+\omega')&&\left|G^R_d(\omega')\right|^2
\end{pmatrix}\right.\\&\left.+\left[G^A_e(\omega'-\omega)G^A_s(\omega'-\omega)\right]^2G^A_P(\omega'-\omega,p-k)\begin{pmatrix}
\left|G^R_d(\omega')\right|^2&&G^A_d(\omega'-\omega)G^A_d(\omega')\\
G^A_d(\omega'-\omega)G^R_d(\omega')&&\left[G^A_d(\omega'-\omega)\right]^2
\end{pmatrix}\right]\,.
\end{split}
\end{align}
Evaluating the same diagrams for the Keldysh component, one obtains the last two pieces of the puzzle:
\begin{align}
\delta\Sigma^{K_1}_E=0\,,
\end{align}
as before, and
\begin{align}\label{eq:loop_end}
\begin{split}
\delta\Sigma^{K_2}_E(\omega,k)=&\frac{i}{2}g_P^4 g_E^2\Omega^4\left(\Omega_s^\text{eff}\right)^2(1-n_V)^2\int\frac{d\omega'}{2\pi}\frac{dp}{2\pi}\left|G^R_e(\omega')G^R_e(\omega'+\omega)G_s^R(\omega')G_s^R(\omega'+\omega)\right|^2\\&\times\delta G^K_P(\omega',p)\delta G^K_P(\omega'+\omega,p+k)\begin{pmatrix}
\left|G^R_d(\omega'+\omega)\right|&& G^A_d(\omega')G^A_d(\omega+\omega')\\
G^R_d(\omega')G^R_d(\omega+\omega')&&\left|G^R_d(\omega')\right|^2
\end{pmatrix}\\&-\!2\!\begin{pmatrix}
2i\left[\Im{\tilde{\Sigma}_E^R(\omega,k)}\right]_{11}&&\!\!\!\!\!\!\left[\tilde{\Sigma}_E^R(\omega,k)+ \tilde{\Sigma}_E^R(-\omega,-k)\right]_{12}\!\\ \!-\!\left[\tilde{\Sigma}_E^R(\omega,k)+ \tilde{\Sigma}_E^R(-\omega,-k)\right]^*_{12}&&\!\!\!\!\!\!2i\left[\Im{\tilde{\Sigma}_E^R(-\omega,-k)}\right]_{11}
\end{pmatrix}.
\end{split}
\end{align}
\end{widetext}
with the same choices for $G_d^R$ in $\Sigma_E^{R/K_2}$ as in $\Sigma_E^{R_1}$. Furthermore, we use $\tilde{\Sigma}_E^R$ as a shorthand notation for the second term in the sum of $\Sigma_E^{R_2}$.\\
As always, throughout this entire theory the bare laser coupling $\Omega_s$ between states $|s\rangle$ and $|d\rangle$ has been replaced by $\Omega_s^\text{eff}=\Omega_s|1+\eta|$, where $\eta$ given by \eqref{eq:chi}
describes the modified conversion rate between $|s\rangle$ and $|d\rangle$ due to the presence of other polaritons, as in section~\ref{sec:Gerry}.
\begin{figure}[htp]
\begin{center}
\includegraphics[width=\columnwidth]{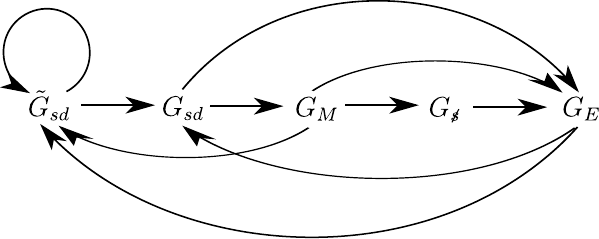}
\caption{Dependency structure and ordering of updates for the self-consistent solution of the Dyson equations \eqref{eq:loop_begin} through \eqref{eq:loop_end} with \eqref{eq:chi} iteratively updating the effective coupling $\Omega_s^\text{eff}$.}
\label{fig:update}
\end{center}
\end{figure}
Contrary to the previous renditions of the self-consistent structure, with the inclusion of $1/L_E$ effects, scattering of probe photons into exchange photons becomes a possibility. Therefore, self-consistence is no longer simply a question of finding the right parameter $\eta$, but actually involves the full frequency and momentum dependent Green's functions $G_E^{R/K}$. As such, the numerical implementation has to find the solution in an iterative manner. Since there is no direct dependence on $\Omega_s$, one can however fix $\Omega_s^\text{eff}$, initialize all Green's functions as bare ones and iterate equations \eqref{eq:loop_begin} through \eqref{eq:loop_end} together with \eqref{eq:chi} until $\eta$ no longer changes, see Fig.~\ref{fig:update}. This once again means that the final value of $\Omega_s$ corresponding to the solution is not known a priori and has to be searched for iteratively, unless the entire phase diagram is calculated. The main advantage of this method lies again in the enhanced convergence that is unaffected by the presence of any phase transition.\\
As we have already discussed in Sec.~\ref{sec:Gerry+ECar}, with the use of the absolute value in the definition of $\Omega_s^\text{eff}$ in the anomalous Green's functions $G_E$ no longer transforms correctly under a global $U(1)$ gauge transformation. Previously this was not a problem for the evaluation of gauge invariant observables. Despite the iterative procedure the backaction at any stage of the iteration for any observable depends only on $|1+\eta|^2$ as is required by gauge invariance. The described self-consistent calculation thus finds the correct value of $|1+\eta|$ and therefore of all normal Green's functions. To also obtain the anomalous components the correct phase of $\eta$ simply has to be restored in the final result.

\bibliography{EIT_biblio}

\end{document}